\documentclass[iop]{emulateapj}

\def\kms{km~s$^{-1}$}

\def\kms{\mbox{km~s$^{-1}$}}
\def\cmc{cm$^{-3}$}
\def\cmq{cm$^{-2}$}

\def\deg{$^{\circ}$}
\def\Msun{\mbox{$M_\odot$}}

\def\Vlsr{$V_{\rm LSR}$}
\def\Vsys{$V_{\rm sys}$}

\def\Vcent{$v_{\rm cent}$}

\def\amm{NH$_3$}

\def\hcop{\mbox{HCO$^+$}}

\def\CO{$^{12}$CO}
\def\co{$^{12}$CO}
\def\tCO{$^{13}$CO}
\def\CeO{C$^{18}$O}
\def\NtwoH{N$_2$H$^+$}
\def\HtCOp{H$^{13}$CO$^+$}

\def\dVres{\mbox{$\Delta v_{\rm res}$}}

\def\dVint{\mbox{$\Delta v({\small \tau\rightarrow 0})$}}

\def\V0{\mbox{$V_{\rm 0}$}}

\def\sigmanth{\mbox{$\sigma_\mathrm{nth}$}}

\def\Tkin{\mbox{$T_{\rm kin}$}}

\def\Tr21{\mbox{$T_{\rm r,21}$}}
\def\Tmb{\mbox{$T_{\rm mb}$}}
\def\Tmbcorr{\mbox{$T_{\rm mb}^{\rm corr}$}}
\def\Tmbcorri{\mbox{$T_{{\rm mb},{\it i}}^{\rm corr}$}}
\def\Tmbcorrms{\mbox{$\Delta T_\mathrm{mb}^\mathrm{corr}$}}

\def\Tmb{\mbox{$T_{\rm mb}$}}

\def\Tex{\mbox{$T_{\rm ex}$}}
\def\meanTex{\mbox{$\langle T_{\rm ex}\rangle$}}
\def\meanTex{\mbox{$\langle T_{\rm ex}\rangle$}}

\def\Tbg{\mbox{$T_{\rm bg}$}}

\def\Reff{\mbox{$R_\mathrm{eff}$}}

\def\sgm{$\sigma$}

\def\MLTE{\mbox{$M_{\rm LTE}$}}

\def\gf{\mbox{GF\,9-2}}

\def\thetahpbw{$\theta_{\rm HPBW}$}

\def\RadpSimSoltn{$\rho (r)\propto r^{-2}$}

\def\lesssim{\mathrel{\hbox{\rlap{\hbox{\lower4pt\hbox{$\sim$}}}\hbox{$<$}}}}
\def\gtrsim{\mathrel{\hbox{\rlap{\hbox{\lower4pt\hbox{$\sim$}}}\hbox{$>$}}}}

\slugcomment{Accepted by ApJ Main Journal on 2014 July 16}

\shorttitle{A Dynamically Collapsing Core and Precursor of a Core in a Filament}
\shortauthors{R. S. Furuya et al.}

\begin{document}

\title{A Dynamically Collapsing Core and a Precursor of a Core in a Filament Supported by Turbulent and Magnetic Pressures}

\author{Ray S. FURUYA\altaffilmark{1}}
\affil{Center for General Education, the University of Tokushima}
\email{rsf@tokushima-u.ac.jp}

\author{Yoshimi KITAMURA\altaffilmark{2}}
\affil{Institute of Space and Astronautical Science, Japan Aerospace Exploration Agency}
\email{kitamura@isas.jaxa.jp}

\and

\author{Hiroko SHINNAGA\altaffilmark{3}}
\affil{Chile Observatory, National Astronomical Observatory of Japan}
\email{hiroko.shinnaga@nao.ac.jp}

\altaffiltext{1}{Minami Jousanjima-Machi 1-1, Tokushima, Tokushima 770-8502, Japan}
\altaffiltext{2}{Yoshinodai 3-1-1, Sagamihara, Kanagawa 229-8510, Japan}
\altaffiltext{3}{Osawa 2-21-1, Mitaka, Tokyo 181-8588, Japan}

\begin{abstract}
To study physical properties of the natal filament gas
around the cloud core harboring an exceptionally young low-mass 
protostar \gf, 
we carried out $J=1-0$ line observations of \CO, \tCO, and \CeO\ molecules
using the Nobeyama 45\,m telescope.
The mapping area covers $\sim 1/5$ of the whole filament.
Our \tCO\ and \CeO\ maps clearly demonstrate
that the core formed at the local density maxima of the filament, and
the internal motions of the filament gas are
totally governed by turbulence with Mach number of $\sim$2.
We estimated the scale height of the
filament to be $H\,=\,0.3\sim 0.7$\,pc, yielding the central density of 
$n_\mathrm{c}\,=\,700\sim 4200$ \cmc.
Our analysis adopting an isothermal cylinder model shows that 
the filament is supported by 
the turbulent and magnetic pressures against the radial and axial collapse
due to self-gravity.
Since both the dissipation time scales of the turbulence and the transverse 
magnetic fields can be comparable to the free-fall time of the filament gas of 
$10^6$ years,
we conclude that the local decay of the supersonic turbulence made
the filament gas locally unstable, hence making the core collapse.
Furthermore, we newly detected a gas condensation with velocity width enhancement
to $\sim$0.3 pc south-west of the \gf\ core.
The condensation has a radius of $\sim$0.15\,pc and an LTE mass of $\sim$5 \Msun.
Its internal motion is turbulent with Mach number of $\sim$3, 
suggestive of a gravitationally unbound state.
Considering the uncertainties in our estimates, 
however, we propose that the condensation is a precursor of a cloud core
which would have been produced by the collision of the two gas components
identified in the filament.
\end{abstract}


\section{Introduction}
\label{s:intro}
Filamentary molecular clouds, filaments, 
are now considered to be one of the evolutionary stages for
the interstellar medium to evolve from molecular clouds to stars.
Analyzing dark globular filaments identified in optical images,
\citet{Schneider79} pointed out that a filament would
fragment into equally spaced condensations of gas and dust due to its gravity.
Indeed molecular line observations in the 1980s showed
that filaments contain such condensations, dense molecular cloud cores, 
which are now known to be the immediate sites of star formation.
This hierarchical structure is typically seen in nearby low-mass
star forming regions such as the Taurus molecular cloud \citep[e.g.,][and references therein]{Palla02}.
Recently our knowledge about filaments is
significantly improved by systematic
far-infrared (FIR) imaging survey of the thermal emission from the interstellar dust
using {\it Herschel} Space Observatory \citep[][and references therein]{Andre10}.
FIR images taken with {\it Herschel} demonstrated that molecular clouds 
have ubiquitous networks of filaments which are highly likely 
produced through the interaction among gravity, interstellar turbulence and magnetic field.
These observations clearly suggest that in general dense cores originate in 
the fragmentation process of filaments.
Another important results from the {\it Herschel} observations is that filaments are
omnipresent even in non-star forming cloud
complexes \citep[e.g.,][]{Men10}.\par

On the other hand, theoretical studies have long discussed that molecular clouds 
collapse into sheet-like clouds which are formed through a compression process
such as cloud-cloud collision, and that the sheet-like clouds fragment to filamentary
clouds which will form dense cloud cores \citep[e.g., ][]{Miyama84}.
Considering balance between self-gravity and
pressure gradient in radial direction of a cylinder,
\citet{Stodolkiewicz63} and \citet{Ostriker64} found
that a filament radially collapses
if the line mass (the mass per unit length) of the filament exceeds the critical value of 
$2c^2_\mathrm{s}/G$ where $c_\mathrm{s}$ 
is the isothermal sound speed and $G$ is the gravitational constant.
Stability of a cylinder in equilibrium against axial perturbations
was studied for the isothermal incompressible cylinder composed
of polytropic gas by \citet{Chandrasekhar53}:
they found critical wave numbers against the axial fragmentation, 
but such an unstable cylinder can be
stabilized by the axial magnetic fields of 
$(0, \,0, \,B_\mathrm{z})$ in the cylindrical coordinate.
\citet{Stodolkiewicz63} studied how stability of the compressible gas depends
on the magnetic fields: 
the critical wavelength becomes longer for $(0, \,B_\mathrm{\phi}, \,0)$,
whereas does short for $(0, \,0, \,B_\mathrm{z})$.
In contrast, \citet{Nagasawa87} 
showed that critical wavelengths do not change with a uniform $(0, \,0, \,B_\mathrm{z})$,
but the growth rates of the unstable modes are suppressed,
which differs from the conclusion drawn from the Stodolkiewicz's calculations.
As summarized in \citet{Larson85},
all the studies showed that a filament breaks into ``clumps" with a separation of 
about 4 times the diameter of the cylinder.
\citet{Inutsuka97} 
investigated the axial fragmentation and radial collapse of a cylinder in more detail; 
they showed that merging and clustering of newly formed ``clumps" occur 
within a time scale of the
fragmentation process of the natal filaments.\par

Although details remain a matter of debate, 
several unstable modes for the filament fragmentation are
proposed by modern numerical simulations 
\citep[see reviews by e.g.,][and references therein]{MacLow04, Andre09}.
These simulations may be categorized by the roles of 
magnetic fields and turbulence being considered.
In principal, a complex interplay between magnetic fields plus turbulence and 
self-gravity of the cloud are believed to control the duration of core formation.
One scenario is the ``fast mode" with weak magnetic field 
\citep[e.g.,][]{Padoan99,Padoan01,Hartmann01,Klessen00,Klessen05, Krumholz05, Krumholz07}.
The opposite is the ``slow mode" with strong magnetic field
\citep[e.g.,][]{Allen00, Elmegreen07, Nakamura05, Nakamura08}.
The difference between the ``fast" and ``slow" modes is whether or not a core
forms within free-fall time of the natal gas.
All the 3D simulations predict formation of network of filaments, 
but their complexity in velocity fields differs significantly \citep[see review by][]{Andre09}.
In general, the ``slow mode" expects quiescent ambient 
velocity field controlled by magnetic field, whereas the ``fast mode" predicts 
large scale supersonic velocity field.
In other words, physical properties of the low density gas
in the filament are considered to determine the formation mechanism 
and evolution of the dense cores.\par

To link physical properties of such low density gas to our
knowledge about collapse of a cloud core, 
we performed a detailed study of 
the dense cloud core \gf\ which is also known as 
L1082C \citep[e.g.,][and references therein]{Bontemps96, Caselli02}.
This core contains
an extremely young low-mass protostar \citep[][hereafter paper I]{rsf06}
whose circumstellar materials are probably responsible for the
IRAS point source PSC\,20503+6006 \citep[e.g.,][]{Ciardi98}.
The core has been identified by several molecular lines such as 
CS (2--1) \citep{Ciardi00}, 
\NtwoH\ (1--0) (Caselli et al. 2002; paper I), \HtCOp\ (1--0), \amm (1,1), and CCS $4_3-3_2$ lines (paper I).
The core is located in the GF\,9 filament ($d\,=$ 200 pc) which
is the ninth filament cataloged in \citet{Schneider79}.
Using the optical image \citet{Schneider79} estimated a total
length of the filament to be 1.25\degr, corresponding to 4.4\,pc,
although the filament curves as a portion of a shell.
The filament contains almost equally spaced seven dense cores
identified in the \amm (1,1) lines
\citep[][hereafter paper II]{rsf08}.
The physical properties of the filament were subsequently studied by
near-infrared (NIR) extinction \citep{Ciardi00}, and 
optical and NIR polarization observations \citep{PB06}.
In particular, \citet{PB06} revealed that there exists 
a well aligned large scale magnetic field, 
and claimed that the field is almost perpendicular to the axis of the filament
as the first order approximation.\par

The protostar harbored in the \gf\ core has not developed an extensive molecular outflow
(Bontemps et al. 1996; paper I).
This fact yields a rare opportunity to investigate 
core collapse conditions free from the disturbance by the outflow.
Our previous studies towards the core using the Nobeyama 45\,m telescope,
CSO 10.4\,m telescope and the OVRO mm-array showed
that the \gf\ core has a radial density profile of 
\RadpSimSoltn\ (paper I).
Furthermore we detected blueskewed profiles in 
optically thick lines of \hcop\ (3--2), (1--0), and HCN (1--0),
suggestive of gas infall motions all over the core 
\citep[][hereafter paper III]{rsf09}.
Modeling the infall spectra, 
we pointed out that the core has undergone its gravitational 
collapse for $\sim 2\times 10^5$ years from an initially unstable state in paper III.
This is because the observed radial density profile was between
those predicted in the runaway collapse scenario 
(Larson 1969; Penston 1969; Hunter 1977; the LPH solution) 
and the quasi-static inside-out collapse scenario \citep{Shu77},
and because the observed velocity field inside the core has a reasonable
consistency with that expected from the former scenario.
However, it was impossible to investigate what had triggered the dynamical
collapse of the core based on the data tracing only high density molecular gas
[$10^5 \lesssim n(\mathrm{H_2})$/\cmc\ $\lesssim 10^6$].
Clearly it required to unveil a close link between the initial conditions
of the collapse and the physical properties of the low density ambient gas in the filament.
We therefore carried out wide-field spectroscopic
observations of the low density 
[$10^3 \lesssim n(\mathrm{H_2})$/\cmc\ $\lesssim 10^4$] gas in the filament
using the Nobeyama 45\,m telescope.\par

Contrary to the previous papers which dealt with the dense core gas, 
this paper presents a detailed observational
study of the low density natal filament gas surrounding the dense core \gf.
The organization of this paper is as follows:
\S2 describes the observations; 
\S3 summarizes the results which were directly derived from the obtained spectra and
maps of the CO emission; 
\S4 describes our analysis of the excitation conditions of the CO lines,
the velocity structure of the filament gas and the calculations of column density; 
\S5 discusses the physical properties of the natal filament gas 
and a scenario of dense core formation; and
\S6 gives a summary of this work.
Finally, details of our analysis presented in \S4 as well as the
associated error calculations are described in Appendix.

\begin{figure}
\begin{center}
\includegraphics[angle=0,scale=.34]{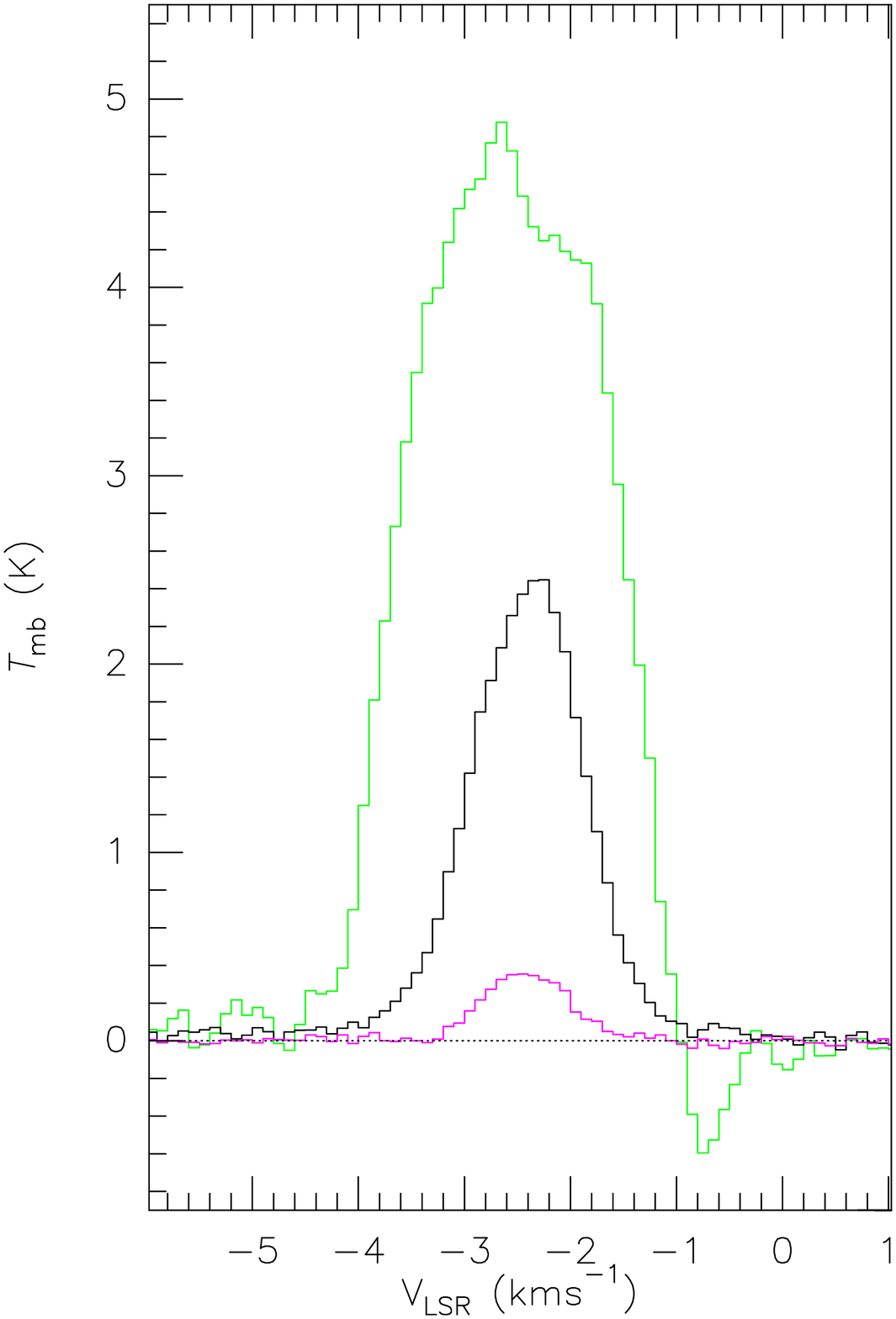}
\caption{Spectra of the $J=1-0$ lines of 
\co\ (green), \tCO\ (black), and \CeO\ (magenta) averaged 
over the whole region observed
in the GF\,9 filament with a beam size (HPBW) of 24\arcsec.
The vertical scale of the plot is the main-beam brightness temperature (\Tmb) 
in K unit.
The velocity resolutions of the spectra are 0.1 \kms.
The RMS noise levels are
760 mK for \co,
360 mK for \tCO, and 
130 mK for \CeO.
The systemic velocity of the \gf\ dense cloud core is 
\Vlsr\ $=\,-2.48$ \kms\ (paper I).
The negative dip seen in the \co\ spectrum around \Vlsr\
$\sim -0.8$ \kms\ is an artifact caused by 
the presence of emission at the off-position (see \S\ref{s:obs}).
\label{fig:sp}}
\end{center}
\end{figure}

\section{Observations}
\label{s:obs}

Using the Nobeyama Radio Observatory 
(NRO)\footnote{Nobeyama Radio Observatory 
is a branch of the National Astronomical Observatory of Japan, 
National Institutes of Natural Sciences.}
45\,m telescope,
we carried out On-The-Fly (OTF) mapping observations
\citep{Sawada08} of the $J=$1--0 transitions of 
$^{12}$C$^{16}$O [rest frequency ($\nu_{\rm rest}$) =\,115271.202\,MHz],
$^{13}$C$^{16}$O ($\nu_{\rm rest}$ =\,110201.353\,MHz), and
$^{12}$C$^{18}$O ($\nu_{\rm rest}$ =\,109782.173\,MHz)
towards a portion of the GF\,9 filament.
Our OTF mapping was performed over a square region of 
800\arcsec\ $\times$ 800\arcsec\ centered on the position of the 3\,mm continuum source 
(R.A. $=\,20^{\rm h}51^{\rm m}$29\farcs827,
Decl. $=$60\deg 18\arcmin 38\farcs06 in J2000; paper I).
Notice that the two adjacent dense cores of 
GF\,9-3 (angular separation of 910\arcsec, 
corresponding to 0.88\,pc to the east; paper II) and
GF\,9-1 (740\arcsec\ $=$ 0.71\,pc to the west) are not included
in the observed area.
The observations were done in an LST range of 17$^h$ -- 2$^h$ over 7 days
in 2008 March.
We used the 25 Beam Array Receiver System (BEARS), and 
configured auto-correlators (ACs) as a backend, 
yielding a velocity resolution
(\dVres ) of 0.039 \kms\ for the \tCO\ and \CeO\ lines in an 8\,MHz bandwidth mode.
For the \co\ line, we employed a coarse velocity resolution mode of
a 16\,MHz bandwidth, providing \dVres\ of 0.074 \kms.
At 112\,GHz, the mean beam size (\thetahpbw ) of the 25 beams was 
estimated to be 14\farcs9 by the observatory.
The mean main-beam efficiencies ($\eta_{\rm mb}$) for the 25 beams
were 40$\pm$2\% at 110\,GHz,
and 32$\pm$2\% at 115\,GHz.
All the spectra were calibrated by the standard chopper wheel method, 
and were converted into main-beam brightness temperature
(\Tmb ) by dividing by $\eta_{\rm mb}$.
The uncertainty in our intensity calibration is $\sim$\,15\%. 
The telescope pointing was checked every 1.2 hrs by observing the SiO $J=1-0$ $v=1$ and 
$v=2$ maser lines, and was found to be accurate to less than 3\arcsec.
The data reduction was done using the NOSTAR package \citep{Sawada08}.
We produced 3-dimensional (3D) data cubes for the lines
with an effective spatial resolution in FWHM of 24\arcsec, 
and an effective velocity resolution of 0.10 \kms.
For producing the 3D cubes, 
we adopted a pixel size of 12\arcsec,
and spatially smoothed the original data by convolving 
a Gaussian function with FWHM $=$ 18\farcs9 to obtain
the final effective resolution of 24\arcsec,
while we applied no smoothing along the velocity axis.

\begin{figure}
\begin{center}
\includegraphics[angle=0,width=4.2cm]{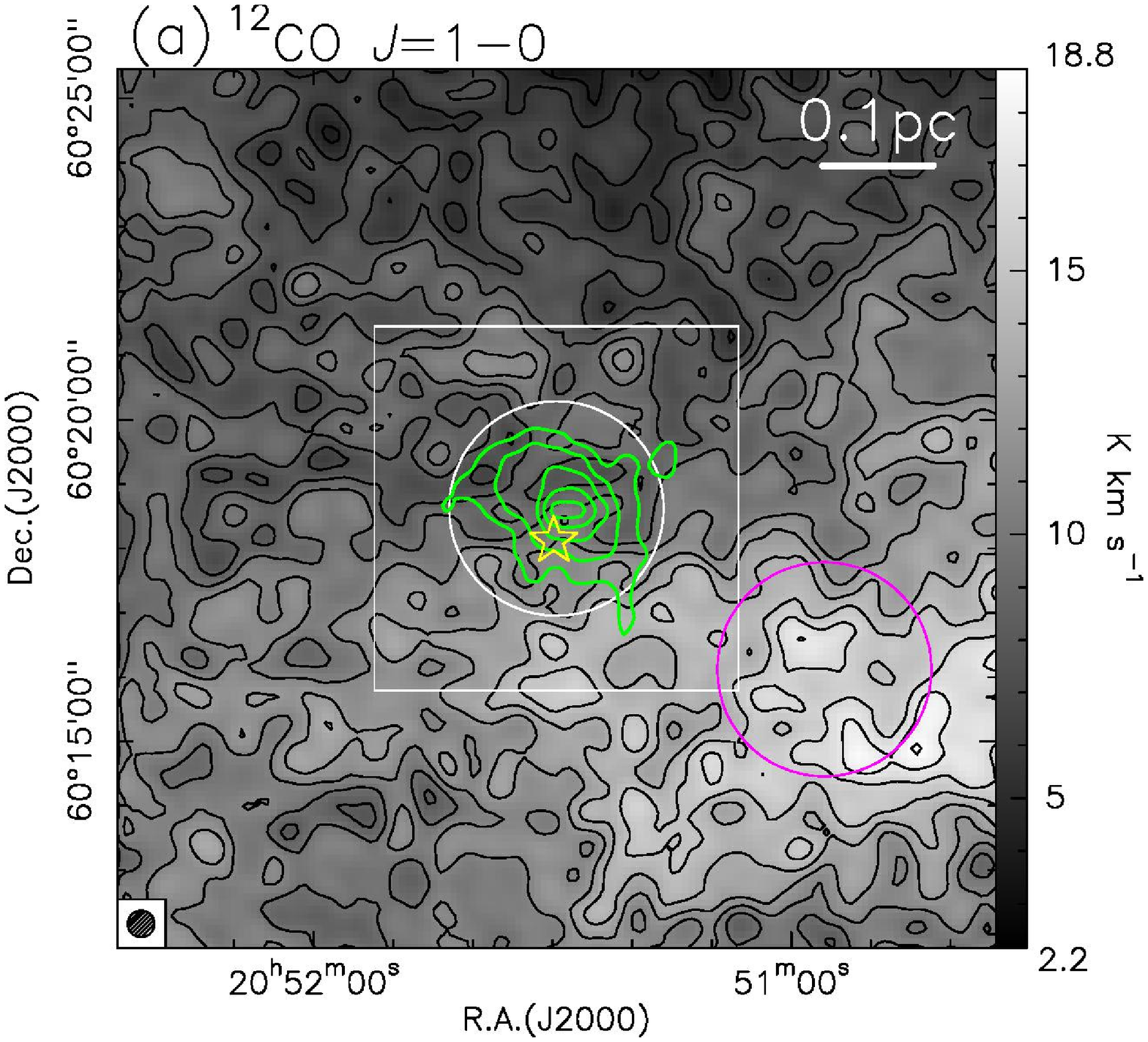} 
\includegraphics[angle=0,width=4.2cm]{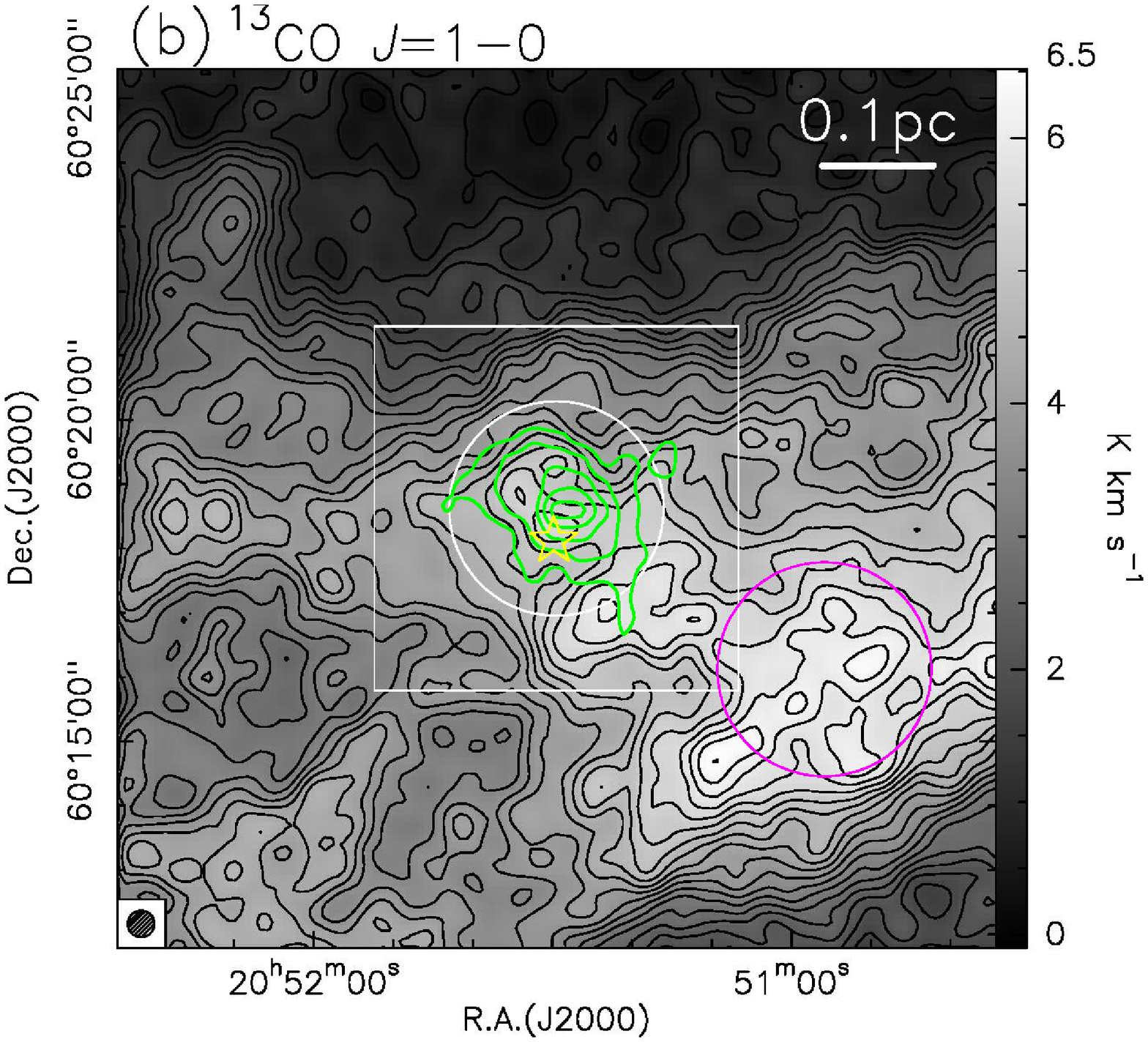} 
\includegraphics[angle=0,width=4.2cm]{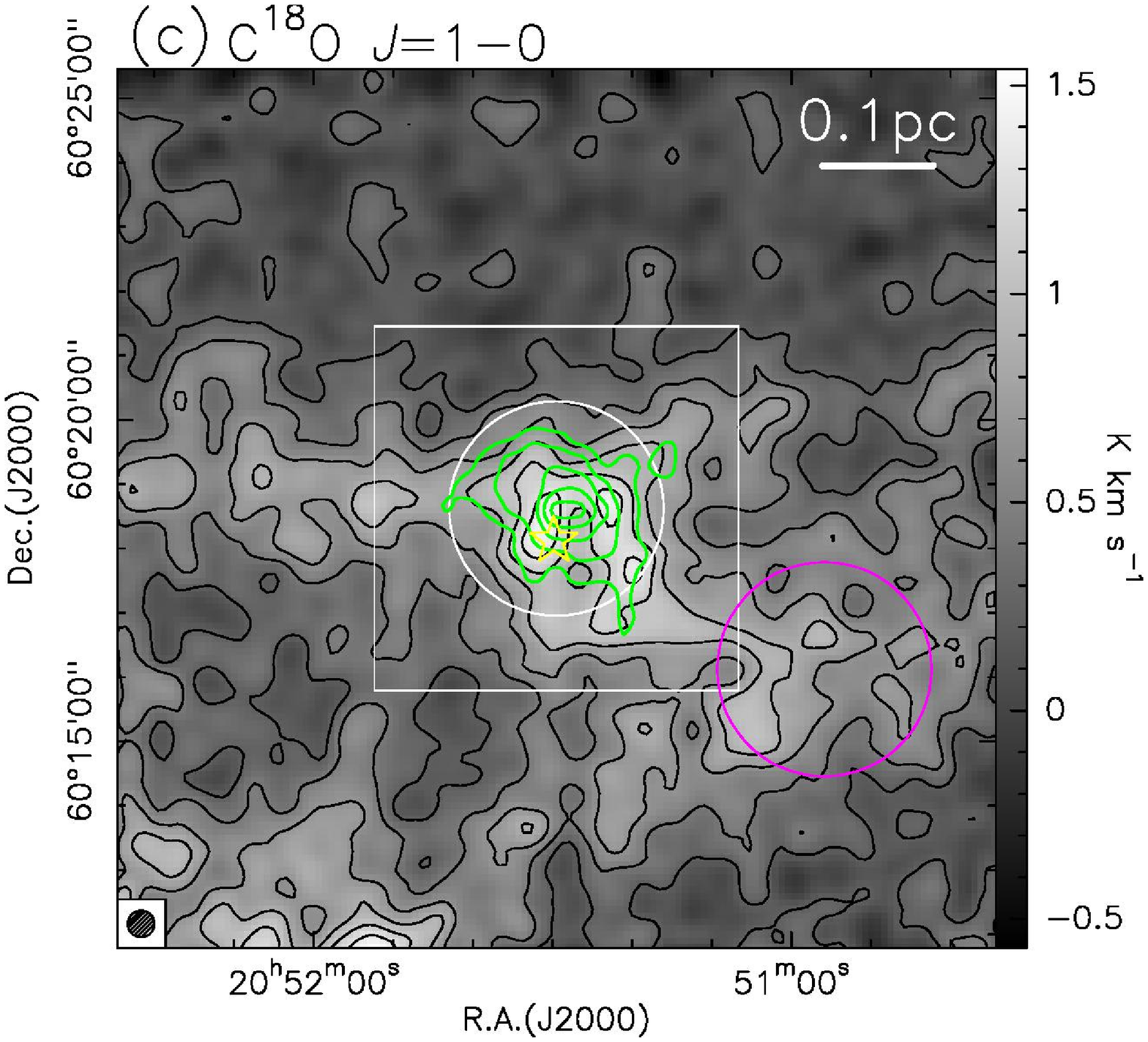} 
\includegraphics[angle=0,width=4.2cm]{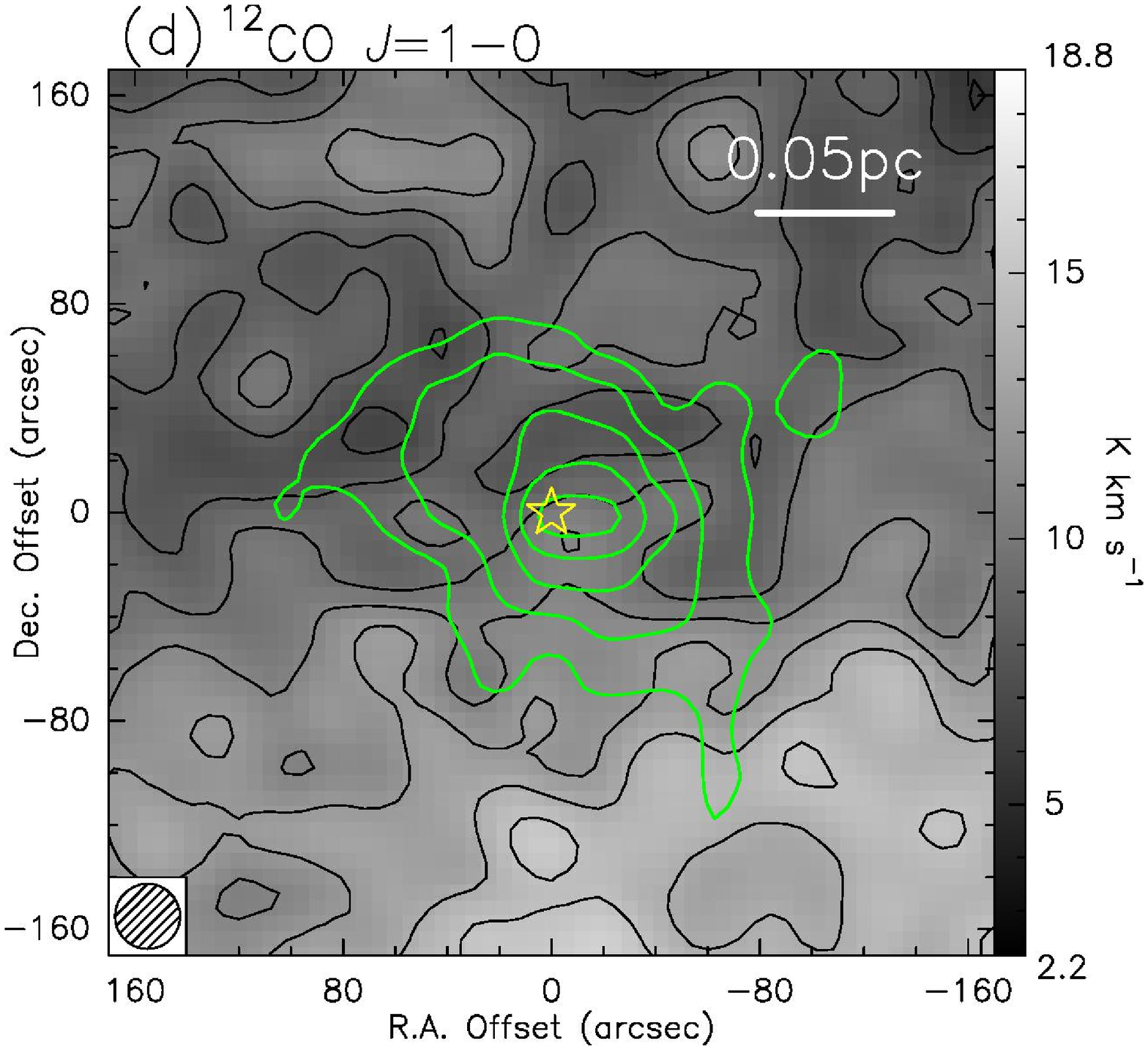} 
\includegraphics[angle=0,width=4.2cm]{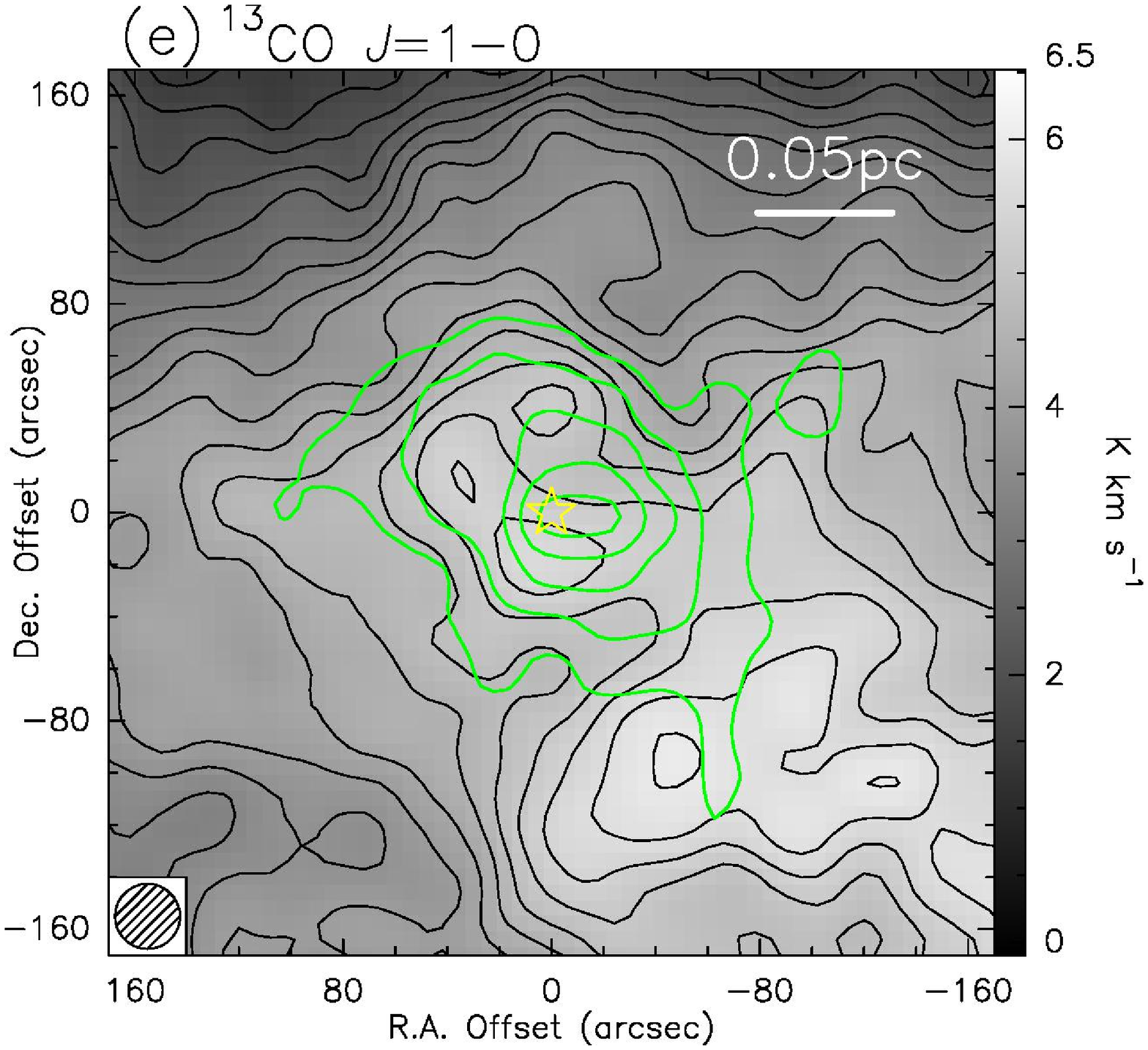}
\includegraphics[angle=0,width=4.2cm]{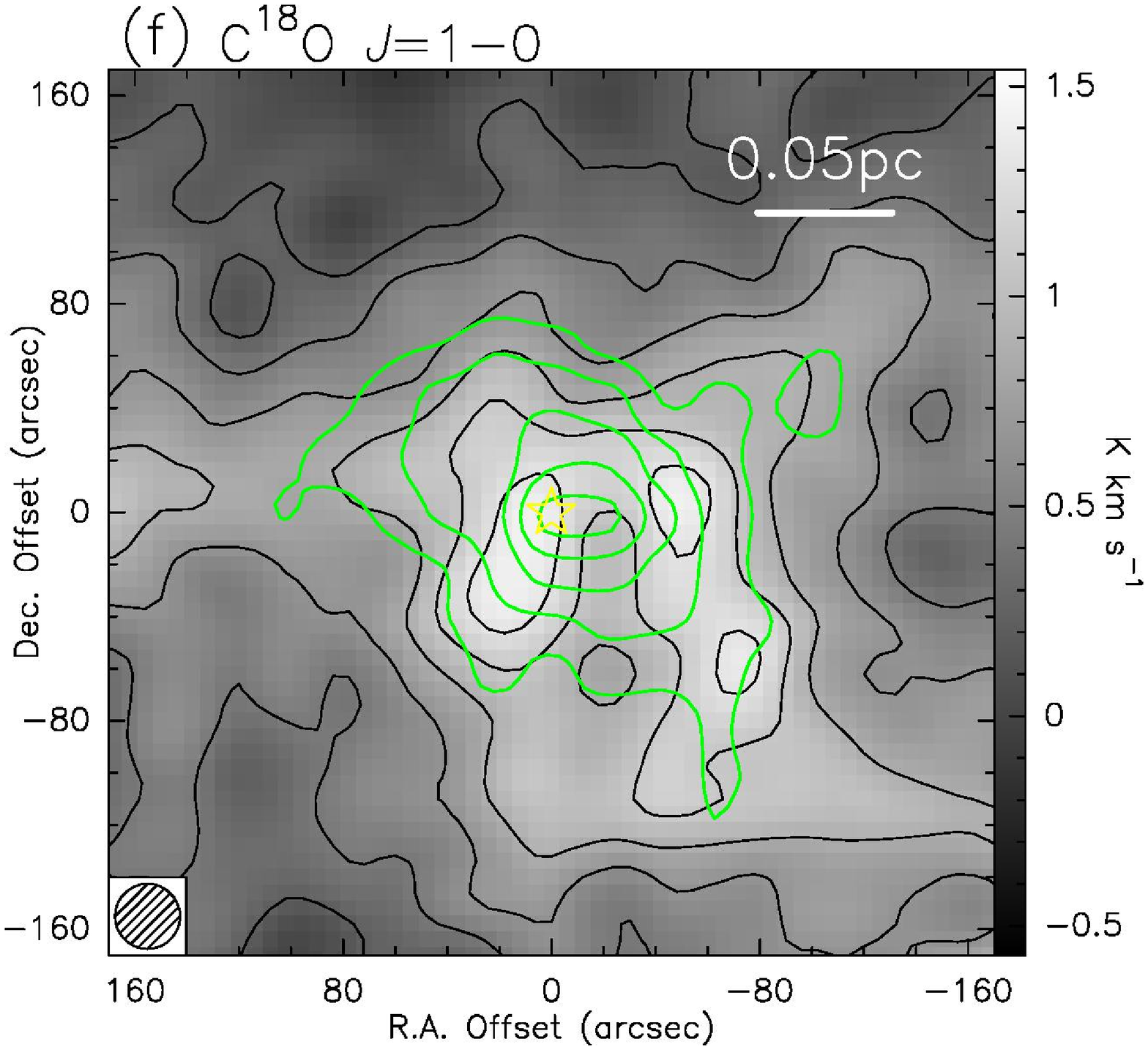}
\caption{\small 
Total integrated intensity maps (greyscale plus thin contour) of the
\co\ (1--0) (left),
\tCO\ (1--0) (middle) and
\CeO\ (1--0) (right) lines of the GF\,9 filamentary cloud
towards the dense cloud core \gf.
The green contours represent the total map of the 
\HtCOp\ (1--0) emission 
(paper I).
The upper panels show all the observed area (800\arcsec$\times$800\arcsec)
in the celestial coordinates, while the lower panels present the close-up views
of the central 320\arcsec$\times$320\arcsec\ region 
in the offset coordinates with respect to the 3\,mm continuum position
(R.A. $=\,20^{\rm h}51^{\rm m}$29\farcs827,
Decl. $=$60\deg 18\arcmin 38\farcs06 in J2000; paper I) designated by the yellow star.
The white boxes in the upper panels indicate the area shown in the lower panels.
The two circles in the upper panels are the areas where the emission were integrated to 
obtain the spectra shown in Figure \ref{fig:spcomparison}.
The hatched circle at the bottom left corner of each panel 
indicates the effective spatial resolution of 24\arcsec, and
the scale bars of 0.1 and 0.05\,pc are indicated at the top right corners 
in the upper and lower panels, respectively.
The \co, \tCO, and \CeO\ emissions are integrated 
over the LSR-velocity ranges of
$-4.2\leq$ \Vlsr/\kms\ $\leq -0.70$, 
$-3.9\leq$ \Vlsr/\kms\ $\leq -0.80$, and
$-3.2\leq$ \Vlsr/\kms\ $\leq -1.6$, respectively, 
using the original data cubes with the 
velocity resolution of 0.10 \kms (\S\ref{s:obs}).
The \co\ emission is detected all over the observed region, and
its intensity ranges
between 2.16 and 18.8 \,K \kms, corresponding to the 5.0$\sigma$ and 43.4$\sigma$ levels,
respectively.
We therefore plot the \CO\ contours starting from the 43$\sigma$ level
down to the 4$\sigma$ level 
at 3$\sigma$ intervals.
The contours for the \tCO\ and \CeO\ lines start from the 3$\sigma$ levels 
with the 3$\sigma$ intervals.
Here the RMS noise (the 1$\sigma$) levels of the images are
0.43 K$\cdot$\kms\ for \CO,
0.094 K$\cdot$\kms\ for \tCO, and 
0.072 K$\cdot$\kms\ for \CeO.
The noise level in each map was estimated from a noise map created by 
integrating emission-free channels
over the same velocity width as in the total emission map.
\label{fig:totmaps}}
\end{center}
\end{figure}

\section{Results}
\label{s:results}

Figure \ref{fig:sp} shows $J=1-0$ transition spectra of \co, \tCO, and 
\CeO\ molecules in \Tmb\ obtained by averaging all the data.
The \co\ line is the brightest with the peak \Tmb\ of $\sim$5\,K, and seems to 
have a self-absorption feature over the LSR-velocity (\Vlsr) range 
of $-2.6$ to $-1.8$ \kms.
The \tCO\ and \CeO\ lines show single-peaked profiles.
These peaks fall in the \Vlsr\ range where 
the \co\ line shows the self-absorption 
and the peak velocities agree with the systemic velocity (\Vsys) of the
\gf\ dense core  (\Vlsr\ $=-2.48$ \kms; paper I).
Averaging the \co\ and \tCO\ spectra in a circle with a 50\arcsec\ 
diameter towards the core,
we verified consistency with those taken with 
the FCRAO 14\,m telescope \citep{Ciardi98}
in terms of the peak intensities and the spectral shapes.\par

\begin{figure}
\begin{center}
\includegraphics[angle=0,width=9.4cm]{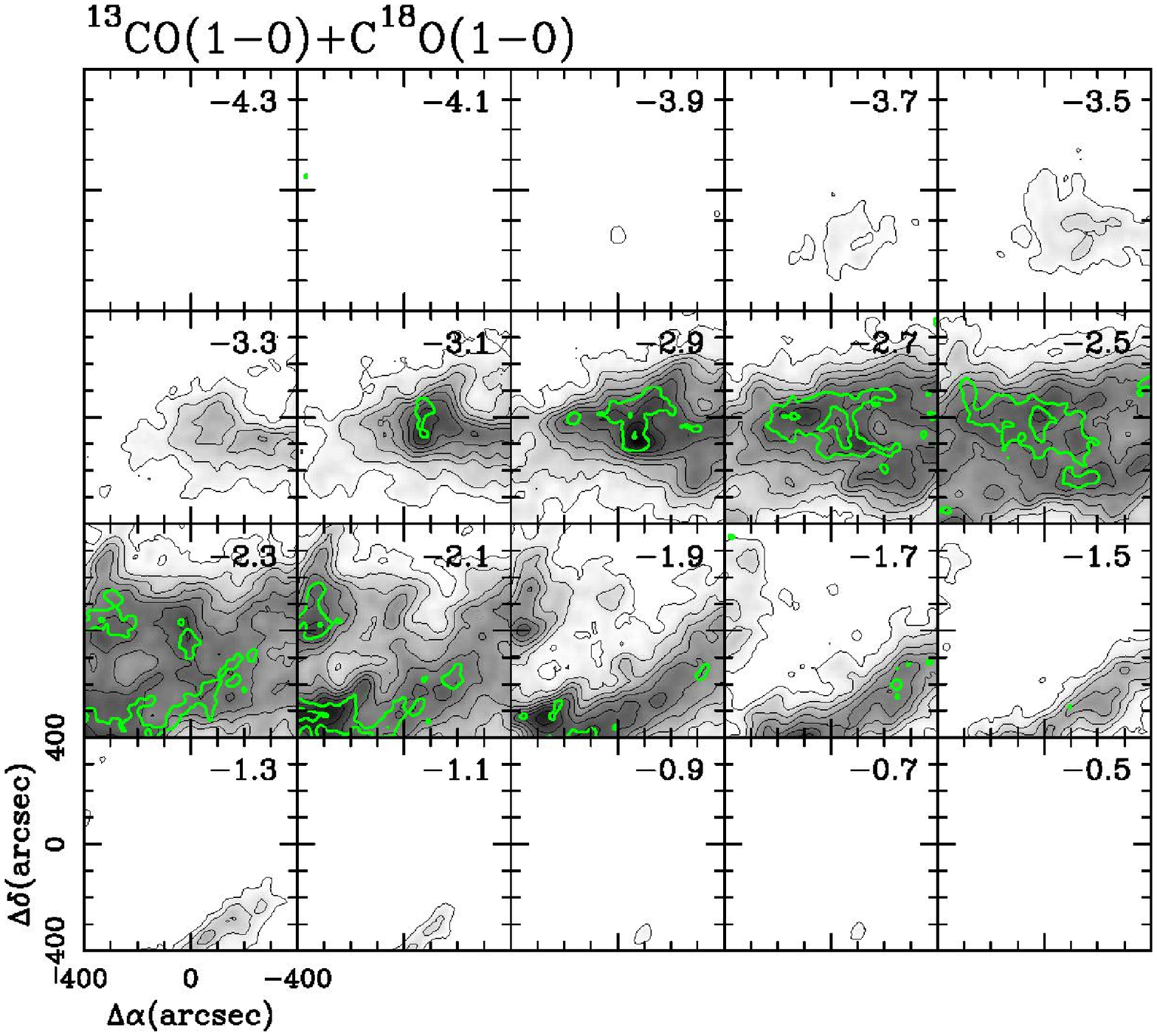}
\caption{Velocity channel maps of 
the \tCO\ (1--0) (greyscale plus black contours) and \CeO\ (1--0) (green contours) emission.
All the contours are plotted at 5\sgm\ intervals }
starting from the 5\sgm\ levels
where \sgm\ = 300 mK and 100 mK in \Tmb\ 
for the \tCO\ and  \CeO\ lines,  respectively.
Each channel map is averaged over a 0.2 \kms\ bin 
whose central LSR-velocity is shown in unit of \kms\ at the top right corner of each panel.
\label{fig:chmaps_13co}
\end{center}
\end{figure}

We present total integrated intensity maps of the three CO isotopologues
in Figure \ref{fig:totmaps} where the total map of a dense
gas tracer, \HtCOp\ (1--0) emission (paper I), 
is overlaid to assess spatial relations
with the \gf\ dense cloud core.
The \co\ emission is detected all over the mapped area,
and becomes intense to the south-west of the core.
On the other hand, the \tCO\ emission clearly represents 
a portion of the GF\,9 filamentary 
dark cloud \citep{Schneider79}, which is elongated along the east-west direction.
The \tCO\ emission becomes intense towards south-west,
as seen in the \co\ map.
In the central region, the \CeO\ emission 
is elongated along northeast-southwest (Figure \ref{fig:totmaps}f),
which agrees with the elongation of the core traced by 
e.g., the \HtCOp\ (1--0), \NtwoH\ (1--0) (paper I),
and \tCO\ (1--0) \citep{Ciardi98} lines.
These results clearly indicate that the \CeO\ line probes 
the gas with a medium density between those traced by the \tCO\ and \HtCOp\ lines.
Consequently, we have a seamless data set over a
volume density range of $10^2\lesssim n(\mathrm{H_2})/$\cmc $\lesssim 10^6$.\par

\begin{figure}
\begin{center}
\includegraphics[angle=0,width=10.2cm]{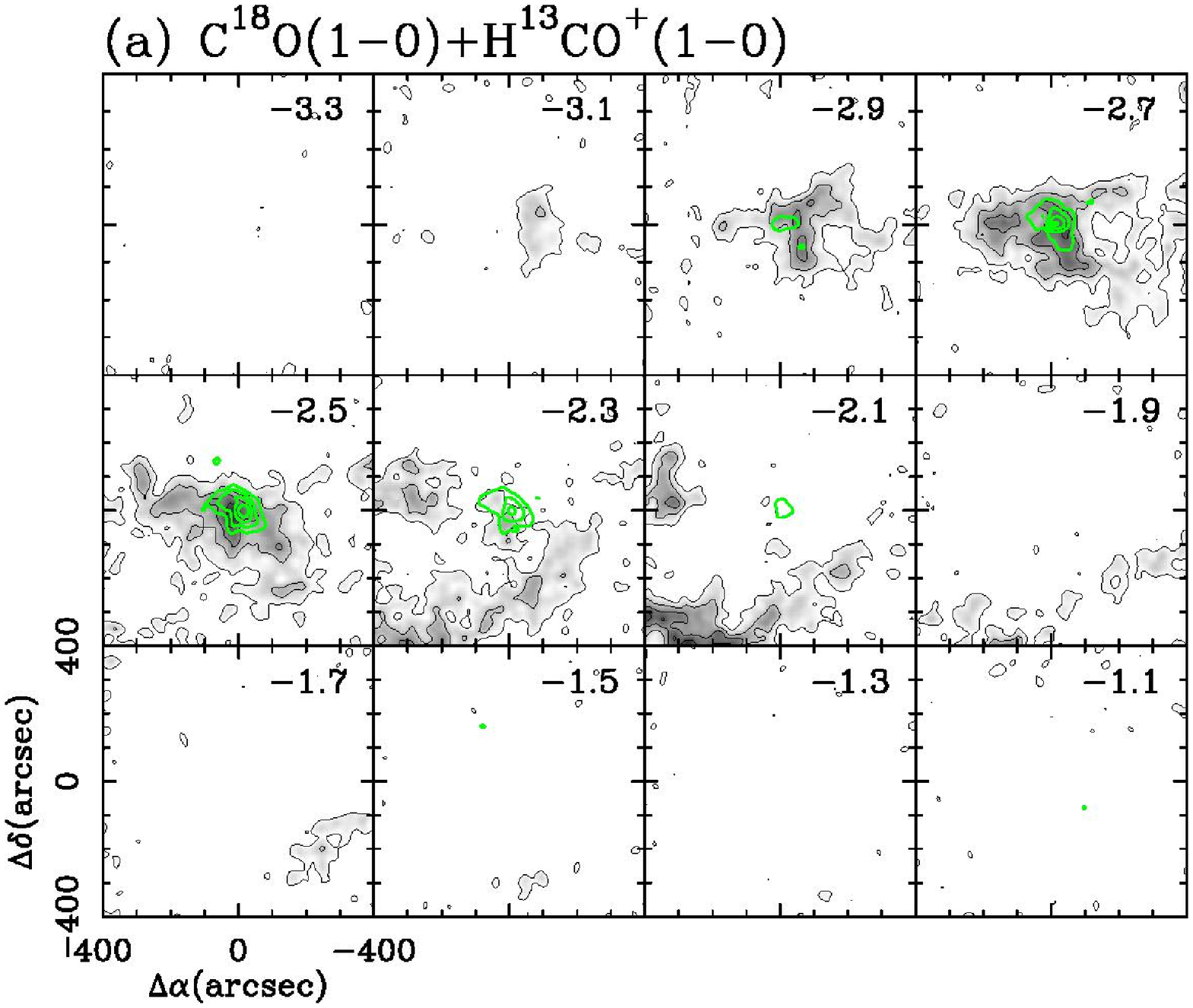}
\includegraphics[angle=0,width=8.9cm]{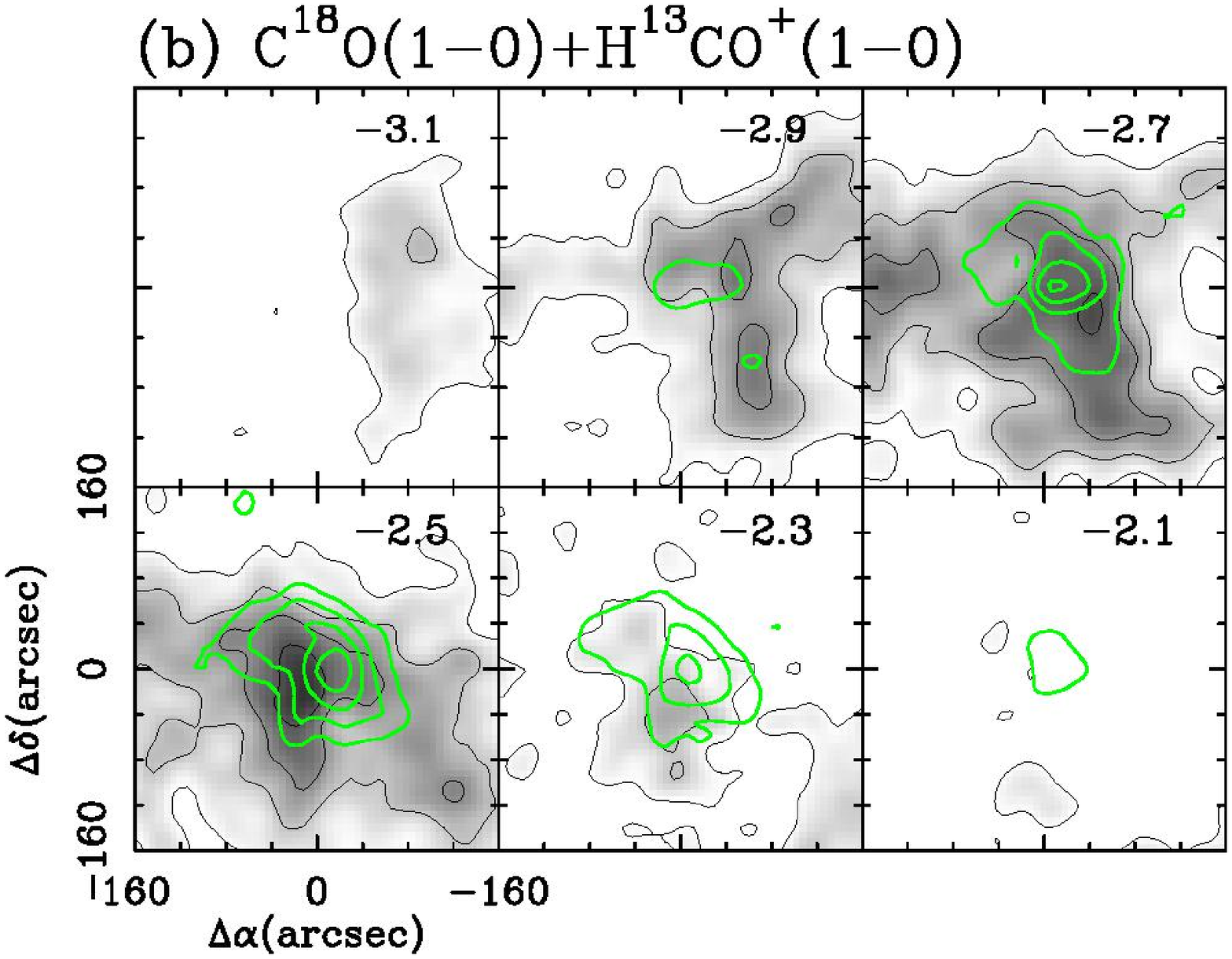}
\caption{\small 
Velocity channel maps of the
\CeO\ (1--0) (greyscale plus black contours) and \HtCOp\ (1--0) (green contours) emission
for (a) whole the observed area and (b) the central 320\arcsec$\times$320\arcsec\ region.
The \HtCOp\ line data are taken from paper I.
All the contours are plotted at 3\sgm\ intervals
 starting from the 3\sgm\ levels
where \sgm\ = 100 mK and 220 mK in \Tmb\ 
for the \CeO, and \HtCOp\ lines,  respectively.
Each channel map is averaged over a 0.2 \kms\ bin 
whose central LSR-velocity is shown in unit of \kms\ 
at the top right corner of each panel.
}
\label{fig:chmaps_c18o}
\end{center}
\end{figure}

Velocity channel maps with a resolution of 0.2 \kms\ are shown in 
Figures \ref{fig:chmaps_13co} and \ref{fig:chmaps_c18o}
where we overlaid the channel maps for the higher-density gas tracer 
with green contours on those for the lower-density gas tracer
with black contours.
First, the large-scale east-west structure 
is mainly seen in the \tCO\ panels of $\sf -3.9$ $\lesssim$ \Vlsr /\kms\ $\lesssim -2.3$.
This velocity range includes the 
\Vsys\ of the GF\,9-2 dense core ($-2.48$ \kms).
Hereafter, we refer to the east-west structure seen in the velocity 
range as ``Component 1".
Second, the Component\,1 is likely to have 
sub-structures with scales of $\lesssim$0.1 pc 
(100$\arcsec$ corresponds to 0.097 pc at $d\,=$ 200 pc), one of which is the \gf\ core.
Third, one finds an overall agreement between the spatial 
distributions of the \tCO\ and \CeO\ emission, 
and between the \CeO\ and \HtCOp\ emission in each velocity channel.
Note that the peak position of the \CeO\ emission does not coincide with the 
\HtCOp\ peak in the \gf\ core 
(see \Vlsr\ $=\, -2.7$ and $-2.5$ \kms\ panels in Figure \ref{fig:chmaps_c18o}b).
Fourth, the \tCO\ channel maps show that,
in addition to the Component 1,
there exists another spatially
coherent structure in the southern parts of the velocity panels of
$-2.1\lesssim$ \Vlsr /\kms\ $\leq$ $\sf -0.7$.
This structure, ``Component 2", is elongated along 
the southeast-northwest direction, 
and shows intense \tCO\ emission towards the south-east.
Its local maxima is found at the bottom left corners of the
velocity channels of \Vlsr\ $=\, -2.1$ and $-1.9$ \kms\ (Figure \ref{fig:chmaps_13co}).
For the \CeO\ line, the Component 2 is mainly identified in
the panels of $-2.3\lesssim$ \Vlsr /\kms\ $\leq$ $\sf -1.7$.
Last, the gas seen in the \Vlsr\ $=-2.3$ \kms\ panel is
highly likely emanated from the two components.
Despite the fact that the two components are not well separable in the velocity space, 
we define the boundary velocity between 
the dual components as $-2.2$ \kms\ in \Vlsr.
This is because the 
{\CeO\ emission associated with the} \gf\ core is not recognized in the channel maps 
redward of 
\Vlsr\ $\gtrsim\, -2.1$ \kms\ (Figure \ref{fig:chmaps_c18o}a).\par

\begin{figure}
\begin{center}
\includegraphics[angle=0,width=5.7cm]{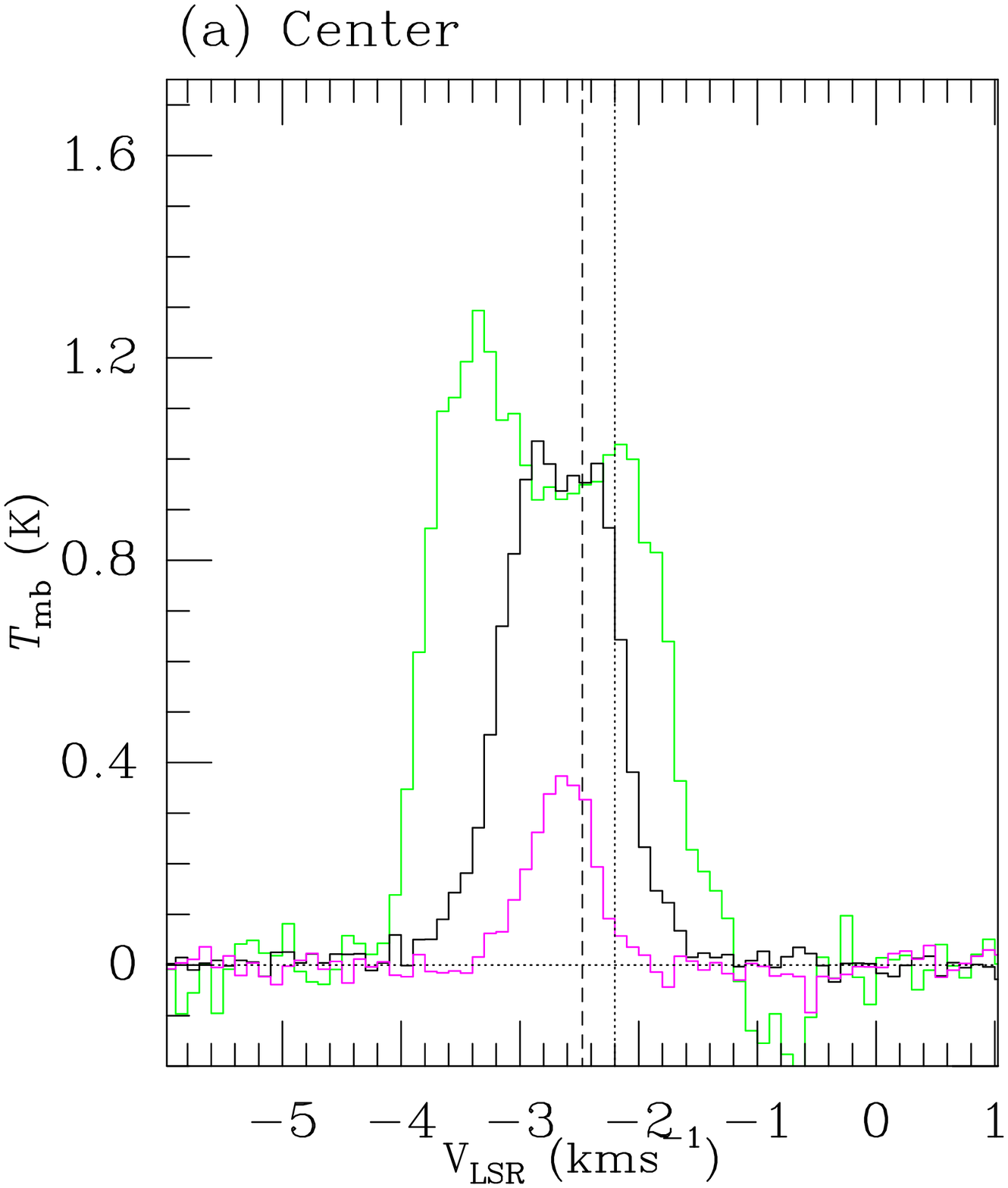}
\includegraphics[angle=0,width=5.7cm]{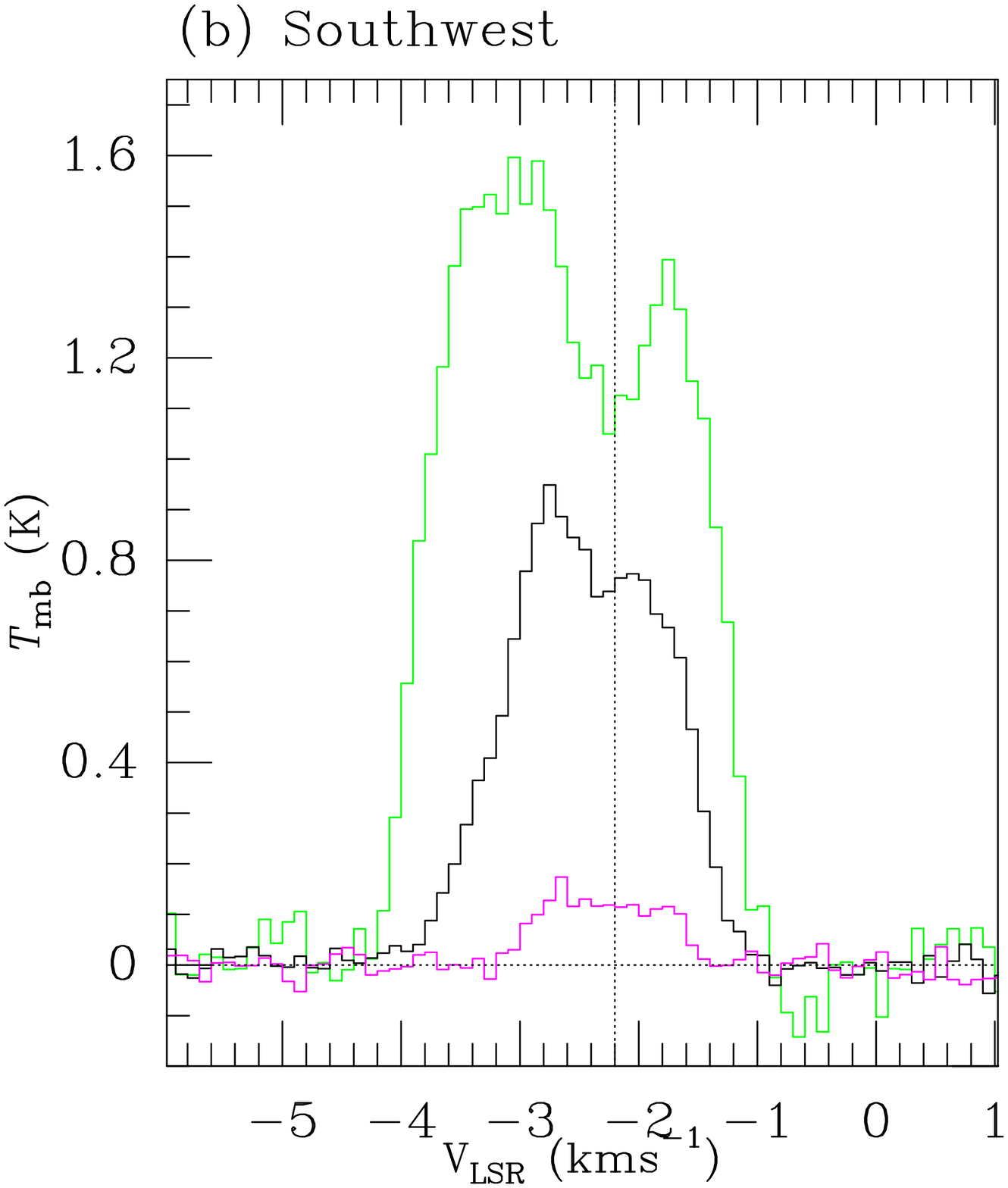}
\caption{Spectra of the $J=1$--0 lines of the \CO\ (green),
\tCO\ (black), and \CeO\ (magenta) obtained by averaging the emission
over the two circles with a 100\arcsec\ radius shown in Figure \ref{fig:totmaps}a - c.
The vertical scale is the \Tmb\ in K the same as in Figure \ref{fig:sp}.
Velocity resolutions of the spectra are 0.1 \kms.
The panel (a) represents the spectra towards the core center
while the panel (b) does those at 
$(\Delta\alpha, \Delta\delta)\,=\,(-250\arcsec, -150\arcsec)$
with respect to the 3\,mm continuum position.
The vertical dashed line at \Vlsr\ $=\,-2.48$ \kms\ in the panel (a) shows the
systemic velocity of the cloud (\Vsys; paper I), and the vertical dotted lines
in the two panels indicate \Vlsr\ $=\,-2.2$ \kms\ which is 
the velocity boundary between the two gas components (see \S\ref{ss:mommaps}).
}
\label{fig:spcomparison}
\end{center}
\end{figure}

Since we detected significant
\tCO\ emission to the south-west of the core
(e.g., Figure \ref{fig:totmaps}b),
we compared two spectra obtained towards the core and the south-western region
(Figure \ref{fig:spcomparison}).
These spectra were obtained by averaging the emission inside the circles
shown in Figures \ref{fig:totmaps}a - \ref{fig:totmaps}c.
Figure \ref{fig:spcomparison} shows the following features: 
(i) the \CO\ emission in the south-west is stronger than in the center,
(ii) a comparison between the two \tCO\ spectra suggests that dividing the filament gas
into the two components by \Vlsr\ $=-2.2$ \kms\ seems to be reasonable,
(iii) the \CeO\ emission towards the core is 
approximately twice as intense as the emission towards the
south-west, and 
(iv) towards the core center, the LSR-velocity of the \CeO\ peak agrees with those
of the central dips seen in the \CO\ and \tCO\ spectra (Figure \ref{fig:spcomparison}a).
However these LSR-velocities are shifted by $\sim -0.2$ \kms\
with respect to the \Vsys.\par

We show the two integrated intensity maps of the Components 1 and 2 in 
Figure \ref{fig:mapcomparison}
by adopting \Vlsr $=\,-2.2$ \kms\ as the boundary velocity.
The \tCO\ emission in the velocity range of
$-4.0 \leq$\Vlsr/\kms $\leq -2.2$ (Figure \ref{fig:mapcomparison}a)
is clearly associated with the dense core gas, 
which reconciles with the fact that 
the \CeO\ emission seen in each panel center of 
$-3.2 \leq$\Vlsr/\kms $\lesssim -2.2$
(Figure \ref{fig:chmaps_c18o}a)
is predominantly associated with the core.
In contrast, the \tCO\ gas in Figure \ref{fig:mapcomparison}b is not
associated with the dense core.
As seen in the \Vlsr $=-2.3$ \kms\ panels of Figures \ref{fig:chmaps_13co}
and \ref{fig:chmaps_c18o}a as well as
the spectrum in Figure \ref{fig:spcomparison}b, 
it is impossible to separate completely the two components.
We consider that 
they are partially overlapping in the line-of-sight and, 
and are partially interacting with each other.
In \S\ref{s:analysis} we analyze the data by showing their boundary, 
instead of separating them.

\section{Analysis}
\label{s:analysis}
\subsection{Producing Opacity-Corrected \tCO\ $J=1-0$ Spectra}
\label{ss:TmbcorrSP}

In order to discuss the velocity structure of the filament gas, 
we corrected the cube data for the line broadening 
due to optical thickness \citep[e.g.,][]{Phillips79}, 
and made opacity-corrected \tCO\ spectra.
This required us to 
estimate velocity width of the \tCO\ line in the limit of $\tau \rightarrow 0$,
\dVint, where $\tau$ denotes the \tCO\ optical depth.
The choice of the \tCO\ is due to the large optical depth 
of the \co\ emission,
and due to the absence of the \CeO\ emission in the low density region
of the filament.
As described in Appendices \ref{ass:rteq} and \ref{ass:tauTex},
we utilized the three CO isotopologues to estimate the optical
depth and excitation temperature (\Tex) of the \tCO\ line. 
We multiplied a factor of $\tau/(1-e^{-\tau})$ to 
the observed \Tmb\ of the \tCO\ line to obtain the 
opacity-corrected \Tmbcorr\ at each LSR-velocity
[see Eq.(\ref{eqn:Tmbcorr}) in Appendix \ref{ass:errTmbcorr}].
After obtaining the opacity-free spectrum 
such as shown in Figure \ref{fig:spanalysis}e, 
we generated a 3D data cube 
by keeping the effective spatial resolution 
of 24\farcs0 and the velocity resolution of 0.1 \kms\ (\S\ref{s:obs}).
The opacity-free \tCO\ cube data were analyzed 
as described in the subsequent subsections, and our
usage of the \co\ and \CeO\ line data are 
limited to estimating $\tau$ and \Tex\ of the \tCO\ emission.

\begin{figure}
\begin{center}
\includegraphics[angle=0,scale=.36]{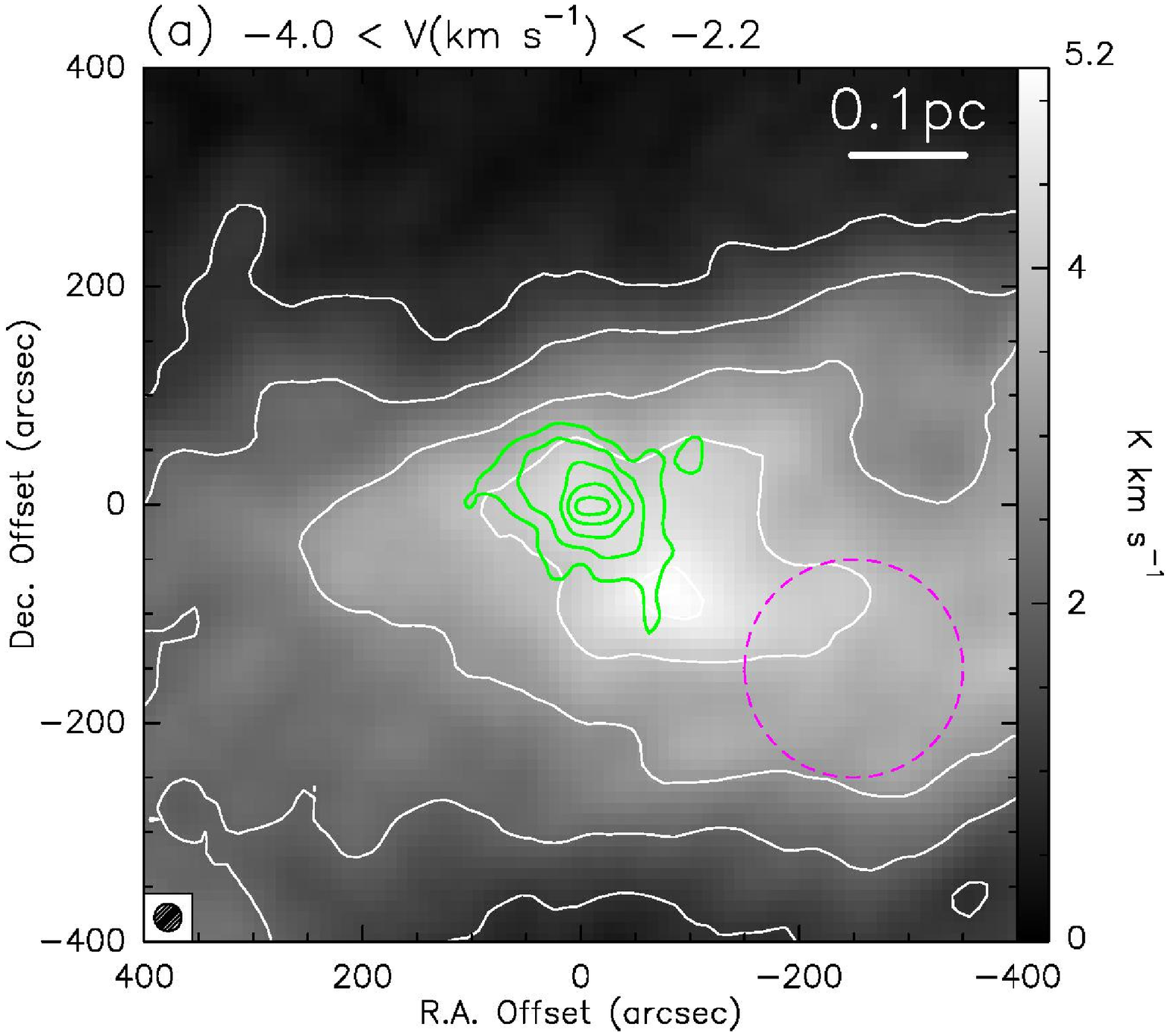}
\includegraphics[angle=0,scale=.36]{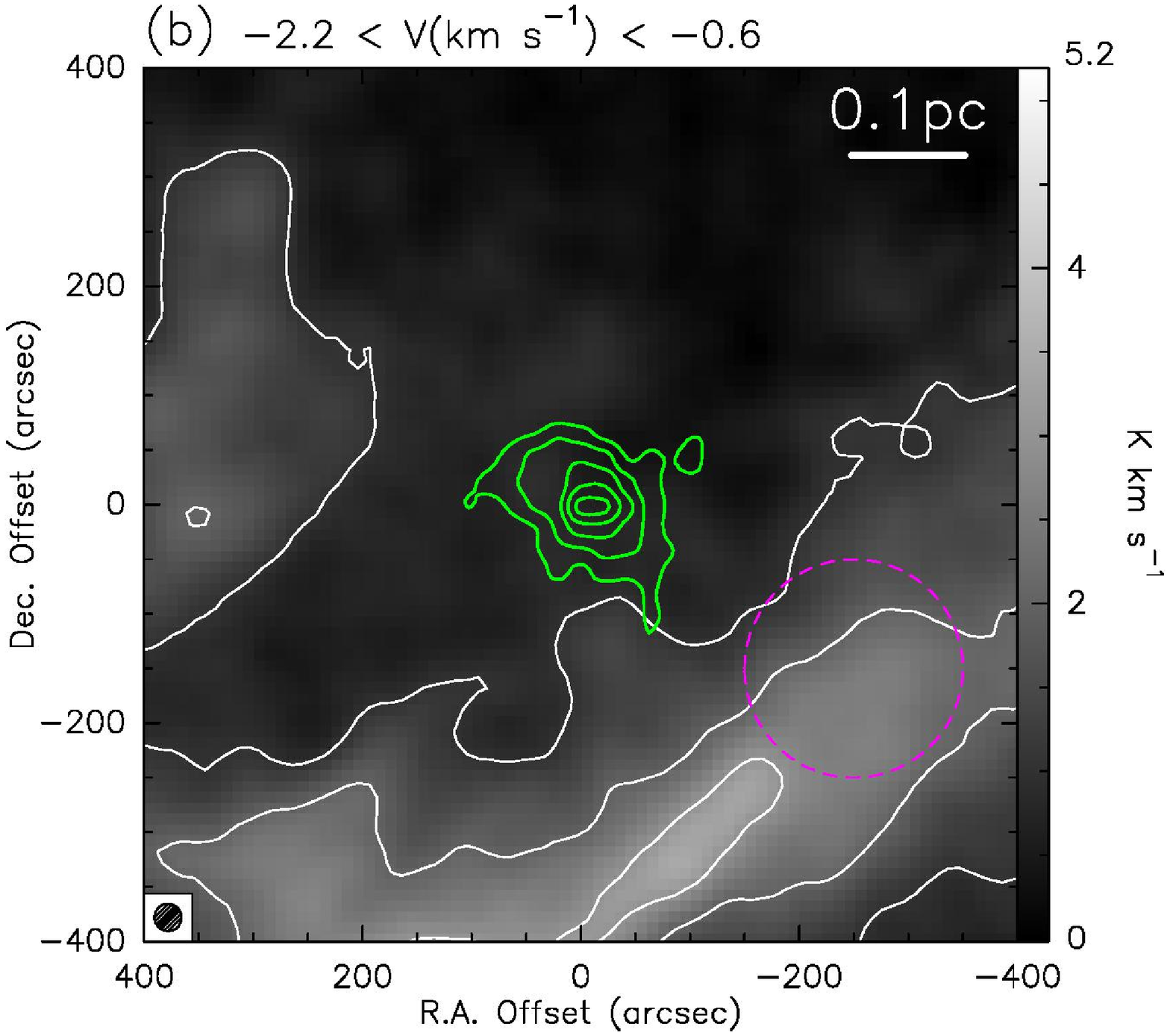}
\caption{Integrated intensity maps of the \tCO\ $J=$1--0 emission in the unit of K$\cdot$\kms.
The maps in panels 
(a) and (b) 
are integrated the emission 
in \Tmb\ over the LSR-velocity ranges of
$-4.0\leq$ \Vlsr/\kms\ $\leq -2.2$ and 
$-2.2\leq$ \Vlsr/\kms\ $\leq -0.6$, respectively (see Figure \ref{fig:spcomparison}).
Note that the intensity ranges in the two maps are the same.
The white contours have the intervals of 1.0 K$\cdot$\kms\ starting from the
1.0 K$\cdot$\kms\ level which corresponds to the 12\sgm\ and 8\sgm\
levels in the maps (a) and (b), respectively.
The peak intensity of 5.2 K$\cdot$\kms\ in the panel (a) is found at 
$(\Delta\alpha, \Delta\delta)\,=\,(-76\arcsec, -78\arcsec)$
with respect to the 3\,mm source position, while
that of (b) (3.4 K$\cdot$\kms) was measured at $(-86\arcsec, -326\arcsec)$.
The green contours show the \HtCOp\ total map in the same fashion 
as for Figure \ref{fig:totmaps}.
The magenta dashed circles show 
the area where the CO isotopologue emission are integrated to
obtain the spectra shown in Figure \ref{fig:spcomparison}b.
}
\label{fig:mapcomparison}
\end{center}
\end{figure}

\subsection{Maps of Optical Depth and Excitation Temperature}
\label{ss:tauTex}

\begin{figure}
\begin{center}
\includegraphics[angle=0,width=8.9cm]{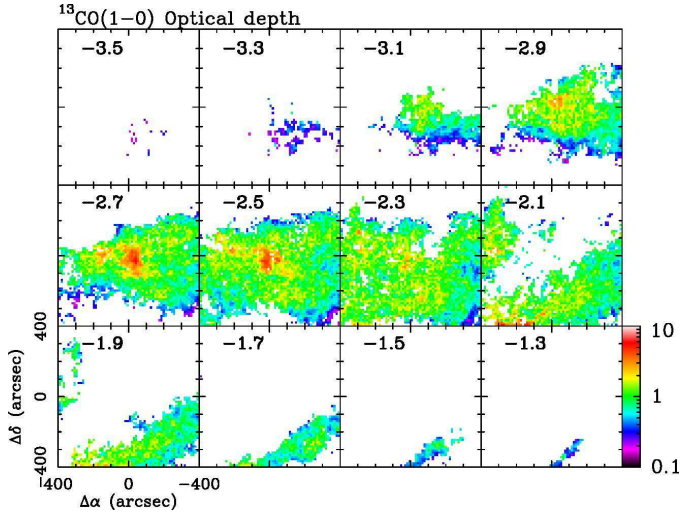} 
\caption{Velocity channel maps of the optical depth
of the \tCO\ (1--0) emission with velocity intervals of 0.2 \kms.
The central LSR-velocity of each channel 
in \kms\ is shown at the top left corner of each panel.
The optical thickness is shown in logarithmic scale
(see the color bar).
The maximum optical depth of 10.0$\pm$3.2 is measured 
near the southern edge of the
\Vlsr\ $=\,-2.1$ \kms\ panel, 
while the second highest value of
5.8$\pm$0.9 was
measured towards the \gf\ core at \Vlsr\ $=\,-2.7$ \kms.
\label{fig:taumap}}
\end{center}
\end{figure}

\begin{figure}
\begin{center}
\includegraphics[angle=0,width=9.3cm]{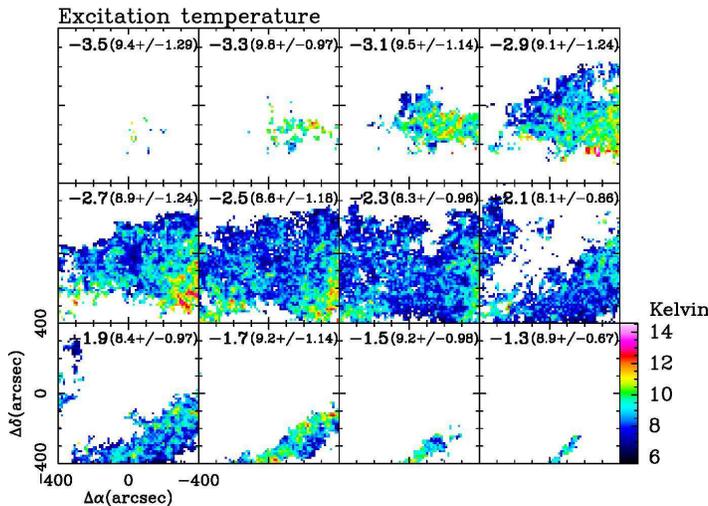} 
\caption{Velocity channel maps of the excitation temperature 
of the \tCO\ (1--0) emission in K with velocity resolution of 0.2 \kms.
The central LSR-velocity of each velocity channel in 
\kms\ is shown at the top left corner of each panel.
The two numbers in the
parenthesis with the \texttt{+/-} sign 
indicate the mean and standard deviation of the excitation temperature
in each panel in unit of K.
Notice that the excitation temperature is shown in 
linear scale (see the color bar).
\label{fig:Texmap}}
\end{center}
\end{figure}

Figures \ref{fig:taumap} and \ref{fig:Texmap} present velocity channel 
maps of the $\tau_\mathrm{^{13}CO}$ and \Tex\ of the \tCO\ (1--0) line emission, respectively. 
Because of our 3$\sigma$ level detection threshold
(Appendix \ref{ass:tauTex}) and
the quality assessment of the results (Appendix \ref{as:filters}),
we have no data points 
blueward of the \Vlsr$=\, -3.7$ \kms\ channel and 
redward of the \Vlsr$=\, -1.1$ \kms\ channel
where the weak \tCO\ emission is seen in Figure \ref{fig:chmaps_13co}.
Figure \ref{fig:taumap} shows that the \tCO\ emission is moderately 
opaque with $\tau_\mathrm{^{13}CO}\sim 1$ in the ambient regions,
and becomes optically thick towards the core.
We measure a mean $\tau_\mathrm{^{13}CO}$ of 0.78$\pm$0.32 and 
a mean \Tex\ of 8.4$\pm$1.0 K 
for all the data shown in Figures \ref{fig:taumap} and \ref{fig:Texmap}, respectively.\par

Towards the position of the dense core,
we have $\tau$ and \Tex\ 
estimates blueward of \Vlsr\ $=-2.3$ \kms,
but no estimates redward of \Vlsr $=-2.1$ \kms.
The highest optical depth in the \gf\ core region is measured 
at the local peak of 5.8$\pm$0.9 at \Vlsr $=-2.7$ \kms,
leading to the \CeO\ optical depth of 1.1 with the solar abundance ratio of [\tCO]/[\CeO] $=$ 5.5.
We measure a mean $\tau_\mathrm{^{13}CO}$ of 1.31$\pm$0.30 and
a mean \Tex\ of 8.0$\pm$0.7 K for the region enclosed 
by the 3$\sigma$ level contour of the \HtCOp\ emission.
Inside the 50\% intensity contour with respect to the peak intensity 
for the \NtwoH\ (1--0) emission (paper I), 
a mean \Tex\ is calculated to be 7.9$\pm$0.6 K
which agrees with the \Tex\ $=$ 9.5$\pm$1.9 K derived from the \NtwoH\ line 
in the previous work within the uncertainties.
In addition, for a circular region with a radius of 0.1\,pc 
centered on the \gf\ core,
we measure a mean \Tex\ of 8.0$\pm$0.1\,K
which agrees with the \Tex\ value expected 
from the Monte Carlo simulation by
\citet{Ciardi00} (see their Figure 9).
In conclusion, the calculated \Tex\ velocity channel maps 
have reasonable consistency with the previous results.
Assuming that the gas is in LTE, 
we consider that the \Tex\ represents 
the kinetic temperature (\Tkin) of the gas.\par

In Figure \ref{fig:Texmap}, we found regions where
\Tex\ is elevated up to $\sim\,14$\,K to the south-west of the core in the
\Vlsr$=\,-2.9\sim -2.5$ \kms\ panels and the $\,-1.7$ \kms\ one,
which belong to the Components 1 and 2, respectively.
It is also interesting that the velocity channels between them
(i.e., $-2.3\leq$ \Vlsr/\kms $\leq -1.9$) do not have
such high temperature regions.
These features will be discussed in \S\ref{sss:cf_scenario_sw}.\par

\begin{figure}
\begin{center}
\includegraphics[angle=0,scale=.37]{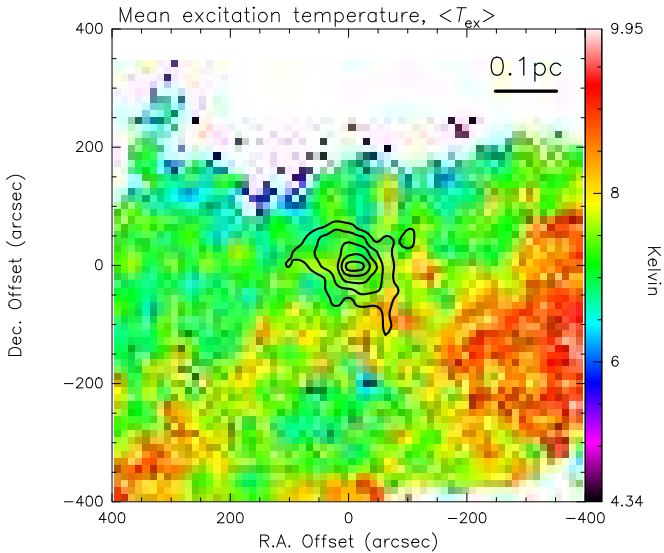}
\caption{Plot of mean excitation temperature obtained by
averaging the \Tex $(v)$ spectra at each pixel position across the velocity axis
(see Figure \ref{fig:Texmap}).
The contours represent the total map of the \HtCOp\ (1--0)
emission in the same manner as for Figure \ref{fig:totmaps}.
The 0.1 pc scale is indicated at the top right corner, 
and the color bar is on the right-hand side of the panel.
\label{fig:meanTex}}
\end{center}
\end{figure}

In order to calculate the column density of the \tCO\ molecules,
we subsequently produced a mean \Tex, \meanTex, map
from the \Tex\ data cube. 
The \meanTex\ map is shown in Figure \ref{fig:meanTex}.
We averaged \Tex$(v)$ values along the velocity axis where the
\Tmbcorr$(v)$ intensities
exceed our detection threshold of the 3\sgm\ level 
(see Figures \ref{fig:spanalysis}d and \ref{fig:spanalysis}e).
We obtained the mean value for \meanTex\ of 7.5$\pm$1.0\,K
over the presented region, 
and adopted this value as the
representative \Tkin\ for the region.

\subsection{Maps of Total Intensity, Centroid Velocity, and Velocity Width}
\label{ss:mommaps}

\begin{figure}
\begin{center}
\includegraphics[angle=0,scale=.33]{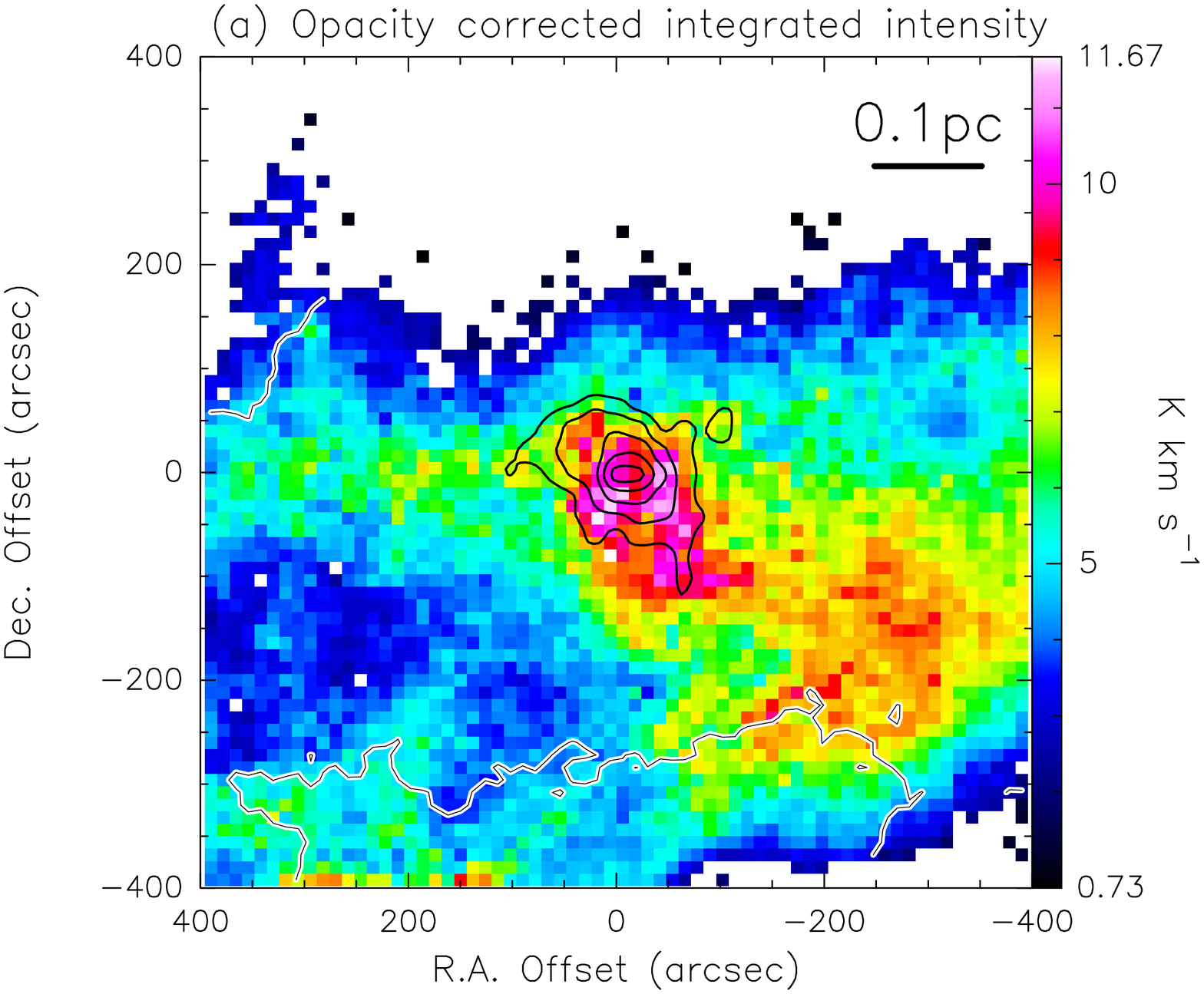}
\includegraphics[angle=0,scale=.33]{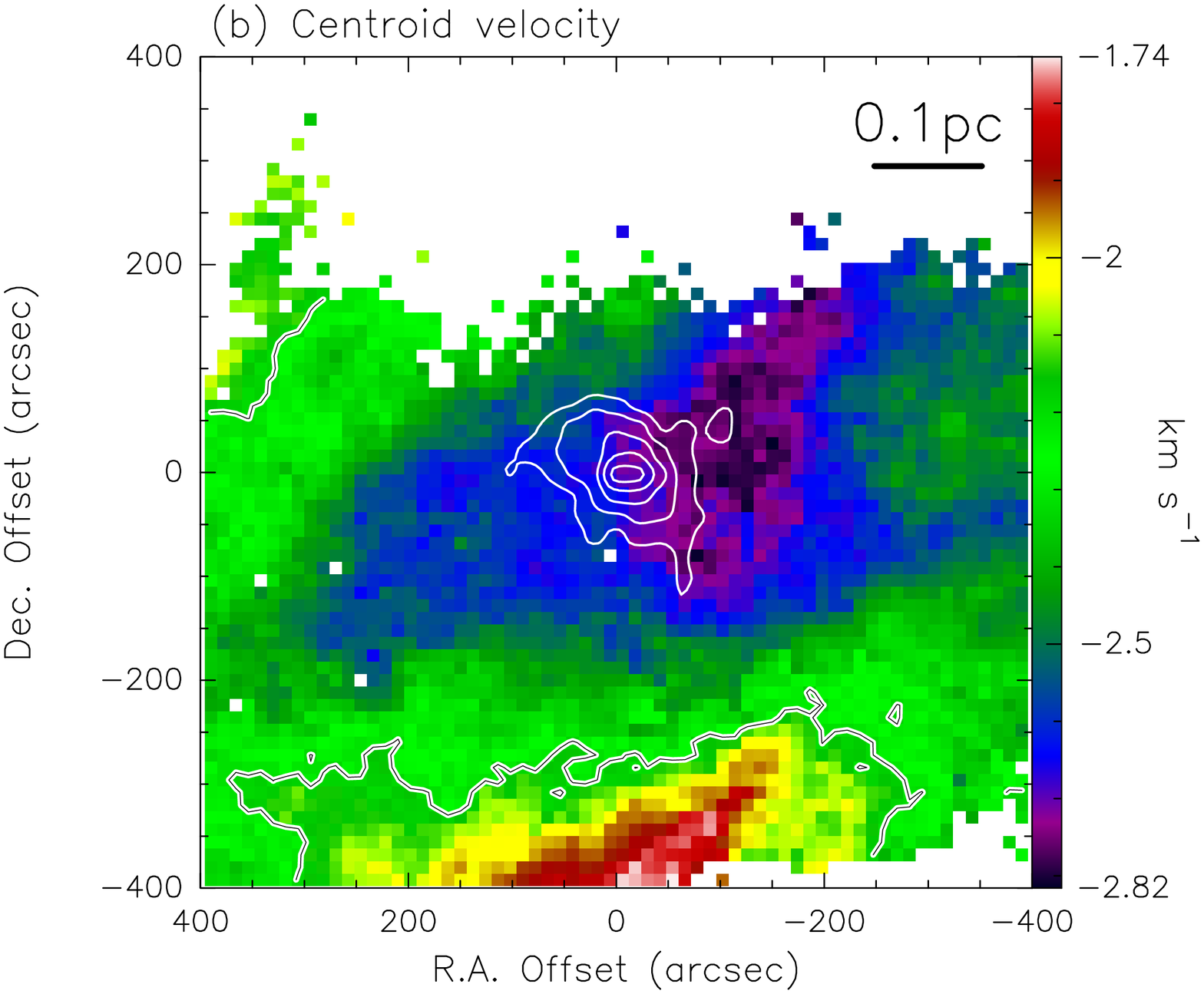}
\includegraphics[angle=0,scale=.33]{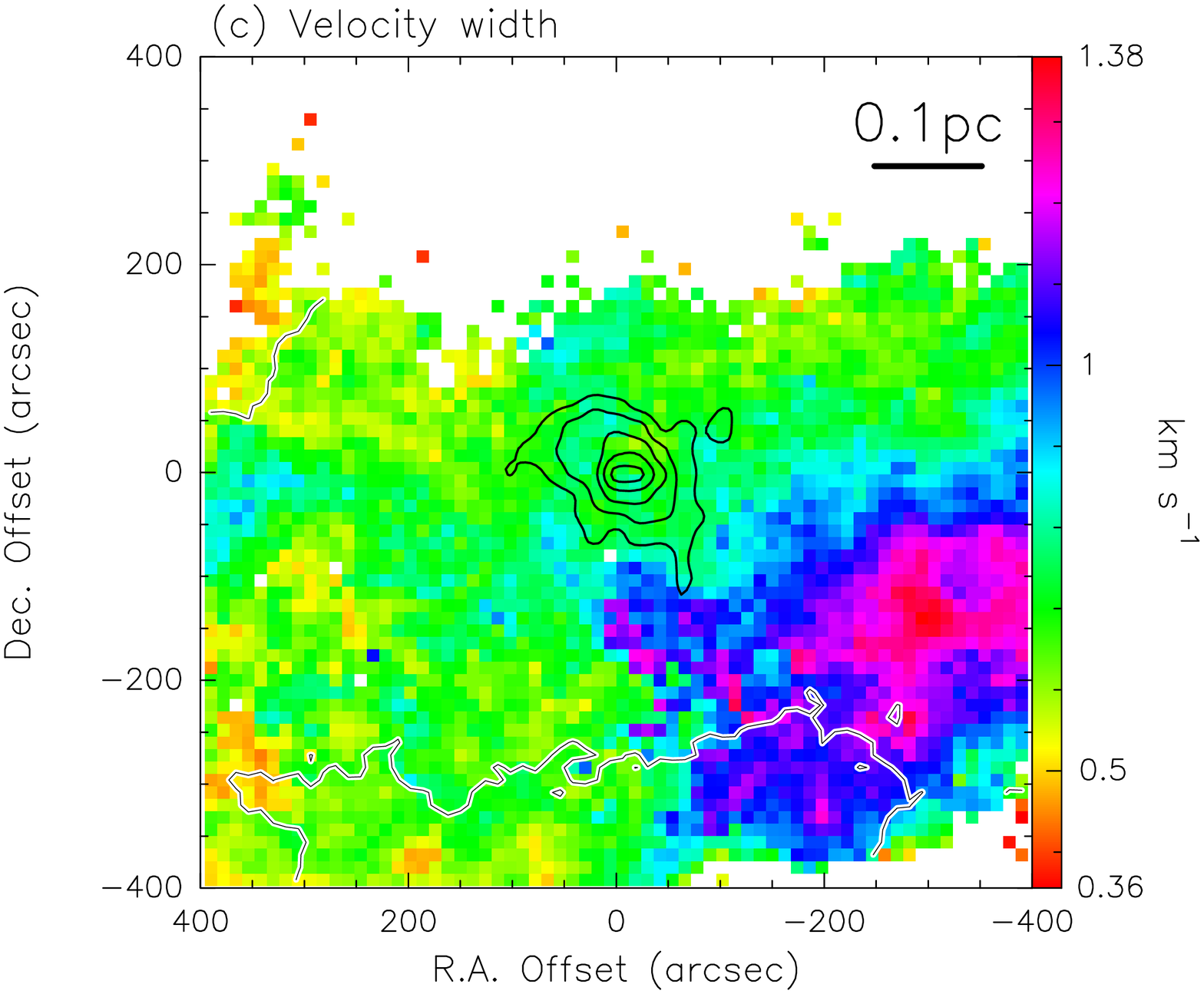}
\caption{\small Maps of (a) the total intensity in K \kms, 
(b) centroid velocity in \kms, and 
(c) velocity width (FWHM) in \kms\ produced from  
the opacity-corrected \tCO\ (1--0) spectra (\S\ref{ss:TmbcorrSP}).
The pixel sizes of these maps are 12\arcsec, i.e., half of the effective beam size
(\S\ref{s:obs}).
The color bar on the right-hand side of each plot shows 
the plotted range in linear scale.
The 0.1 pc-scale at the distance of 200\,pc 
is shown at the top right corner of each panel.
The contours at each map center represent the \HtCOp\ total map as shown in Figure \ref{fig:totmaps}.
The black-in-white contours seen in the south and the north-east of the panel
(b) represent the \Vcent\ $=\,-2.2$ \kms\ contours
which are also shown in the panels (a) and (c).
See \S\ref{s:results} for the reasoning.
\label{fig:mommaps}}
\end{center}
\end{figure}

The color images in Figure \ref{fig:mommaps} represent 
total integrated intensity, intensity-weighted mean velocity, 
i.e., centroid velocity (\Vcent), 
and velocity width (FWHM) maps produced from 
the opacity-corrected \tCO\ spectra.
After calculating the zeroth, first, and second order
moments as well as their errors at each spatial position,
we produced the maps without any smoothing 
(Appendices \ref{as:errspmom} and \ref{as:filters}).
Since we are interested in the natal cloud of the dense core,
we indicate the boundary between the Components 1 and 2 by the black-in-white contours
which correspond to 
\Vcent\ $=\,-2.2$ \kms\ in Figure \ref{fig:mommaps}b.\par

Figure \ref{fig:mommaps}a shows us that 
the morphology of the \tCO\ bright region is principally 
similar to the distribution of the \CeO\ emission 
(Figures \ref{fig:totmaps}c and \ref{fig:totmaps}f),
and that the dense core traced by the \HtCOp\ emission 
has formed in the inner densest part of the filamentary cloud.
Figure \ref{fig:mommaps}b shows that
the Component 1 gas is confined in a narrow velocity range of
$\sim$\,0.6 \kms\ ($-2.8\lesssim$\,\Vcent/\kms $\lesssim -2.2$)
which is about twice the isothermal sound velocity
($c_\mathrm{s}^2\,=\,8\ln 2\,\frac{~kT_\mathrm{kin}~}{\mu m_\mathrm{H}}$ where
$\mu$ denotes the mean molecular weight of 2.33 for [He] = 0.1 [H]).
The measured \Vcent\ of the immediate surroundings of the dense core
shown in the blue-to-dark magenta
(corresponding to $-2.8 \lesssim$\,\Vcent/\kms $\lesssim -2.6$)
has reasonable consistency with the
velocity structure of the dense core 
traced by the \HtCOp\ (1--0) and \NtwoH\ (1--0) line observations
(see Figure 12 in paper I).
As we argued in \S\ref{s:results} with Figures \ref{fig:spcomparison} and \ref{fig:mapcomparison}, 
the two gas components cannot be separated completely,
Figure \ref{fig:mommaps}b demonstrates that the main body 
of the Component 2 gas is seen in \Vlsr\ $\gtrsim\,-2$ \kms.
Another important result from Figure \ref{fig:mommaps}b is that
neither systematic motions of the gas caused 
by any activities of YSOs, 
such as a well-collimated typical molecular outflow, 
nor any sharp discontinuity of the line-of-sight velocity, e.g., 
due to shock fronts, can be recognized.
These velocity features strongly suggest that both the Components 1 and 2 
belong to the GF\,9 filament \citep{Schneider79}
where seven dense cores and three possible candidates are 
located at regular intervals of $\sim$0.9\,pc (paper II).\par

Figure \ref{fig:mommaps}c shows that the velocity width is
enhanced up to $\sim 1.4$ \kms\ in the fourth quadrant of the map.
The extent of this region is almost the same as the south-western intensity enhancement 
seen in Figure \ref{fig:mommaps}a.
The mean velocity width for the fourth quadrant is 
$\Delta v_\mathrm{FWHM}\,=\, 0.96 \pm$0.24 \kms,
while the mean value for the first to third quadrants is 0.69$\pm$0.12 \kms.
Both the mean values exceed the isothermal sound velocity width in FWHM
of 0.38$\pm$0.13 \kms\ for \Tkin\ = 7.5$\pm$1.0\,K (\S\ref{ss:tauTex}).

\subsection{Column Density Map and LTE Mass}
\label{ss:Ncol}

After obtaining the maps of the total integrated intensity, 
$\int T_{\rm mb}^{\rm corr}(v)dv$ (Figure \ref{fig:mommaps}a), 
and the mean excitation temperature \meanTex\ (Figure \ref{fig:meanTex}),
one can readily calculate \tCO\ column density, $N_\mathrm{^{13}CO}$
(Appendix \ref{as:Ntot}).
Here we used the \meanTex\ map (Figure \ref{fig:meanTex}) 
instead of the \Tex\ channel maps (Figure \ref{fig:Texmap}).
Considering the \Tex\ range of $5.6\leq$ \Tex /K$\leq 13.6$ (see Figure \ref{fig:Texmap}),
the dependency of $T_\mathrm{ex}$ in the $N_\mathrm{^{13}CO}$
calculations is smaller than the other errors 
[see Eqs.(\ref{eqn:Ntot}) and (\ref{eqn:Ntot_err}) in Appendix \ref{as:Ntot}].
Furthermore, the \Tex$(v)$ spectra are
generally ``flat" with respect to the LSR-velocity (see Figure \ref{fig:spanalysis}d),
justifying the usage of the \meanTex\ value as the representative \Tex\ at each map pixel.\par

\begin{figure}
\begin{center}
\includegraphics[angle=0,scale=.38]{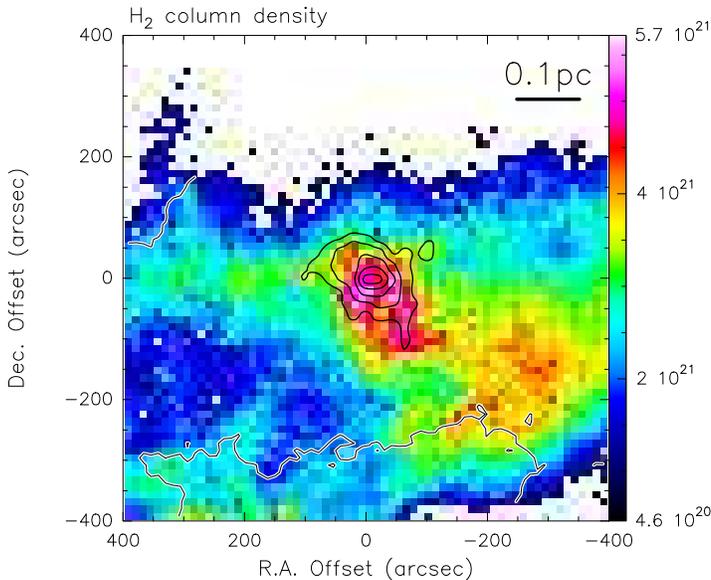}
\caption{Map of the total column density of molecular hydrogen ($N_\mathrm{H_2}$)
derived from the \tCO\ emission in unit of \cmq.
The color scale is shown in logarithmic
as shown in the color bar.
The 0.1 pc scale is indicated at the top right corner.
The black contours at the center represent the \HtCOp\ total map
as shown in Figure \ref{fig:totmaps}.
The black-in-white contours in the south and the north-east of the core correspond to 
the \Vcent\ $=\,-2.2$ \kms\ contours as shown in Figure \ref{fig:mommaps}.
\label{fig:Ncol}}
\end{center}
\end{figure}

Figure \ref{fig:Ncol} presents a map of 
the molecular hydrogen column density ($N_\mathrm{H_2}$)
that is converted from the \tCO\ column density by 
adopting the \tCO\ fractional abundance of $X$(\tCO)$\, =\,2\times 10^{-6}$ 
\citep[e.g.,][]{Dickman78, Frerking82, Lequeux05}.
The $N_\mathrm{H_2}$ map clearly demonstrates that 
the filament has a high-column density region in the center where
the \HtCOp\ core is observed.
In addition, the map also shows that the filament gas has a 
column density range of 
$20.7\lesssim \log N_\mathrm{H_2}/\mathrm{cm}^{-2} \lesssim 21.8$.
We measure a mean $N_\mathrm{H_2}$ of 
$(2.7\pm 0.9)\times 10^{21}$ \cmq\ for the Component 1 and
$(2.4\pm 0.7)\times 10^{21}$ \cmq\ for the Component 2.
We calculate LTE-masses (\MLTE) 
of $24\pm 10$ \Msun\ and $17\pm 7$ \Msun\ for the Components 1 and 2,
respectively, with $d=$\,200 pc.\par

We checked consistency with the previous work by \citet{Ciardi00}.
We measured the maximum $N_\mathrm{^{13}CO}$ of $1.2\times 10^{16}$ \cmq\ 
at the center of the \gf\ core.
This agrees with the value that can be read from Figure 6a in 
\citet{Ciardi00} ($1.4\times 10^{16}$ \cmq).
It should be noted that 
\citet{Ciardi00} 
used $d=$\,440 pc, and estimated \MLTE\ of
$53\pm 8$ \Msun\ for a rectangular region of 480\arcsec$\times$600\arcsec\
centered on the \gf\ core.
Towards the same rectangle, 
we measured a mean $N_\mathrm{^{13}CO}$ of
$\,(5.3\pm 2.2)\times 10^{15}$ \cmq.
With $d=$\,440 pc, we recalculated
\MLTE\ of $55\pm 23$ \Msun\ from the mean $N_\mathrm{^{13}CO}$.
Therefore our analysis with $d=$\,200 pc 
has good consistency with the previous one with $d=$\,440 pc.
\par

A hint to discuss the origin and nature of the filament may be
obtained from a histogram of the H$_2$ column density and its cumulative 
distribution shown in Figure \ref{fig:Ncol_hist}.
The histogram is well approximated by a log-normal probability distribution function
with a standard deviation of 0.15, 
indicating that the turbulence formed the density structure 
of the filament gas \citep[see e.g.,][]{Ostriker01, Padoan02, Ossenkopf02, Nakamura07, Kainulainen09}.
In addition, Figure \ref{fig:Ncol_hist}b shows that 
50\% of the filament materials exist in the higher column density region of 
$N_\mathrm{H_2}\gtrsim 2.5\times 10^{21}$ \cmq, 
which is similar to the results obtained in Taurus 
\citep{Goldsmith08}.
The order of the derived column density 
($N_\mathrm{H_2}\sim 10^{21}$ \cmq) corresponds to 
the visual extinction of $A_\mathrm{v}\sim 1$
which is comparable to those measured from the optical images
\citep{Schneider79, PB06}.

\section{Discussion}
\label{s:discussion}
In this section, we discuss the nature of the filament gas to shed light on
the initial conditions of the dynamical collapse of the \gf\ cloud core which
harbors the exceptionally young low-mass protostar (papers I and III). 
On the basis of all the results and analysis, 
we conclude that the core is physically associated with the Component 1 filament
(\S\ref{s:results} and \S\ref{ss:mommaps}).
This is because the dense core is located at the local density maxima of the Component 1 
and the LSR-velocity of the core falls in the velocity range of the Component 1. 
Although it is impossible to separate the two components completely, 
the Component 2 is not the natal gas of the \gf\ core.\par

\subsection{Turbulent Motions of the Gas in the Component 1}
\label{ss:ambient}

We present a map of the ratio between the non-thermal velocity dispersion 
(\sigmanth ) and the sound velocity $c_\mathrm{s}$ in Figure \ref{fig:sigmanthmap}
to examine the degree of turbulence in the Component 1.
The ratio is obtained from, 
\begin{equation}
\left(\frac{\sigmanth}{c_\mathrm{s}}\right)^2 = \left\{\frac{~\dVint~}{c_\mathrm{s}\sqrt{8\ln 2}}\right\}^2 - 
\left(\frac{\sigma_\mathrm{thm}}{c_\mathrm{s}}\right)^2,
\label{eqn:sigmanth_ciso}
\end{equation}
where $\sigma_{\rm thm}^2$ is given by $\frac{kT_k}{m_{\rm ^{13}CO}}$,
and $m_\mathrm{^{13}CO}$ denotes the molecular mass of \tCO.
For calculations of $\sigma_\mathrm{thm}$,
we used the $\langle T_\mathrm{ex}\rangle$ map (Figure \ref{fig:meanTex}).
Over the Component 1, we obtained a mean ratio of 
$\langle\sigmanth/c_\mathrm{s}\rangle =\,$ 
2.1$\pm$0.50, corresponding to $\langle\sigmanth\rangle =\,0.34\pm$\,0.080 \kms.
The observed non-thermal contribution must be simply attributed to the random turbulent
motions of the gas
because we did not identify any systematic motions of the filament gas
(\S\ref{ss:mommaps}).\par

In the following, we deal with the Component 1 as a single filament, 
and we do not consider an interpretation proposed by 
\citet{Hacar11} that a supersonic turbulent filament is mimicked by
``twisted" subsonic filaments.
In summary, the internal motion of the filament gas is
governed by the supersonic turbulence.
Note that our conclusion differs from that by \citet{PB06}
who inferred the absence of turbulence in the filament based on their
polarization maps.

\begin{figure}
\begin{center}
\includegraphics[angle=0,scale=.45]{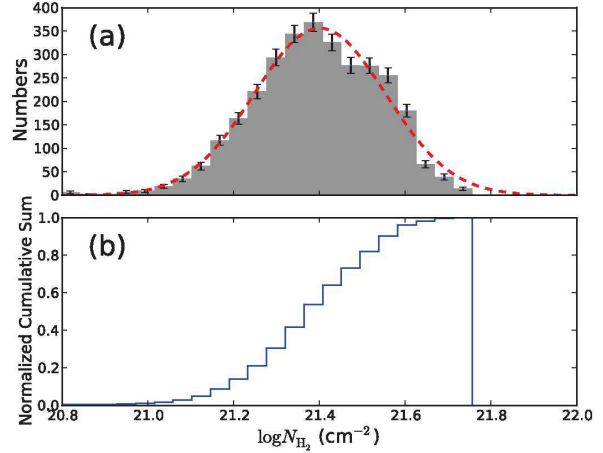}
\caption{
(a) Histogram of the molecular hydrogen column densities
($N_\mathrm{H_2}$) shown in Figure \ref{fig:Ncol}.
The dashed red curve indicates 
the best-fit log-normal function with a mean of 21.4,
i.e., 2.52$\times 10^{21}$ \cmq\ 
(corresponding to the 50\% percentile in the panel b)
and a standard deviation of 0.15.
(b) Plot of normalized cumulative sum calculated from the histogram.
\label{fig:Ncol_hist}}
\end{center}
\end{figure}

\begin{figure}
\begin{center}
\includegraphics[angle=0,scale=.38]{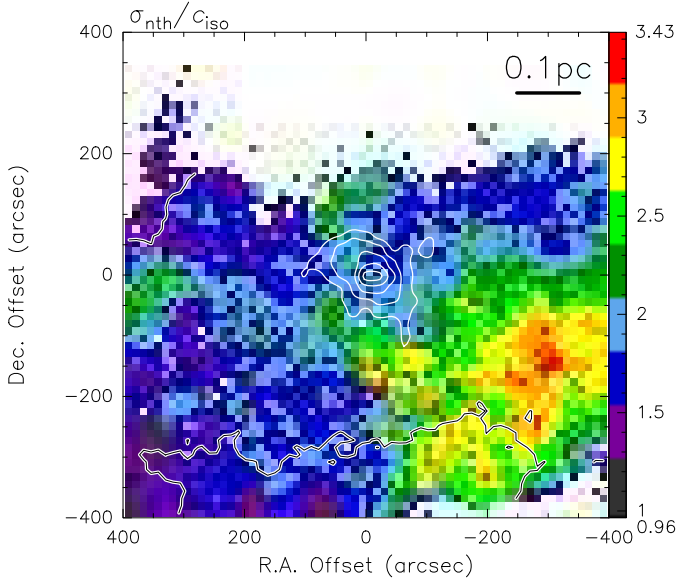}
\caption{Map of the ratio between the non-thermal velocity dispersion
and the local sound speed. See \S\ref{ss:ambient} for details.
The color bar shows the ratio in linear scale.
The 0.1-pc scale at the distance of 200\,pc is shown 
at the top right corner of the panel.
The white contours at the center represent the \HtCOp\ total map
as shown in Figure \ref{fig:totmaps},
and the black-in-white contours
in the south and the north-east of the core indicate 
the \Vcent\ $=\,-2.2$ \kms\ contours as in shown Figure \ref{fig:mommaps}.
\label{fig:sigmanthmap}
}
\end{center}
\end{figure}

\subsection{Is the Component 1 Radially Supported by the Supersonic Turbulence?}
\label{ss:ostriker_analysis}

Next we assess whether or not the filament is dynamically supported by the supersonic turbulence 
against radial collapse. This yields the first step 
to discuss how the gravitational collapse of the
GF\,9-2 core (paper I) was triggered in the supersonic turbulent filament.
For this purpose, we examine the stability of the Component 1 gas
based on a model for an isothermal cylinder in hydrostatic equilibrium
\citep{Stodolkiewicz63, Ostriker64}.
Although the mean \Tex\ map  shows that
the gas temperature in the south-western region is enhanced up to $\sim 10$\,K
(Figure \ref{fig:meanTex}),
we assume that the filament is isothermal with 7.5\,K (\S\ref{ss:mommaps})
as the first order of approximation.
The model gives the radial density distribution of,
\begin{equation}
\rho(\bar{\omega}) = \rho_\mathrm{c}
\left\{1+\left(\frac{\bar{\omega}}{H}\right)^2\right\}^{-2}.
\label{eqn:rho_cylinder}
\end{equation}
Here $\bar{\omega}$ is the radial distance from the cylinder axis, and
$H$ is the scale height given by,
\begin{equation}
H = \sqrt{\frac{2c_\mathrm{s}^2}{\pi G\rho_\mathrm{c}}} \sim \lambda_\mathrm{J},
\label{eqn:scalehight}
\end{equation}
where $\rho_\mathrm{c}$ is the central density of the filament
and $\lambda_\mathrm{J}$ the Jeans length at $\rho_\mathrm{c}$.
Therefore, the critical line mass
($m_\mathrm{line, crit}$),
above which the cylinder radially collapses,
is calculated by,
\begin{equation}
m_\mathrm{line, crit} = \int^{\infty}_0 2\pi \bar{\omega} \rho(\bar{\omega}) d\bar{\omega} = 
\frac{~2c_\mathrm{s}^2~}{G},
\label{eqn:linemass}
\end{equation}
which does not depend on $\rho_\mathrm{c}$.
To apply this model to the Component 1 where the supersonic turbulence dominates
the gas kinematics,
we replace $c_\mathrm{s}^2$ with $c_\mathrm{s}^2 + \sigma_\mathrm{nth}^2$,
and define an {\it effective critical line-mass} by,
\begin{equation}
m_\mathrm{line, crit}^\mathrm{eff} = 
\frac{~2~}{G}\frac{kT_\mathrm{kin}}{\mu m_\mathrm{H}}
\left\{1+\left( \frac{~\sigma_\mathrm{nth}~}{c_\mathrm{s}}\right)^2\right\}.
\label{eqn:Mline}
\end{equation}
This equation represents the maximum {\it line-mass} that 
can be supported by the total internal pressure of the filament.
We stress that the $\sigma_\mathrm{nth}$ term represents
the non-thermal pressures due to predominantly the supersonic turbulence (\S\ref{ss:ambient})
and probably the magnetic field as well (described in \S\ref{ss:magfield}),
although it is impossible to separate them.
Using $\langle T_\mathrm{kin}\rangle \,=\, 7.5\pm 1.0$ K (\S\ref{ss:mommaps}) 
and $\langle\sigma_\mathrm{nth}/c_\mathrm{s}\rangle\, = 2.1\pm 0.50$ (\S\ref{ss:ambient}),
we calculated an effective critical mass ($M_\mathrm{crit}^\mathrm{eff}$) of 
$51^{+32}_{-22}$ \Msun\ for the Component 1 by 
multiplying $m_\mathrm{line, crit}^\mathrm{eff}$ 
of  $41^{+35}_{-26}$ \Msun\ pc$^{-1}$
by the observed length of the filament (0.77\,pc; see Figure\,\ref{fig:totmaps}).
Comparing the $M_\mathrm{crit}^\mathrm{eff}$ value to the \MLTE\
of $24\pm 10$ \Msun\ (\S\ref{ss:Ncol}),
we suggest that 
the filament is highly likely 
gravitationally stable with respect to radial collapse
owing to the turbulent support.
Note that the filament cannot be supported only by
the thermal pressure because the LTE mass is larger than 
$m_\mathrm{line, crit}\, \times$ 0.77\,pc $\sim$ 9 \Msun\
[see Eq.(\ref{eqn:linemass})].\par

\subsection{Does the Component 1 Fragment Axially?}
\label{ss:axial_fragmentation}
\subsubsection{Estimate of the Filament Scale Height from the Column Density Map}
\label{sss:Hestimate}
To discuss the axial fragmentation of the Component 1, 
we estimated the scale height of the filament 
from the column density map (Figure \ref{fig:Ncol}) 
using the Stodolkiewicz-Ostriker model (\S\ref{ss:ostriker_analysis}). 
A column density profile of an isothermal cylinder 
supported by the thermal and turbulent pressures
can be written from Eq.(\ref{eqn:rho_cylinder}) as,
\begin{equation}
N_\mathrm{H_2}(r) = \frac{c_\mathrm{s}^2+\sigma_\mathrm{nth}^2}{\mu m_\mathrm{H}GH}
\left\{1+\left(\frac{r}{H}\right)^2\right\}^{-\frac{3}{2}},
\label{eqn:Ncylinder}
\end{equation}
where $r$ is the projected distance from the central axis of the cylinder in the plane of the sky.
We applied this equation to the Component 1 gas in the column density map (Figure \ref{fig:Ncol}).
In the model fitting
we adopted P.A. $=\,-90\arcdeg$ for the cylinder axis, 
which was forced to pass the position of the 3\,mm continuum source, and
$\langle\sigma_\mathrm{nth}\rangle \,=\,2.1 c_\mathrm{s} = $\,0.34 \kms,
setting only $H$ as a free parameter.
This is because we could not well determine the P.A. value 
and the location of the central axis as free parameters
owing to the insufficient spatial coverage of the observations.\par

\begin{figure}
\begin{center}
\includegraphics[angle=0,scale=.34]{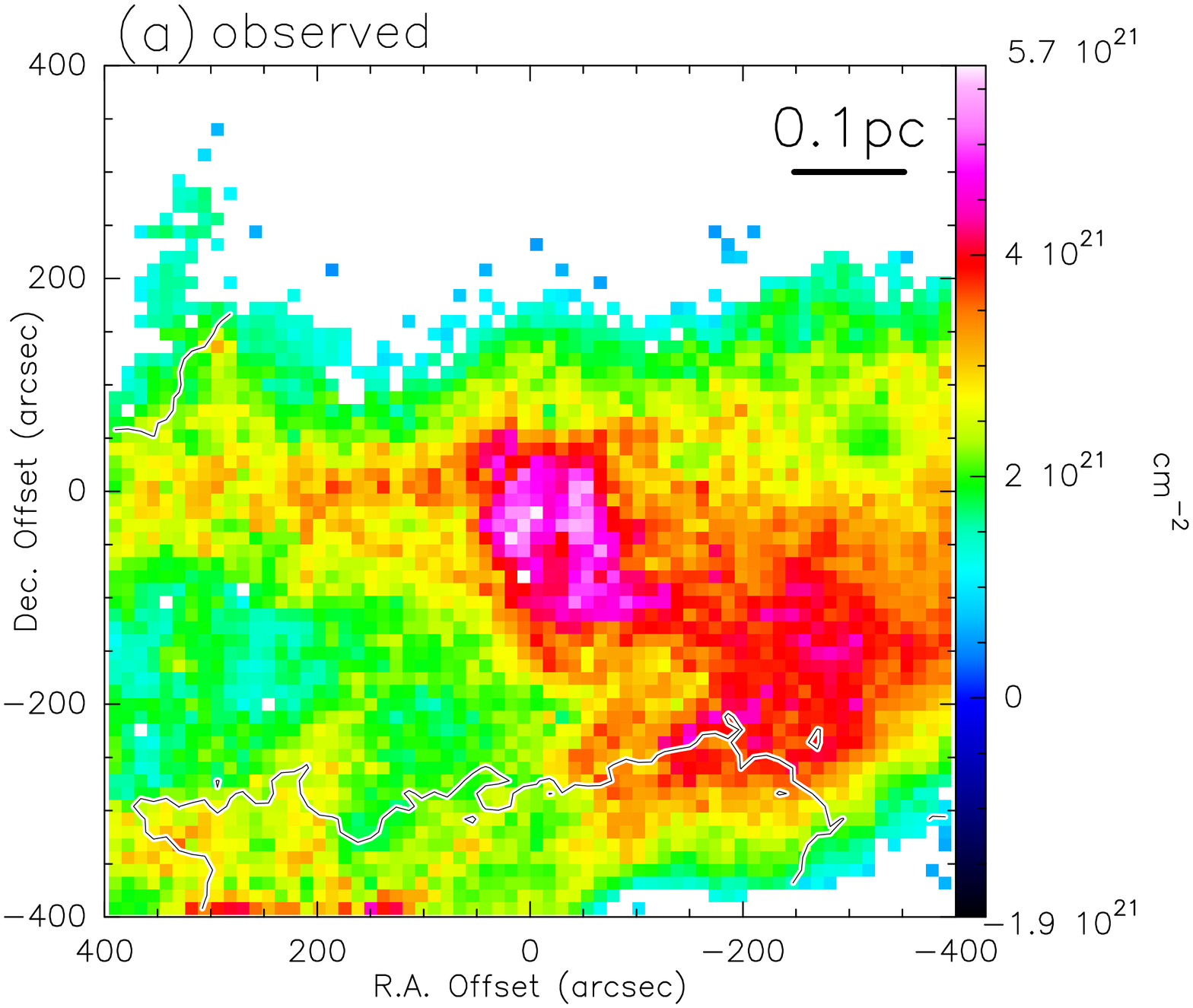}
\includegraphics[angle=0,scale=.34]{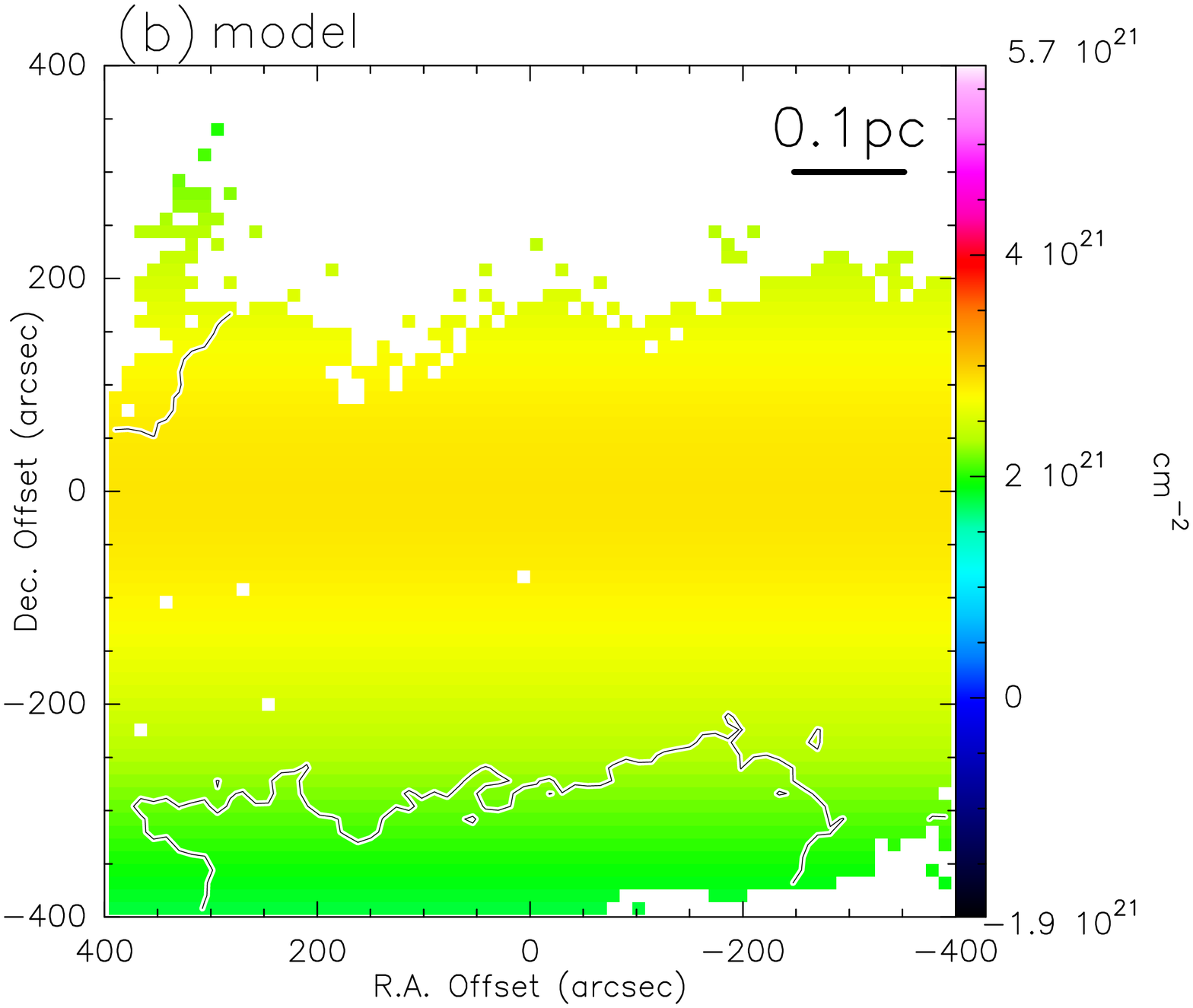}
\includegraphics[angle=0,scale=.34]{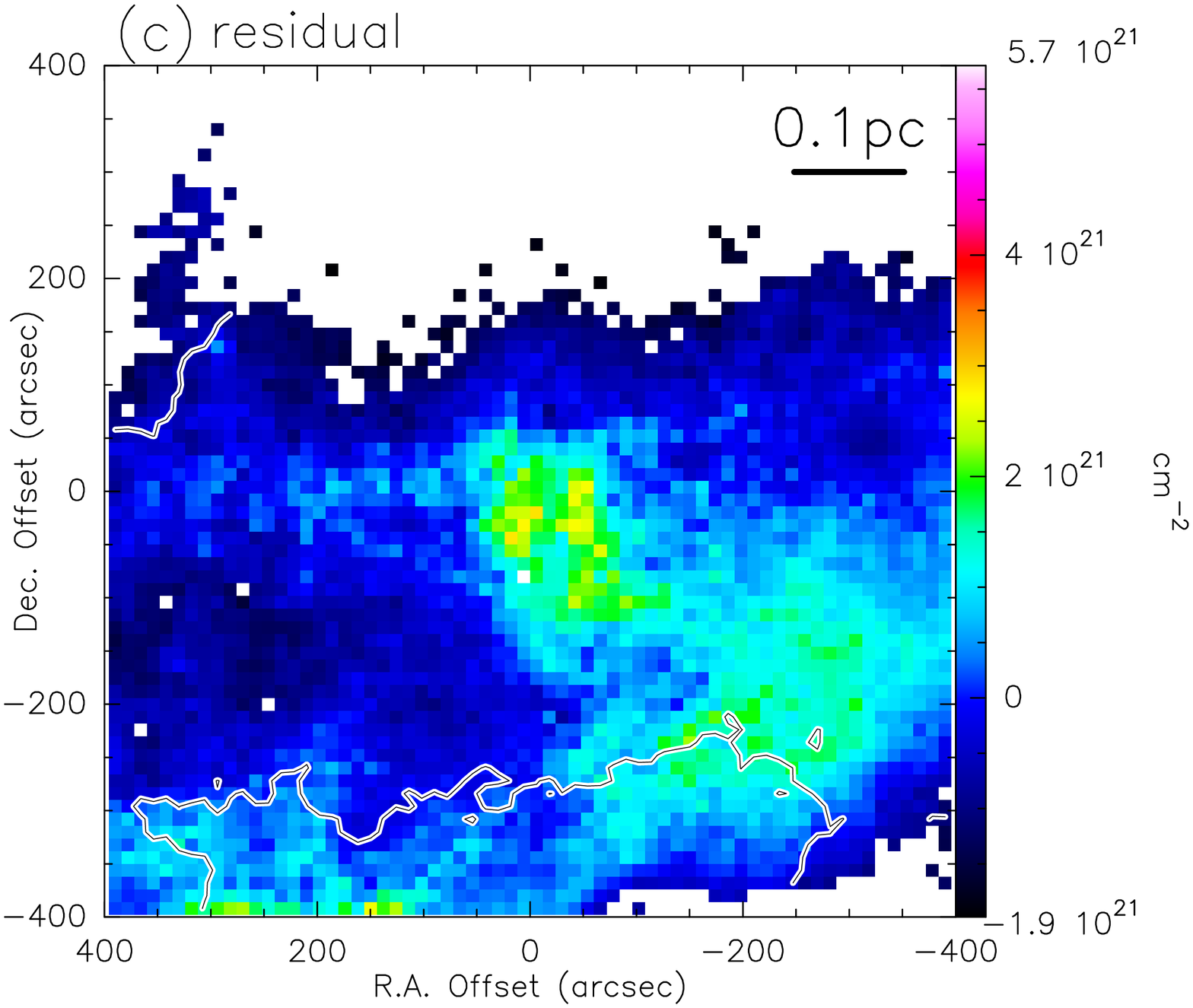}
\caption{Model fitting to the molecular hydrogen column density of the Component 1.
The panels (a), (b), and (c) show the observed column density
(Figure \ref{fig:Ncol}), model, and residual maps, respectively.
Note that all the color maps are shown in the fixed common range of the column density
(see the color bars).
See \S\ref{ss:ostriker_analysis} for details.
\label{fig:ostriker_mdl}
}
\end{center}
\end{figure}

\begin{figure}
\begin{center}
\includegraphics[angle=0,scale=.32]{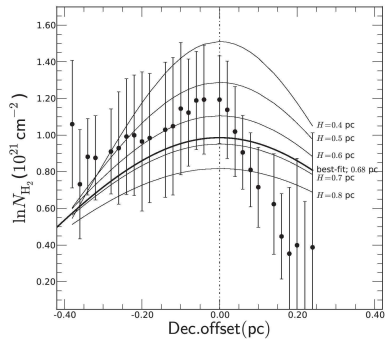}
\caption{
Radial column density profile of molecular hydrogen 
averaged along the central axis for the ``Component 1" filament.
The negative and positive sides of the Dec. offset axis, respectively, 
correspond to the south and north sides
of the $N_\mathrm{H_2}$ map (Figure \ref{fig:ostriker_mdl}a).
The thick curve presents the best-fit profile of 
the Stodolkiewicz-Ostriker cylinder model and
the thin curves show the model profiles with the $H$ values between 0.4 and 0.8\,pc.
See \S\ref{sss:Hestimate} and Appendix \ref{as:H_errors}.
\label{fig:NmolRadP}}
\end{center}
\end{figure}

Figure \ref{fig:ostriker_mdl} presents maps of the column density, the best-fit model, 
and the residual for the Component 1.
Figure \ref{fig:NmolRadP} shows the radial profile of the column density 
averaged along the R.A. direction, 
and we obtained the best-fit value of 
$H\,=\,0.68\pm 0.04$\,pc 
by considering the uncertainty in the column density 
(see Appendices \ref{as:Ntot} and \ref{as:H_errors}).
We notice that the derived width of $2H\,=\,1.4$\,pc is significantly larger than the
observed width of the filament of
$\sim$\,600\arcsec\ (see e.g., Figure \ref{fig:mommaps}),
corresponding to 0.58\,pc.
Our analysis may be also affected by the small dynamic range 
of the column density of
20.9\,$\lesssim \log N_\mathrm{H_2}/\mathrm{cm}^2\,\lesssim$\,20.8
(Figures \ref{fig:Ncol} and \ref{fig:Ncol_hist})
and possibly by the asymmetric structure of the filament 
(see Figure \ref{fig:NmolRadP}) which exists all over the observed region.
If we underestimated the peak column density, 
then the $H$ value would become small according to Eq (\ref{eqn:Ncylinder}).

The best-fit $H$ value leads to the filament width in FWHM to be
$1.5H \,=\,1.0\pm 0.03$\,pc which is significantly larger than the typical width
of $\sim 0.1$\,pc derived from the {\it Herschel} survey towards
the IC\,5146 \citep{Arzoumanian11} and B211/213 \citep{Palmeirim13} regions.
In order to assess whether or not the difference is caused by that
between the adopted models, we reanalyzed our column density map
using the Plummer-like cylinder model
\citep[see e.g., Eq.(B1) in][and references therein]{Palmeirim13}
instead of using Eq.(\ref{eqn:Ncylinder}).
Notice that the Plummer-like function \citep{Plummer1911, Nutter08} 
has three free parameters, while 
the Stodolkiewicz-Ostriker-like cylinder that we adopted has only one, i.e., $H$.
With the Plummer-like function we estimated the filament width in FWHM to be 0.38$\pm$0.06\,pc,
which is still significantly larger that those by {\it Herschel}.
We therefore conclude that the difference between the {\it Herschel} filaments
and the GF\,9 one is real.
One possible cause is that 
the CO lines in our study could not trace the higher density regions detected 
by the {\it Herschel} survey with the dust continuum emission.
Alternative cause is that
there might exist an intrinsic difference between the GF\,9 and the regions that {\it Herschel} surveyed.\par

Considering all the results as well as their limitations,
we adopt fiducial values of 
$H\,=0.3\sim 0.7$\,pc for further discussion (Table \ref{tbl:summary}).
Here the lower limit of 0.3\,pc is obtained from the
map ($2H \,=\,0.58$\,pc) while the upper limit is from
the best-fit value. 
The estimated range
of the scale height leads to
the $\rho_c$ value in Eq.(\ref{eqn:rho_cylinder})
of $(3\sim 16)\times 10^{-21}$ g \cmc\ using the relation in our model of,
\begin{equation}
\rho_c = \frac{2}{~\pi GH^2~}\left(c_\mathrm{s}^2+\sigma_\mathrm{nth}^2\right).
\end{equation}
The resultant $\rho_c$ value corresponds to 
$n_\mathrm{c}(\mathrm{H}_2) =\,800\sim 4200$ \cmc
which has reasonable consistency with the critical density required to excite
the \tCO\ $J=1$--0 transition.
It should be noticed that 
the above central density characterizes the tenuous filament gas 
surrounding the dense core \gf\ (not the dense core itself).\par

If the cylinder model is appropriate to the description of our data,
a tenuous core-like structure is identified
in the residual $N_\mathrm{H_2}$ map (Figure \ref{fig:ostriker_mdl}c).
The position and extent of the ``tenuous core" agree fairly well with 
those of the dense core observed in the
\NtwoH\ (1--0), \HtCOp\ (1--0), CCS $4_3-3_2$ and \amm\ (1,1) lines (paper I).
We therefore speculate that it is a low-density envelope 
surrounding the dynamically infalling dense core (paper III).
The LTE mass of the ``tenuous core" is calculated to be 
$M_\mathrm{TEC}\sim $\,0.8 \Msun\
where TEC denotes the tenuous envelope of the core.
Furthermore, we can estimate a ``core formation efficiency" for the \gf\ dense cloud core of
$\eta_\mathrm{\small CF}\,=\,M_\mathrm{LTE}$/($M_\mathrm{LTE}$ + $M_\mathrm{TEC}$) 
$=$ 1.3 \Msun/(1.3 \Msun + $\sim 0.8$\Msun) $\gtrsim\, 60\%$ where
\MLTE\ is the mass of the dense core traced by the \NtwoH\ and \HtCOp\ lines (paper I).\par

\subsubsection{Gravitational Stability of the Component 1 against Axial Perturbations}
\label{sss:ostrikermodel_results}
Theoretical studies by e.g., \citet{Nagasawa87} and \citet{Inutsuka97}
showed that the minimum wavelength 
($\lambda_\mathrm{min}$) which
makes an infinite filament unstable against perturbations along the major axis
is given by $\lambda_\mathrm{min}\,\sim\,4H$,
regardless of the presence of the magnetic field along the axis
[the definition of $H$ in \citet{Nagasawa87} differs by 
$1/\sqrt{8}$ times of those in e.g., \citet{Inutsuka92}.].
The line mass of the filament can be calculated from the
LTE-mass (\S\ref{ss:Ncol}) and the observed length as
24 \Msun / 0.77\,pc $=$\,31 \Msun\ pc$^{-1}$.
Therefore the critical ``clump mass" produced by the axial perturbation may be 
31 \Msun\ pc$^{-1}\times 4H\,=\,37\sim 87$ \Msun.
It is clear that the estimated range of the clump masses is one order of magnitude
larger than the LTE-masses of the \amm\ cores 
(\MLTE\ $=\,2\sim 8$ \Msun: paper II).
We therefore consider that 
the filament is gravitationally stable against the axial perturbations probably 
owing to the transverse magnetic field \citep{PB06}
which is reanalyzed in \S\ref{ss:magfield}.\par

Alternatively, the GF\,9 filament may have narrow subfilaments 
\citep{Hacar11,Hacar13}
with $H\sim$\,0.1\,pc and $n_\mathrm{c} \sim 10^4$ \cmc. 
In fact, such a sub-structure may be recognized in the \CeO\ total intensity map of 
Figure \ref{fig:totmaps}c.
If this is the case, the axial fragmentation of the subfilament 
explains the observed spatial intervals of $\sim$0.9\,pc between the \amm\ cores
(paper II) because of (4--8)$H\, =\, 0.4 \sim 0.8$\,pc.
However, this scenario has a caveat that the expected ``clump masses" given by
$m_\mathrm{line, crit}^\mathrm{eff}\times$ (4--8) $H\,=\, 16 \sim 40$\, \Msun\
are larger than the LTE-masses of the \amm\ cores.\par

In summary, our analysis suggests that 
the natal gas of the core is in dynamical equilibrium
with respect to both the radial and axial collapses.
Na\'ively speaking, this inference does not seem to be reconciled with the presence
of the dynamically collapsing core (papers I and III). 
Before resolving this issue, we assess the role of the large-scale magnetic field 
existing in the filament.\par

\subsection{Role of the Magnetic Field in the Gas Surrounding the GF\,9-2 Cloud Core}
\label{ss:magfield}

\begin{figure}
\begin{center}
\includegraphics[angle=0,scale=.41]{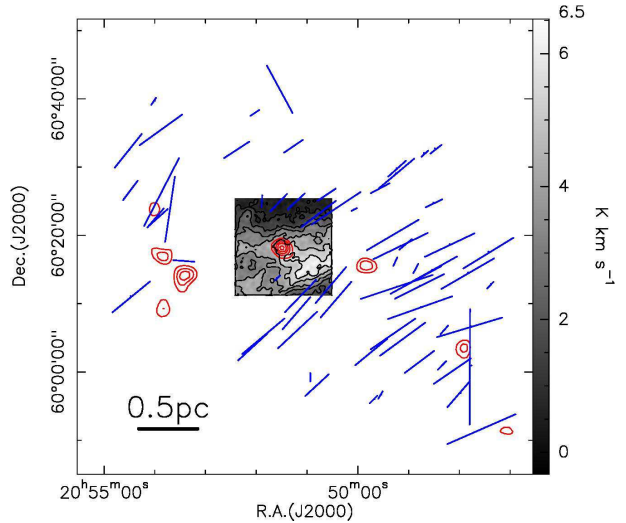}
\caption{
An overlay of the optical polarization map (blue line segments)
plotted using the data shown in Table 1 of \citet{PB06} 
on the total integrated intensity maps of the \amm\ ($J,~K$)\,$=$\,(1,1) 
(red contours; paper II) and the \tCO\ (1--0) 
(grey scale plus contours; see Figure \ref{fig:totmaps}a; 
{starting from the 10\sgm\ level with the 10\sgm\ intervals}) emission.
The red contours start from the 6\sgm\ levels with the 3\sgm\ intervals 
as shown in paper II.
The 0.5\,pc scale is shown at the bottom left corner of the panel.
The grey-scale bar at the right-hand side shows the integrated intensity
of the inserted \tCO\ map.
\label{fig:polmap}
}
\end{center}
\end{figure}

Figure \ref{fig:polmap} shows a comparison between the optical polarization map
taken from \citet{PB06} and the distribution of the dense cores, including the candidates,
seen in the \amm\ emission (paper II).
Since the directions of the optical polarization angles are thought to be 
parallel to the magnetic field, 
\citet{PB06} pointed out that
the magnetic field is almost perpendicular to the filamentary dark cloud \citep{Schneider79}.
Such a large-scale configuration is also found in the Taurus molecular cloud
\citep[see e.g.,][and references therein]{Chapman11}
suggesting that the filament formed 
through compression of the diffuse gas by external pressure
along the magnetic field
or/and through gas accretion due to self-gravity along the magnetic field.
However, 
this scenario should be reexamined in more detail because
the optical polarization
angles in the vicinity of the \gf\ core are inclined by $\sim 45\degr$
with respect to the Component 1 filament axis.
It is also interesting that the polarization angles are almost parallel to the
elongation of the Component 2 gas.\par

An alternative mechanism to produce such a large-scale configuration is that the observed field 
is the projected component of a toroidal field.
Because of no firm evidence,
we do not discuss this interpretation here.
On the other hand, the magnetic field in the filament should be 
far from good alignment at small scales owing to micro-turbulence
\citep[e.g.,][]{White77, Heyer08}.
However we can not discuss this issue
because the angular resolution of the 
polarization map is too coarse to assess the role of the magnetic field
at scales smaller than the core size of $\sim$0.1\,pc.\par

It is well known that not only supersonic turbulence but also magnetic field prevent 
interstellar gas from collapsing due to its self-gravity.
Applying the Chandrasekhar \& Fermi method \citep{CFmethod53},
\citet{PB06} estimated the magnetic field strength
in the plane of the sky, $|\vec{B}_\mathrm{pos}|$.
They derived $|\vec{B}_\mathrm{pos}|$ of $170\pm 56\,\mu$G for the ``Core"
which was observed by the CS (2--1) emission \citep{Ciardi00}.
To calculate $|\vec{B}_\mathrm{pos}|$, they used Eq. (4) in \citet{Crutcher04} of
$|\vec{B}_\mathrm{pos}|\approx 9.3 \sqrt{n(\mathrm{H_2}})
~\frac{\Delta v_\mathrm{\scriptsize FWHM}~}{\delta\phi}~\mu\mathrm{G}$
where $\delta\phi$ is the dispersion in a polarization angle in degrees.
Since we have the revised mean velocity width of 
$\Delta v_\mathrm{\scriptsize FWHM}\,=\, 0.69\pm 0.12$ \kms\ (\S\ref{ss:mommaps}) and
the number density of $n_\mathrm{c}(\mathrm{H_2}) \,=\,800\sim 4200$ \cmc\
(\S\ref{ss:ostriker_analysis}), 
we recalculated $|\vec{B}_\mathrm{pos}|$ of $55\pm 30$ $\mu$G.
Here we assumed that the $\delta\phi$ value of $5.9^\circ$
derived from the infrared data towards the CS (2--1) core
\citep{PB06} holds in the region we observed.\par

With the magnetic field strength, we estimate the magnetic pressure of
$\log P_\mathrm{mag}\,=\,\log\left( B^2/8\pi\right) \,=\,-10.6\sim -9.5$
which is larger than the effective internal pressure of the filament given by 
$\log P_\mathrm{eff}\equiv\,
\log\left\{\rho\left(c_\mathrm{s}^2 + \sigma_\mathrm{nth}^2\right)\right\}
\,=\,-11.3\sim -10.6$.
Considering the ratio of
$P_\mathrm{eff}/P_\mathrm{mag}\,=$\,0.03$\sim$0.2
(Table \ref{tbl:summary}),
we speculate that the significant magnetic pressure can 
support the filament against the axial fragmentation.\par

Subsequently, we examine whether or not the large scale 
magnetic field can support the \gf\ core through a comparison of,
\begin{eqnarray}\nonumber
\frac{
~\left|\frac{~M~}{~\Phi~}\right|_\mathrm{GF\,9-2}}
{\left|\frac{~M~}{~\Phi~}\right|_\mathrm{critical}}
\,=\, 
\frac{
\frac{~M_\mathrm{LTE}~}{~|\vec{B}_\mathrm{pos}|\,R_\mathrm{eff}^2~}
}{\frac{~M_0~}{~B_0R_0^2~}}
\,=\, \\ 
\frac{
~\frac{(1.3\pm 0.5)\,M_\odot }{{\bf (55\pm 30)} \mu\mathrm{G} \cdot (0.032\,\mathrm{pc})^2}~
}{ \left(\frac{5}{9G}\right)^{1/2}} 
\,\sim\,
\frac{{(4.8^{+9.9}_{-2.9})}\times 10^3}{2900}
\,\sim\, 2_{-1}^{+3}
\label{eqn:magflux}
\end{eqnarray}
where the denominator is the critical mass-to-magnetic-flux ratio for a uniform spherical cloud,
and the numerator is the ratio for the \gf\ dense core traced by the
\HtCOp\ and \NtwoH\ lines (the \MLTE\ and \Reff\ values are taken from Table 8 in paper I),
which exists in the large-scale magnetic field (Figure \ref{fig:polmap}).
Although the non-dimensional coefficient in the relation of 
$\left|\frac{~M~}{~\Phi~}\right|_\mathrm{cr}\propto \frac{1}{\sqrt{G}~}$
differs in various models 
\citep[see e.g.,][and references therein]{Lang80, Shu92, Lequeux05, Bodenheimer11},
such a difference is negligible compared to the
uncertainty in the estimate of the magnetic field strength (\S\ref{ss:magfield})
and that in the \MLTE\ estimate due to the uncertain fractional abundances of the tracers.
The above comparison suggests that the \gf\ core is in magnetically
super critical state, which does not contradict
the fact that the core is dynamically collapsing
because no magnetic fields can halt such collapse once it has begun.

\begin{deluxetable}{lclc}
\tabletypesize{\scriptsize}
\tablecaption{Physical Properties of the Filament Gas Around the GF\,9-2 Core\label{tbl:summary}}
\tablewidth{0pt}
\tablehead{
\colhead{Property} & \colhead{Value} & \colhead{Unit} & \colhead{Section} 
}
\startdata
$T_\mathrm{gas}$\tablenotemark{a} & 7.5$\pm$1.0 & K & \S\ref{ss:tauTex}  \\
$\langle N_\mathrm{H_2}\rangle$\tablenotemark{b} & (2.7$\pm$0.9)$\times 10^{21}$ & \cmq\ & \S\ref{ss:Ncol} \\
$\langle\sigma_\mathrm{nth}/c_\mathrm{s}\rangle$\tablenotemark{c} & 2.1$\pm$0.5 & ... & \S\ref{ss:ambient} \\
\MLTE/$M_\mathrm{crit}^\mathrm{eff}$ & 0.5$\pm$0.2 & ... & \S\ref{sss:Hestimate} \\
$H$\tablenotemark{d} & $0.3\sim 0.7$ & pc & \S\ref{sss:Hestimate} \\
$n_\mathrm{c}(\mathrm{H_2})$\tablenotemark{e} & $~\,800\sim 4200$ & \cmc\ & \S\ref{ss:ostriker_analysis} \\
$|\vec{B}_\mathrm{pos}|$\tablenotemark{f} & $55\pm 30$ & $\mu$G & \S\ref{ss:magfield} \\
$P_\mathrm{eff}/P_\mathrm{mag}$ & $0.03\sim 0.2$ & ... & \S\ref{ss:magfield} \\
\enddata
\tablenotetext{a}{Gas temperature}
\tablenotetext{b}{Mean column density of H$_2$ molecules}
\tablenotetext{c}{Mean Mach number defined by the ratio of non-thermal velocity dispersion ($\sigma_\mathrm{nth}$) to the sound velocity ($c_\mathrm{s}$)}
\tablenotetext{d}{Scale height}
\tablenotetext{e}{Central volume density of H$_2$ molecules}
\tablenotetext{f}{Magnetic field strength in the plane of the sky}
\end{deluxetable}

\subsection{Core Formation in the Filament}
\label{ss:cf_scenario}

The discussion so far indicates that the natal gas of the \gf\ core is 
supported by the turbulent and magnetic pressures against its self-gravity. 
In Table \ref{tbl:summary}, we summarize the derived physical properties
of the natal filamentary gas.
Recall that the central $30\arcsec$ region of the dense core,
corresponding to a diameter of $\sim$0.06\,pc at $d =$\,200\,pc, 
is dynamically infalling onto the protostar (papers I and III).
These facts reinforce our assertion that
a dynamically collapsing core formed in a gravitationally stable filament. Such a scenario immediately raises the following questions:
how have the turbulence and the magnetic fields decayed locally 
at the spatial scale of the core?;
can such dissipation determine the initial conditions of the core collapse?
The answer to the former may be
that such a spatial scale was set by the Jeans length 
of the tenuous gas.
As for the latter,
we require that 
the dissipation time scales 
of the  supersonic turbulence ($t^\mathrm{disp}_\mathrm{turb}$) and
the magnetic fields ($t^\mathrm{disp}_\mathrm{mag}$)
are less than the free-fall time ($t_\mathrm{ff}$)
of the gas traced by the \tCO\ line.
After addressing these issues, we discuss the nature of the
south-western condensation.\par

\subsubsection{The GF\,9-2 Core: An Unstable Core in the Stable Filament}
\label{sss:cf_scenario_gf92}
Theoretical studies have predicted the presence of 
gravitationally unstable cores in a stable filament.
For instance, 3D simulations by \citet{Klessen00, Klessen05} showed that
density enhancements caused by strong shocks can produce gravitationally
unstable dense cores,  
despite the fact that parental clouds are being prevented from the global collapse 
because of the turbulent support. 
In other words, a local collapse can occur 
even in the turbulent-supported self-gravitating global medium
\citep{Klessen00}.
Although their calculations did not include the effect of magnetic fields,
their results are reconciled with
the previous low-resolution 2D calculations 
considering magnetic fields\citep{Enrique96}.
Subsequent numerical simulations \citep[e.g., ][]{MacLow99, Ostriker01, Enrique05, Heitsch09}
with magnetic fields
claim that the core formation must have completed within the global free-fall time of the
natal cloud gas. A recent theoretical study by \citet{Leao13} demonstrated
that magnetic flux can be quickly removed from a cloud core by magnetic reconnection diffusion,
which removes the magnetic flux through reconnection of the field lines by the 
co-existing turbulence (not by ambipolar diffusion)
with a time scale as short as $t_\mathrm{ff}$.\par

Contrary to these theoretical predictions,
we propose a simple scenario that 
the supersonic turbulence 
and the magnetic fields
locally decayed at the 
spatial scale comparable to the Jeans length of 
the filament gas:
$\lambda_\mathrm{J} = \,0.3\sim\,0.6$\,pc for \Tkin $=\,$7.5 K and 
$n_\mathrm{H_2}=\,800\sim 4200$ \cmc\ (\S\ref{sss:Hestimate}).
Once the dissipation of the turbulence and the magnetic fields
had occurred locally over a region with size scale of the Jeans length, 
the region lost the support against self-gravity and contracted into
a compact gas clump with 0.1 pc size, i.e., a typical size of a low-mass
star forming cloud core, by 
increasing its density as high as 
$\sim 10^3$ \cmc$\times (\lambda_\mathrm{J}/0.1\,\mathrm{pc})^3 
=\,(3\sim\,22)\times 10^4$ \cmc.
With this number density, 
the \amm\ (1,1) lines (critical density of $n_\mathrm{crit}\sim\,10^4$ \cmc) and 
the \HtCOp\ (1--0) and \NtwoH\ (1--0) transitions ($n_\mathrm{crit}\sim\,10^5$ \cmc)
can be collisionally excited.
Because the latter higher density tracers were detected towards the \gf\ core, 
the above argument may favor the larger side of
$\lambda_\mathrm{J}\,\sim\,0.6$\,pc.
An alternative idea is that 
the core formed as a result of
internal shock in the filament.
However, this is unlikely because the shock compression with a Mach number 
($\cal{M}$) of $\sim 2$ (\S\ref{ss:ambient})
would increase the gas density by a factor of $\cal{M}$$^2$ in the case
of the J-type shock \citep{Spitzer78},
i.e., up to at most $\sim 10^3$ \cmc\ $\times 2^2 < 10^4$ \cmc.
Clearly this is lower than the critical densities of the 
\NtwoH\ and \HtCOp\ lines.\par

Next, we examine whether the relation of 
$t^\mathrm{disp}_\mathrm{turb} \lesssim t_\mathrm{ff}$ holds or not.
Assuming that the core formed through the local dissipation of the supersonic turbulence
with a spatial scale of 
$\lambda_\mathrm{J}\,=\,0.3\sim\,0.6$\,pc,
the dissipation time scale is given by, 
\begin{equation}
t^\mathrm{disp}_\mathrm{turb}\sim \frac{\lambda_\mathrm{J}}{~\sigma_\mathrm{nth}~}
\,\sim\, 10^6
\left(\frac{\lambda_\mathrm{J}}{0.3\sim 0.6\,\mathrm{pc}}\right)
\left(\frac{\sigma_\mathrm{nth}}{0.34\,\mathrm{km\,s^{-1}}}\right)^{-1} ~\mathrm{yrs},
\label{eqn:t_dispturb}
\end{equation}
for the natal Component 1 gas. 
The other time scale to assess the core formation process is the free-fall time of, 
\begin{equation}
t_\mathrm{ff}\,=\,\sqrt{\frac{3\pi}{~32G\rho~}}
\,\sim\,10^6
\left(\frac{n_\mathrm{H_2}}{~800\sim 4200\,\mathrm{cm^{-3}}~}\right)^{-1/2} ~\mathrm{yrs}.
\end{equation}
The two time scales are comparable to each other within large uncertainties,
hence we conclude that $t^\mathrm{disp}_\mathrm{turb}\sim t_\mathrm{ff}$ holds.\par

Last, we attempt to give some constraints on 
the dissipation time scale of the magnetic fields
($t^\mathrm{disp}_\mathrm{mag}$).
The upper limit of $t^\mathrm{disp}_\mathrm{mag}$ may be estimated by the
ambipolar diffusion time scale ($t_\mathrm{AD}$)
for a spherical cloud with a radius of $R\,\sim\,\lambda_\mathrm{J}/2$
assuming a typical ionization degree in the ISM
[see e.g., Eq.(10.5) in \citet{SP05} and Eq.(2.51) in \citet{Bodenheimer11}] as,
\begin{eqnarray}\nonumber 
t^\mathrm{disp}_\mathrm{mag} & \lesssim t_\mathrm{AD}\sim 
10^5-10^8
\left(\frac{n_\mathrm{H_2}}{~800\sim 4200\,\mathrm{cm^{-3}~}}\right)^{\frac{\,3\,}{2}}\times \\
 & \left(\frac{B}{~55\pm 30\,\mu\mathrm{G}~}\right)^{-2}
\left(\frac{R}{~0.2\sim 0.3\,\mathrm{pc}~}\right)^{2} ~\mathrm{yrs}.
\label{eqn:t_ad}
\end{eqnarray}
The lower limit of $t^\mathrm{disp}_\mathrm{mag}$ might be given by the time scale of 
the magnetic reconnection diffusion ($t_\mathrm{MRD}$) 
which should be comparable to or less than $t_\mathrm{ff}$ \citep{Leao13}.
Hence we may have a very robust constraint of
$t_\mathrm{ff} \lesssim t^\mathrm{disp}_\mathrm{mag} \lesssim 100\, t_\mathrm{ff}$.
We, however, speculate that $t^\mathrm{disp}_\mathrm{mag} \sim 100\, t_\mathrm{ff}$
is very unlikely because the Eq.(\ref{eqn:t_ad}) assumes a constant density and
a uniform magnetic field for a quiescent cloud.\par

We therefore discuss that the formation of the \gf\ core was 
triggered by the local dissipation(s) of the supersonic turbulence
and probably the magnetic fields
with the time scale comparable to 
$t_\mathrm{ff}$,
i.e., the ``fast mode'' of star formation (\S\ref{s:intro}).
Namely this scenario naturally produces an unstable core collapsing 
in a runaway fashion.\par

\begin{figure}
\begin{center}
\includegraphics[angle=0,scale=.37]{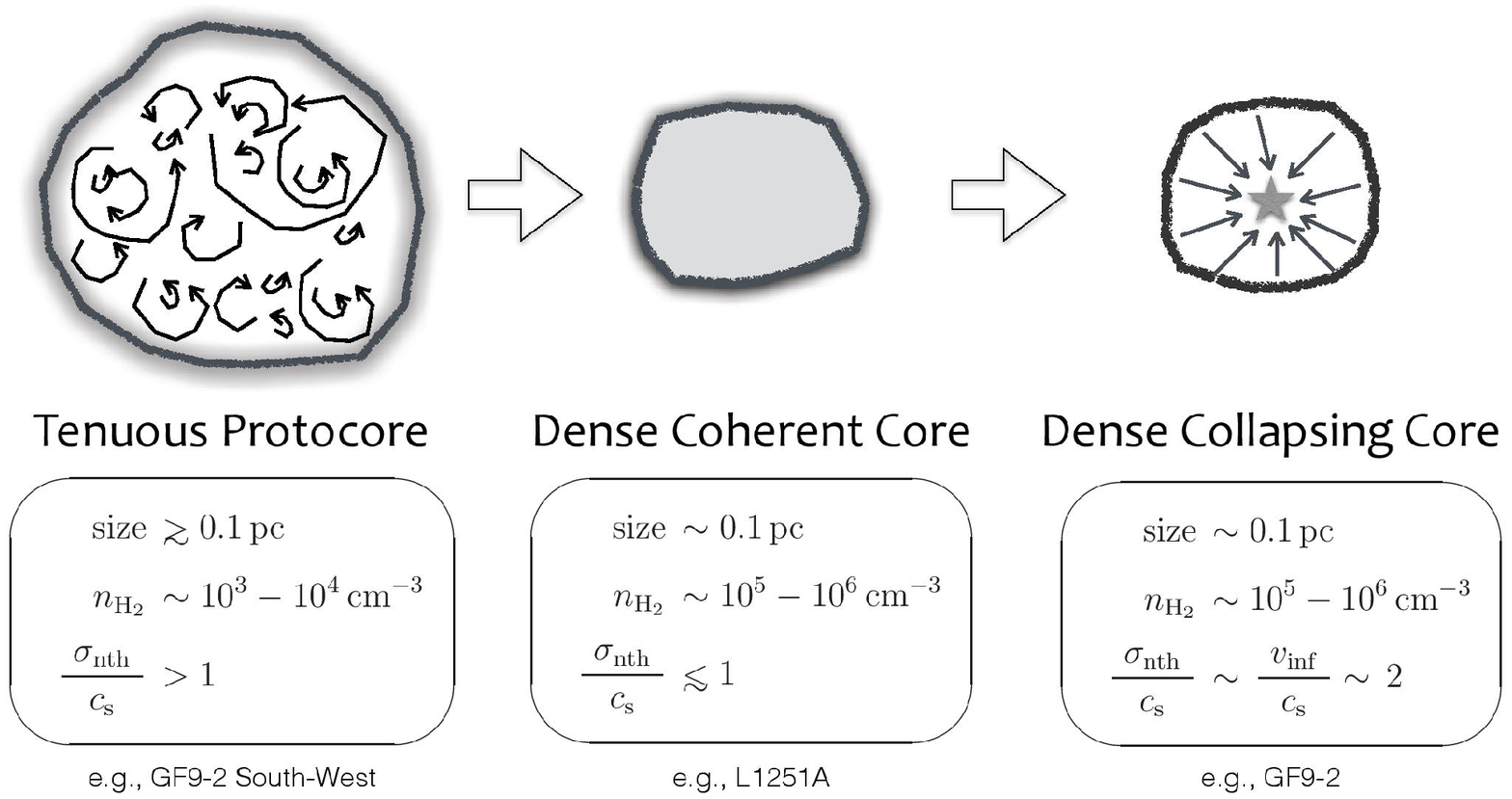}
\caption{
Schematic illustration of the proposed scenario for the evolution of a low-mass 
star-forming cloud core in the ``fast mode" of star formation,
i.e., the runaway collapse scenario \citep{Larson69, Penston69, Hunter77}.
The dense core of L1251\,A is proposed to be a ``coherent core''
which has $\langle\sigmanth/c_\mathrm{s}\rangle\lesssim 1$ 
\citep{Goodman98, Barranco98}.
Here \sigmanth\ denotes non-thermal velocity dispersion,
$c_\mathrm{s}$ isothermal sound velocity, and $v_\mathrm{inf}$ infall velocity of the gas.
See \S\ref{s:intro} and \S\ref{sss:cf_scenario_sw} for details.
\label{fig:scenario}
}
\end{center}
\end{figure}

\subsubsection{The South-Western Condensation with Velocity Width Enhancement: \\
A ``Protocore" Prior To a Dense Core?}
\label{sss:cf_scenario_sw}

We discuss the nature of another well-defined object discovered by our
observations, i.e., the south-western condensation 
where both the temperature and velocity width of the gas are enhanced.
Neither indirect evidence for star formation,
such as the presence of a point-like infrared source, masers, radio continuum emission, 
and a molecular outflow 
nor dense gas emission,
such as \amm (1,1) lines,
has been reported  towards the region
\citep[][paper II]{Ciardi98, Ciardi00, PB06}.
Hence, the \CeO\ (1--0) emission shown in Figure \ref {fig:totmaps}c
is the densest gas tracer detected towards the region.
As mentioned above, 
we reject the possibility that the condensation represents a molecular outflow lobe
driven by the protostar embedded in the \gf\ core for the following reasons.
First, 
we did not detect any high velocity wing emission
in the \CO\ spectrum towards the condensation (see Figure \ref{fig:spcomparison}b),
which is evidence for a well-developed outflow.
Second, our recent interferometric observations of the \CO\ (3--2)
line with the SMA clearly demonstrated the presence of a compact
($\sim 5\times 10^{-3}$\,pc)
molecular outflow
driven by the 3\,mm continuum source in the \gf\ core \citep[][Furuya et al. in prep.]{rsf14}.
Therefore we argue that the south-western condensation may be 
a precursor of a dense cloud core, i.e., {\it protocore}.\par

In order to verify such an interpretation,
we need to assess the dynamical state of the condensation,
hence we estimated the spatial extent of the condensation
using the results from Figures \ref{fig:Ncol} and \ref{fig:Ncol_hist}.
With the 90\% percentile of 
$N_\mathrm{H_2}\,=\, 3.9\times 10^{21}$ \cmq\ 
in Figure \ref{fig:Ncol_hist}
a closed contour is uniquely defined for the condensation.
In Figure \ref{fig:Ncol}, the 70\% percentile of 
$N_\mathrm{H_2}\,=\, 3.6\times 10^{21}$ \cmq\ encloses
both the \gf\ core and the condensation by a single contour.
We therefore judged that these contours should give us a robust
size range which is presented by
an effective radius of $r_\mathrm{eff}\,=$ 0.09$\sim$0.14 pc.
Here we used $r_\mathrm{eff}\,= \sqrt{A/\pi}$ where $A$ is the area 
enclosed by the contour.
We obtained a column density range of 
$(3.6\sim 3.9)\times 10^{21}$ \cmq,
leading to an \MLTE\ range of 1.9$\sim$4.3 \Msun\ 
with the fixed X(\tCO) value.
We also estimated that 
$\Delta v_\mathrm{FWHM}$ ranges $\,1.1\sim 1.2$ \kms\ 
from Figure \ref{fig:mommaps}c
(corresponding to the velocity dispersion range of 0.47$\sim 0.50$ \kms), 
yielding $\langle\sigmanth/c_\mathrm{s}\rangle\sim 3$.
These quantities lead to an energy balance 
based on the Virial theorem
between the supersonic turbulence and the self-gravity for a uniform spherical cloud of,
\begin{equation}
\frac{~E_\mathrm{turb}~}{~|E_\mathrm{grav}|~} =
\frac{
	\frac{~3~}{2}M_\mathrm{LTE}\sigma^2}{~
	\left|-\frac{~3~}{5}\frac{GM_\mathrm{LTE}^2}{r_\mathrm{eff}}\right|~}
\,=\,1\sim 13,
\label{eqn:swbalance_turb}
\end{equation}
suggesting that the condensation may be dispersed by the turbulence.
It should be noted that, if we include the uncertainty in the $X$(\tCO) (a factor of 3)
for the $M_\mathrm{LTE}$ estimate (\S\ref{ss:Ncol}),
the possibility that the condensation is gravitationaly bound
is not completely ruled out.
On the other hand, the south-western condensation 
seems in magnetically subcritical state
because of,
\begin{equation}
\frac{\left|\frac{~M~}{~\Phi~}\right|_\mathrm{protocore}}{\left|\frac{~M~}{~\Phi~}\right|_\mathrm{critical}}
\,=\,
\frac{~400\sim 5000~}{2900}\,=\, 0.1\sim 2,
\label{eqn:swbalance_magflux}
\end{equation}
using Eq.(\ref{eqn:magflux})
for the above \MLTE\ and $r_\mathrm{eff}$ ranges.\par

If the south-western condensation is not gravitationally bound,
it must be a transient object which will eventually disperse by the turbulence
with the time scale of $2r_\mathrm{eff}/\sigma\sim 10^5$ yrs.
We speculate that 
it might be a ``failed core" 
in \citet{Enrique05}, a bi-product from core formation.
Alternatively, if the magnetic pressure overcomes the turbulent one,
the magnetic fields should sustain the condensation.\par

If the condensation is gravitationally bound,
it may be a precursor of a low-mass star-forming dense core, {\it protocore},
which is at an evolutionary phase before the dissipation of the supersonic
turbulence (see Figure \ref{fig:scenario}).
More specifically, we consider that the evolutionary phase of this object is
prior to the coherent core phase,
which is characterized by $\langle\sigmanth/c_\mathrm{s}\rangle\lesssim 1$ 
\citep{Goodman98, Barranco98, Caselli02, Pineda10}.
Such a {\it protocore}
interpretation is consistent with the following two facts.
First, the object is two times
larger than that of a typical low-mass star-forming core.
If the object 
contracts to half its size,
its mean density would increase up to 
$\sim 10^4$ \cmc\ 
which can be traced by e.g., collisionally
excited \amm\ (1,1) emission.
Second, the object is massive ($\sim 5$\Msun) enough to evolve towards 
a typical cloud core, regardless of an $\eta_\mathrm{\small CF}$ value (\S\ref{sss:ostrikermodel_results}).
Since the GF\,9 filament is likely gravitationally stable 
(\S\ref{ss:ostriker_analysis} and \S\ref{ss:axial_fragmentation}), we argue
that the formation of the {\it protocore} was triggered by 
a collision between the Components 1 and 2.
In this scenario, the velocity width enhancement is interpreted 
as the sum of the turbulent motions in the two components and 
the radial velocity difference between them.
The difference is estimated to be $\Delta v_\mathrm{diff}\,\sim 1$ \kms\ from 
Figures \ref{fig:spcomparison}b and \ref{fig:mommaps}b, 
suggestive of a weak collision with $\cal{M}$ $\sim 2-3$.
It is a non-trivial 
issue to estimate analytically how much the gas temperature
is raised (\S\ref{ss:tauTex}) due to the weak collision 
(it is possibly a C-type shock).
Nevertheless, we believe that the cloud-cloud collision scenario
reasonably explains the observed enhancements
of the gas temperature and velocity width.
Similar to the case of the \gf\ core (\S\ref{sss:cf_scenario_gf92}),
an unstable core is anticipated to form under the conditions
with a collision time scale of 
$t_\mathrm{coll}\sim\, 2r_\mathrm{eff}/\Delta v_\mathrm{diff}\sim 10^5$ yrs,
which is comparable to 
$t_\mathrm{ff}
(\sim t_\mathrm{turb}^\mathrm{disp}\sim t_\mathrm{mag}^\mathrm{disp}$) of $10^5$ yrs
for the condensation gas.\par

\section{Summary}
\label{s:summary}
Analyzing the CO isotopologue data taken with the Nobeyama 45\,m telescope,
we have studied the physical properties of the low density ($\sim 10^3$ \cmc ) filament gas surrounding 
the \gf\ dense cloud core which harbors the exceptionally young protostar.
Our main results are summarized as follows.

\begin{enumerate}

\item The  \tCO\ map covering a $\sim 0.78\times 0.78$ pc$^2$ 
square field centered on the core clearly shows the filamentary morphology.
The \CeO\ map demonstrates that the GF\,9-2 core formed at the local intensity maxima
of the filament. 
Solving the optical depths and excitation temperatures of the CO isotopologue lines, 
we made the opacity-free \tCO\ spectra which were used to
produce spectral momentum maps. 

\item On the basis of the spatial and velocity structures of the filament gas, 
we identified two gas components,
one of which is the natal filament of the \gf\ dense core (Component 1).
The centroid velocity map shows
that the Component 1
is confined in a rather narrow LSR-velocity range of $\sim 0.6$ \kms,
and does not show any systematic motions. 
In contrast, the velocity width 
is enhanced up to $\Delta v_\mathrm{FWHM}\,=$ 1.4 \kms\ to the south-west of the core.
A mean velocity dispersion of the south-western region is $\sigma\,=$ 0.47 \kms.

\item Using the optical depth, excitation temperature, and velocity width maps,
we calculated the H$_2$ column density.
The column density map clearly traces the natal filament of the \gf\ core, 
and the core is located at the local column density peak in the filament.
Furthermore, the column density histogram is well described by 
a log-normal function, suggesting that 
the supersonic turbulence governs the density structure of the filament.

\item The mean temperature of the gas in the observed area is 7.5\,K with a
standard deviation of 1.0\,K.
Subtracting the contributions of thermal gas motions 
from the $\Delta v_\mathrm{FWHM}$ map, we made 
a ratio map between the non-thermal velocity dispersion and the local sound speed 
($\sigma_\mathrm{nth}/c_\mathrm{s}$).
The map clearly demonstrates that the natal tenuous filament is 
in a supersonic turbulent state whose Mach number ranges from 0.96 to 3.4 with a mean of 2.1.

\item\label{sum:filament} Considering the turbulent pressure into 
a model of an isothermal cylinder in hydrostatic equilibrium,
we assessed the dynamical stability of the filament.
The maximum mass that can be radially supported by the internal pressures 
(predominantly the turbulent pressure)
is estimated to be $M_\mathrm{crit}^\mathrm{eff}\,=\,51^{+32}_{-22}$ \Msun.
On the other hand, we obtained the LTE mass of $24\pm$10 \Msun,
suggestive of a gravitationally stable state against radial collapse. 
Furthermore, analyzing the column density map on the basis of the isothermal cylinder model,
we estimated a scale height of the filament to be 
$H\,=\,0.3\sim\,0.7$\,pc,  
yielding the central number density of 
$n_\mathrm{c}(\mathrm{H_2}) =\,800\sim 4200$ \cmc.
Since the \amm\ core masses in the filament are smaller than those
expected in the axial unstable modes,
the filament is likely to be gravitationally stable against axial fragmentation as well.

\item With the $\langle\sigma_\mathrm{nth}\rangle$ of 0.34$\pm 0.80$ \kms\ and
the $n_\mathrm{c}(\mathrm{H_2}) =\, 800\sim 4200$ \cmc,
we recalculated the strength of 
the large-scale well-aligned magnetic fields 
in the natal filament gas to be $|B| =\,55\pm 30\,\mu$G
by following 
the Chandrasekhar \& Fermi method in \citet{PB06}.

\item The transverse magnetic field can support the filament against 
the axial collapse, because the magnetic pressure appears to exceed the internal
(thermal and turbulent) gas pressure.  
In contrast, the \gf\ core is in a magnetically 
super critical state through a 
comparison of the mass-to-magnetic-flux ratio of the core
with the theoretical critical value.
This inference agrees with the previous results that the core is
dynamically collapsing.

\item The dissipation time of the supersonic turbulence 
is comparable to the free-fall time of the natal tenuous gas of
$10^6$ yrs within the uncertainties.
Although the dissipation time of the magnetic fields has large uncertainties, 
we consider that the local dissipation(s) of the turbulence and the magnetic fields made
a part of the filament gas unstable,
resulting in the formation of the gravitationally unstable \gf\ core.

\item The south-western condensation,
where the gas temperature and velocity width are enhanced, 
may be a precursor of a low-mass star forming cloud core, {\it protocore},
if it is gravitationally bound.
The formation of the south-western {\it protocore} must have been
triggered by a collision between the two gas components.
Because of $\langle\sigma_\mathrm{nth}/c_\mathrm{s}\rangle\sim 3$, 
the evolutionary stage of the south-western {\it protocore} is highly likely before
the phase of the coherent dense core which is characterized by
$\langle\sigma_\mathrm{nth}/c_\mathrm{s}\rangle\lesssim 1$ \citep{Goodman98}.

\end{enumerate}

\acknowledgments

We sincerely acknowledge the anonymous referee whose
comments significantly helped to improve our data presentation
and the quality of the discussion.
R. S. F. gratefully acknowledges T. Sawada, S. Takahashi, A. Higuchi, and N. Kuno
for their generous help during the observations,
K. Schubert and J. Noumaru for their help in computing,
and P. T. P. Ho, N. Hirano, M. Taffalla, A. Hacar, J. M. Torrelles, 
R. Cesaroni, K. Tomisaka, M. N. Machida, T. Inoue, S. Inutsuka, and Y. Fukui for fruitful discussion.
R. S. F. also acknowledges R. Kawabe, M. Hayashi, H. Takami, N. Arimoto, 
T. Usuda, H. Araki, and Y. Takaishi
for their continuous support and encouragement.
The authors sincerely thank Daniele Galli for a critical reading of the 
manuscript at the last stage of preparation. 
This work was partially supported by Grant-in-Aids 
from the Ministry of Education, Culture, Sports, 
Science and Technology of Japan
(No.\,20740113), 
International Exchange Support Program of 
Foundation for Promotion of Astronomy (Tenmon Zaidan) of Japan, 
the JSPS Institutional Program for Young Researcher Overseas Visits (Wakate Haken), 
the NAOJ grants for domestic universities and institutes (Daigaku Shien), and
AWA Support Center at the University of Tokushima.

{\it Facilities:} \facility{Nobeyama 45\,m telescope}

\appendix
\section{Optical Depth and Excitation Temperature of the \tCO\ $J=1-0$ Line Emission}
\label{as:analysis}

\subsection{Radiative Transfer Equation}
\label{ass:rteq}
The main beam brightness temperature (\Tmb ) of molecular 
line emission emanated from a homogeneous gas cloud is written by,
\begin{equation}
\Tmb (v) = f\{J_\nu(\Tex)-J_\nu(\Tbg)\}[1-\exp\{-\tau (v)\}],
\label{eqn:Tmb}
\end{equation}
where $\tau (v)$ is the optical depth of the line 
as a function of radial velocity, $v$,
$f$ is the beam filling factor,
\Tex\ the excitation temperature of the line, and
\Tbg\ the temperature of the cosmic background radiation.
The function $J_\nu (T)$ is the radiation temperature defined by 
$\frac{h\nu}{k}\frac{1}{\exp(h\nu/kT)-1}$
where $\nu$ is the frequency of the line,
$T$ the gas temperature,
$k$ the Boltzmann constant, and $h$ the Planck constant.

\subsection{Estimate of Optical Depth and Excitation Temperature}
\label{ass:tauTex}

The optical depth and excitation temperature of the \tCO\ (1--0) line
can be estimated by applying Eq.(\ref{eqn:Tmb}) 
to a set of CO isotopologue spectra,
assuming the abundance ratios between \co\ and \tCO,
and between \tCO\ and \CeO.
For simplicity,
we assumed that the \co\ (1--0) line at rest frequency of 
115.271202\,GHz, the \tCO\ (1--0) line at 110.201353\,GHz, and 
the \CeO\ (1--0) line at 109.782173\,GHz are excited 
with a common \Tex\ at each spatial position; we also assumed 
$J_{\rm 115\,GHz}(T ) = J_{\rm 110\,GHz}(T)$ and $f = 1.0$.
Adopting the solar abundance ratios of $\alpha \equiv \rm [^{12}CO]/[^{13}CO ] = 89$ and 
$\beta \equiv \rm [^{13}CO]/[C^{18}O ] = 5.5$ \citep[e.g.,][]{Lang80, Garden91, Lequeux05},
we solved the following equations for $\tau$ and a quantity of
$A$ defined by $J_{\rm 110\,GHz}(\Tex ) - J_{\rm 110\,GHz}(\Tbg )$
at each pixel in the 3-dimensional (3D) space of 
right ascension offset ($\Delta\alpha$), 
declination offset ($\Delta\delta$), and LSR-velocity (\Vlsr);
\begin{equation}
\begin{array}{rl}
T_{12} = & A(1-e^{-\alpha\tau}), \\[6pt]
T_{13} = & A(1-e^{-\tau}), \\[6pt]
T_{18} = & A(1-e^{-\tau/\beta}), \\[6pt]
\end{array}
\label{eqn:solved}
\end{equation}
where $T_{12}$, $T_{13}$, and $T_{18}$ are the main beam brightness
temperatures of the \co, \tCO, and \CeO\ isotopologue lines, respectively.\par

\begin{figure}
\begin{center}
\includegraphics[angle=0,width=5.5cm]{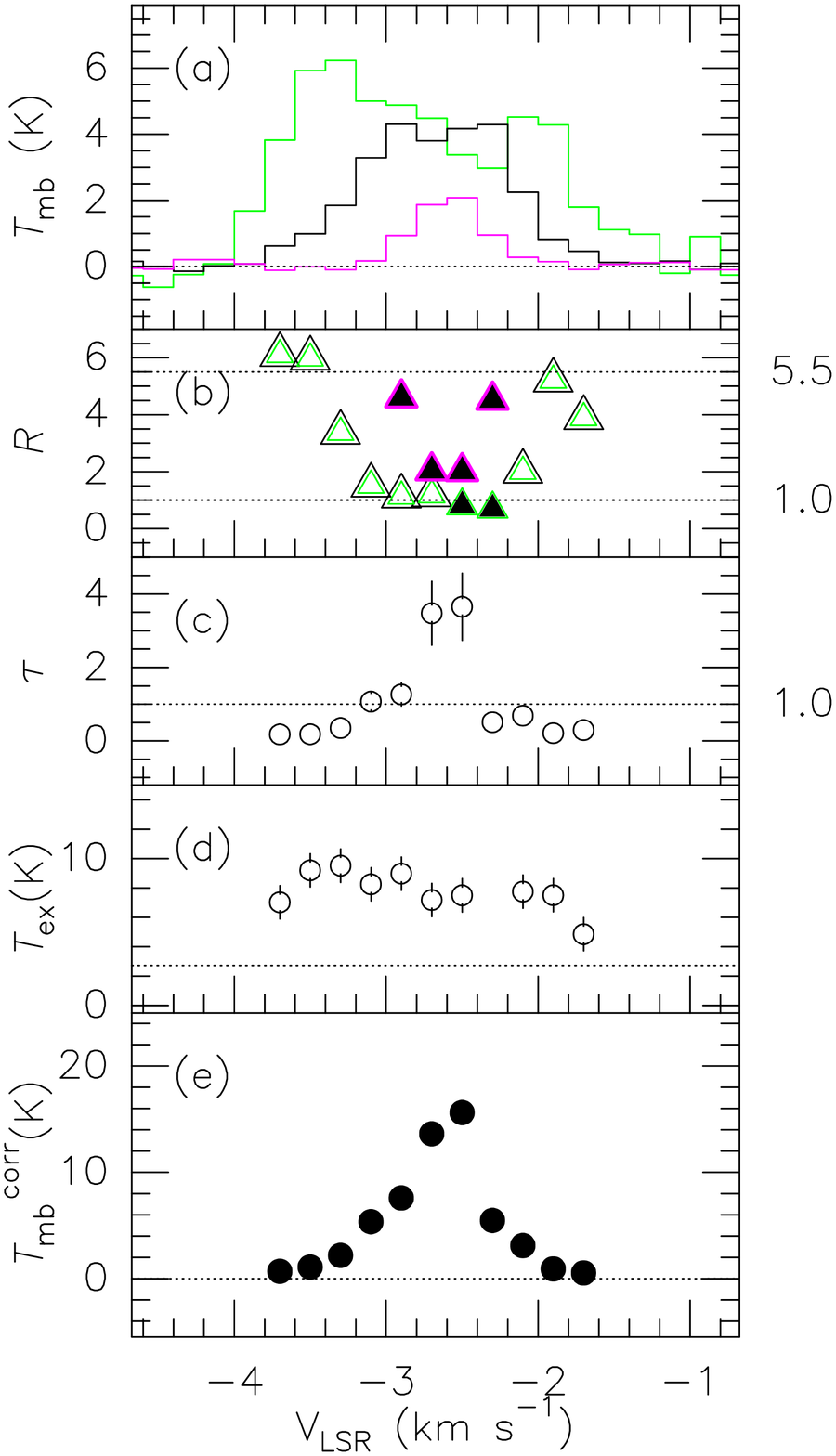}
\caption{{\bf corresponding to Appendix Fig.1. ~}
Example of our analysis 
to estimate 
the \tCO\ optical depth ($\tau$; panel c) and
excitation temperature (\Tex; panel d) for obtaining an
opacity-corrected \Tmb\ spectrum of the \tCO\ line (panel e).
The panel (a) shows the \co\ (green), \tCO\ (black), and \CeO\ (magenta) 
spectra in the \Tmb\ scale
towards the position of the 3\,mm continuum source (paper I).
The panel (b) represents intensity ratios
between the \co\ and \tCO\ lines
[$R_{12/13} = T_{\rm mb}(^{12}{\rm CO})/T_{\rm mb}(^{13}{\rm CO})$;
black-and-green triangles] and
between the \tCO\ and \CeO\ lines
[$R_{13/18} = T_{\rm mb}(^{13}{\rm CO})/T_{\rm mb}(^{18}{\rm CO})$;
magenta-and-black triangles].
See Appendix $\ref{ass:tauTex}$ for details.
\label{fig:spanalysis}}
\end{center}
\end{figure}

In the following, we describe our analysis by 
showing the CO spectra observed towards the \gf\ core center
as an example (Figure \ref{fig:spanalysis}a).
We started the analysis by calculating the line intensity ratios of
$R_{12/13} = T_{12} /T_{13}$ and $R_{13/18} = T_{13} /T_{18}$ at each velocity channel
(Figure \ref{fig:spanalysis}b).
On the basis of these ratios,
we divided all the velocity channels into the following four cases,
\begin{equation}
\begin{array}{rrl}
& {\rm case ~(i):} & 1.0 < R_{12/13} < \alpha {\rm ~and~} T_{18} < 3\Delta T^{18}_\mathrm{rms}, \\[6pt]
& {\rm case ~(ii):} & 1.0 < R_{13/18} < \beta {\rm ~and~} (R_{12/13} > \alpha {\rm ~or~} T_{12} < 3\Delta T^{12}_\mathrm{rms}), \\[6pt]
& {\rm case ~(iii):} & 1.0 < R_{12/13} < \alpha {\rm ~and~} 1.0 < R_{13/18} < \beta, \\[6pt]
\mathrm{and} & & \\[6pt]
& {\rm case ~(iv):} & (0.0 < R_{12/13} < 1.0 {\rm ~or~} R_{12/13} \geq \alpha) {\rm ~and~} (0.0 < R_{13/18} < 1.0 {\rm ~or~} R_{13/18} \geq \beta), \\[6pt]
\end{array}
\label{eqn:cases}
\end{equation}
where $\Delta T^{12}_\mathrm{rms}$ and $\Delta T^{18}_\mathrm{rms}$ denote the RMS noise levels
in \Tmb\ for the \co, and \CeO\ lines, respectively.
As seen in the above, we adopted detection threshold of 3$\sigma$. 
We did not consider the case that only the
\co\ line is detected, as we cannot define an intensity ratio.\par

Case (i) is the category where both the \co\ and \tCO\ lines are detected with $T_{12} > T_{13}$,
but the \CeO\ line does not show significant emission.
Case (ii) corresponds to the category that both the \tCO\ and \CeO\ lines are detected with
$T_{13} > T_{18}$, but the \co\ line is too strong or is not detected.
Case (ii) is often found near the velocity channels where 
the \co\ line suffers self-absorption and the \tCO\ one shows intense
emission (see Figure \ref{fig:spanalysis}a).
For Cases (i) and (ii),
we numerically solved the following equations for $\tau$:
\begin{equation}
\begin{array}{rcll}
R_{12/13} & = & \frac{1-e^{-\alpha\tau}}{1-e^{-\tau}} & {\rm for~~ case ~(i),} \\[6pt]
R_{13/18} & = & \frac{1-e^{-\tau}}{1-e^{-\tau/\beta}} & {\rm for~~ case ~(ii).} \\[6pt]
\end{array}
\label{eqn:ratio}
\end{equation}
We employed the bisection method with an accuracy of 0.001, 
and performed about 10 repetitions for many cases, 
as expected from $1/2^{10}\sim 0.001$.
After obtaining $\tau$, 
we calculated \Tex\ using Eq.(\ref{eqn:solved}).
In the 3D space of
($\Delta\alpha$, $\Delta\delta$, \Vlsr) where we have a total of
$72\times 72\times 150\,=\,7.775\times 10^5$ points,
we obtained solutions at 
34668 points for Case (i) and
784 points for Case (ii), which correspond to
4155 and 632 spatial positions in the ($\Delta\alpha$, $\Delta\delta$) coordinates, respectively.\par

Case (iii) is a category where 
all three lines 
are detected with an intensity 
order of $T_{12} > T_{13} > T_{18}$,
allowing us to utilize the maximum likelihood method on the basis of $\chi^2$.
We searched the best-fit values for $\tau$ and $A$ 
by minimizing the $\chi^2$ value defined by
\begin{equation}
\chi^2 = \sum_{i\,=12,\,13,\,18} 
\left(\frac{~~T_i-T_i^{\rm model}~~}{\Delta T^{i}_\mathrm{rms}}
\right)^2,
\label{eqn:chisq}
\end{equation}
where 
$T_{12}^{\rm model} \,=\, A (1-e^{-\alpha\tau})$,
$T_{13}^{\rm model} \,=\, A (1-e^{-\tau})$, and
$T_{18}^{\rm model} \,=\, A (1-e^{-\tau/\beta})$.
In order to give better initial guesses for the likelihood method analysis,
we solved Eq.(\ref{eqn:ratio}) for $\tau$ and $A$ 
by means of the bisection method
before performing $\chi^2$-fitting.
We calculated the \tCO\ optical depths from 
both the $R_{12/13}$ and $R_{13/18}$,
and adopted the mean value between them as the center of the ``searching area"
in the $A$-$\tau$ plane 
(see Figure \ref{fig:bestfit}) for finding 
the minimum $\chi^2$ value, $\chi^2_{\rm min}$.
In the ($\Delta\alpha$, $\Delta\delta$, \Vlsr) space,
there are 449 points that satisfy the criteria for Case (iii),
corresponding to 328 positions in the ($\Delta\alpha$, $\Delta\delta$) space.
For the 449 points we calculated that
the $\chi^2_{\rm min}$ values have a minimum of 0.10, 
a maximum of 8.0, 
a mean of 0.69, 
a standard deviation of 1.0, 
and a median of 0.29.
Since the obtained $\chi^2_\mathrm{min}$ 
is distributed around unity, 
our estimates of $\tau$ and $A$ are considered to be reasonable.\par

Case (iv) corresponds to the data sets where
either the \co\ or/and \tCO\ line shows self-absorption 
or their intensity ratios are inconsistent with those expected from the given abundance ratios.
We identified 20 points in the 3D space for such a case,
which may be negligible compared to the numbers of the solutions
obtained in the above three cases because
it corresponds to 1.7\% of all the analyzed data points.
Eleven out of the twenty points are found either at the
boundary LSR-velocity between the Components 1 and 2, 
\Vlsr\ $=\,-2.2$ \kms\ defined in \S\ref{s:results}, 
or at the adjacent channels.
Since such anomaly ratios do not allow us to estimate $\tau$ and \Tex,
we estimated $\tau$ from the observed $T_{13}$ value
using an empirical relation between $\tau$ and $T_{13}$ 
obtained from the Case (iii) analysis;~ 
$T_{13}\,=\,(3.95\pm0.02)[1-\exp\{-(1.68\pm0.03)\tau\}]$ for $\tau \geq 0.1$.
After obtaining $\tau$ by this way,
we calculated \Tex\ using Eq.(\ref{eqn:solved}).\par

In summary, we obtained a set of ($\tau$, \Tex) values at a total of 36622
points in the ($\Delta\alpha$, $\Delta\delta$, \Vlsr) space; these are 
34668 points from Case (i),
784 points from Case (ii),
1154 points from Case (iiI), and
20 points from Case (iv).\par

\subsection{Error Estimates for the Optical Depth and Excitation Temperature}
\label{ass:tauTexerr}
In this subsection, we describe error estimates for the optical
depth and excitation temperature of the \tCO\ emission.
Figures \ref{fig:tauerrmap} and \ref{fig:Texerrmap} show observed velocity
channel maps of the uncertainties in
the \tCO\ optical depth and excitation temperature, respectively.
Comparing Figures \ref{fig:taumap} with \ref{fig:tauerrmap} and
Figures \ref{fig:Texmap} with \ref{fig:Texerrmap}, 
we found that the resultant uncertainties in  the $\tau$ and \Tex\ are 15\% 
with respect to their values.

\subsubsection{Error Estimates for Cases (i) and (ii)}
Since the \tCO\ optical depth in Eq.(\ref{eqn:ratio}) 
cannot be explicitly solved,
we estimated the uncertainty in $\tau$ of $\Delta\tau_\mathrm{rms}$, as follows.
In the Case (ii), for example, the ratio $R_{13/18}$ is considered 
to be a function of $\tau$, and thus the following equation holds,
\begin{equation}
(\Delta R^\mathrm{rms}_{13/18})^2 = \left(\frac{\partial R_{13/18}}{\partial\tau}\right)^2(\Delta\tau_\mathrm{rms})^2.
\label{eqn:variance_R1}
\end{equation}
Furthermore, recall that the ratio $R_{13/18}$ is defined by $R_{13/18} = T_{13}/T_{18}$
where the two temperatures of $T_{13}$ and $T_{18}$ have the uncertainties 
of $\Delta T_{13}^\mathrm{rms}$ and $\Delta T^{18}_\mathrm{rms}$, respectively. Therefore, we have,
\begin{equation}
(\Delta R^\mathrm{rms}_{13/18})^2 = 
\left( \frac{\partial R_{13/18}}{\partial ~T_{13}}\right )^2 \left(\Delta T_{13}^\mathrm{rms}\right)^2 +
\left( \frac{\partial R_{13/18}}{\partial ~T_{18}}\right )^2 \left(\Delta T_{18}^\mathrm{rms}\right)^2.
\label{eqn:variance_R2}
\end{equation}
Here the covariance, 
$\langle\Delta T_{13}^\mathrm{rms}\cdot \Delta T_{18}^\mathrm{rms}\rangle $, 
was set to be zero because the two variables are independent of each other.
Combining Eqs.(\ref{eqn:variance_R1}) and (\ref{eqn:variance_R2}),
we can write the desired $\Delta\tau_\mathrm{rms}$ by
\begin{equation}
(\Delta\tau_\mathrm{rms})^2 = \left(\frac{\partial R_{13/18}}{\partial\tau}\right)^{-2}
\left\{
\left( \frac{\partial R_{13/18}}{\partial ~T_{13}}\right )^2 \left(\Delta T_{13}^\mathrm{rms}\right)^2 +
\left( \frac{\partial R_{13/18}}{\partial ~T_{18}}\right )^2 \left(\Delta T_{18}^\mathrm{rms}\right)^2
\right\}
\label{eqn:variance_tau}
\end{equation}
where 
$\frac{\partial}{\partial\tau}R_{13/18} =
\frac{e^{-\tau}}{1-\exp(-\tau/\beta)}-
\frac{e^{-\tau/\beta}(1-e^{-\tau})}{\beta\{1-\exp(-\tau/\beta)\}^2}$,
$\frac{\partial}{\partial ~T_{13}}R_{13/18}=\frac{1}{T_{18}}$ and
$\frac{\partial}{\partial ~T_{18}}R_{13/18}=-\frac{T_{13}}{T_{18}^2}$.
After obtaining $\Delta\tau_\mathrm{rms}$, 
we subsequently calculated the uncertainty in \Tex\ of 
$\Delta T^\mathrm{ex}_\mathrm{rms}$ through the relationship from Eq.(\ref{eqn:solved}) as follows,
\begin{equation}
\left(T_{13} - T_{13}^{\rm model}\right)^2 = 
\left( \frac{\partial ~T_{13}}{\partial \tau}\right )^2 \left(\Delta\tau_\mathrm{rms}\right )^2 +
\left( \frac{\partial ~T_{13}}{\partial T_{\rm ex}}\right )^2 \left(\Delta T_\mathrm{ex}^\mathrm{rms}\right )^2
\label{eqn:variance_Tex}
\end{equation}
where
$\frac{\partial T_{13}}{\partial \tau} = Ae^{-\tau}$ and
$\frac{\partial T_{13}}{\partial T_{\rm ex}} = \left(\frac{h\nu}{kT_{\rm ex}}\right)^2
\frac{\exp(h\nu/kT_{\rm ex})\{1-\exp(-\tau)\}}{\{\exp(h\nu/kT_{\rm ex})-1\}^2}$.
Replacing $\Delta\tau_\mathrm{rms}$ with Eq.(\ref{eqn:variance_tau}),
the desired $\Delta T_\mathrm{ex}^\mathrm{rms}$ can be explicitly written as a function of 
$T_{13}$, $T_{18}$, $\Delta T_{13}^\mathrm{rms}$, $\Delta T_{18}^\mathrm{rms}$, 
\Tex, and $\tau$.

\subsubsection{Error Estimates for Case (iii)}
\label{ass:caseiii_err}
\begin{figure}
\begin{center}
\includegraphics[angle=-90,scale=.25]{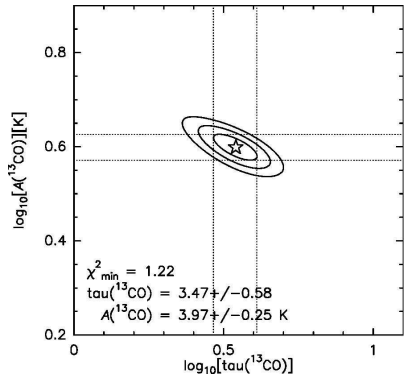}
\caption{{\bf corresponding to Appendix Fig.2. ~}
Contour map showing the ``confidence region boundaries" for the
case (iii) in our analysis as a function of the optical depth
and the quantity $A$, defined in Appendix \ref{ass:tauTex}, 
of the \tCO\ emission.
This plot represents an example for a velocity channel of
\Vlsr $=\,-2.6$  \kms\ for the spectra shown in Figure \ref{fig:spanalysis}.
The central star indicates the position where the minimum $\chi^2$, $\chi^2_{\rm min}$, 
of 1.22 is found.
The three contours show the levels
of $\chi^2_{\rm min}+2.30$, $\chi^2_{\rm min}+6.18$, and $\chi^2_{\rm min}+11.8$,
corresponding to the confidence levels of 68.3\%, 95.45\%, and 99.73\%, respectively.
The interval between the two dashed lines on each axis gives the 1\sgm\ error
for the corresponding parameter.
\label{fig:bestfit}}
\end{center}
\end{figure}

\begin{figure}
\begin{center}
\includegraphics[angle=0,width=8.4cm]{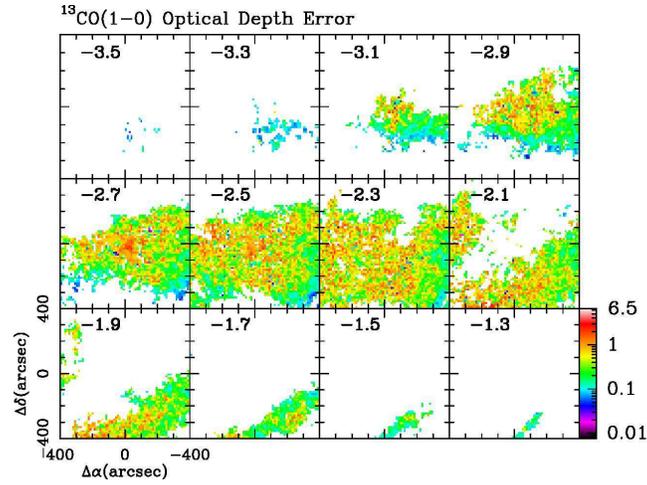}
\caption{{\bf corresponding to Appendix Fig.3. ~}
Velocity channel maps of the errors 
of the \tCO\ optical depth maps shown in Figure \ref{fig:taumap}.
The central LSR-velocity of each velocity channel in 
\kms\ is shown in each panel.
See Appendix \ref{ass:tauTexerr} for details.
\label{fig:tauerrmap}}
\end{center}
\end{figure}

\begin{figure}
\begin{center}
\includegraphics[angle=0,width=8.4cm]{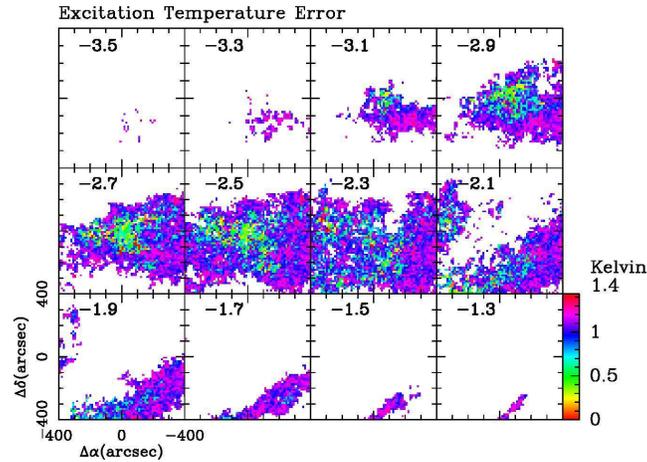}
\caption{{\bf corresponding to Appendix Fig.4. ~}Velocity channel maps of the errors 
of the \tCO\ excitation temperature maps shown in Figure \ref{fig:Texmap}.
Notice that the error is shown in linear scale.
See Appendix \ref{ass:tauTexerr} for details.
\label{fig:Texerrmap}}
\end{center}
\end{figure}

For the Case (iii) where we performed the maximum likelihood analysis,
the uncertainties of the best-fit parameters, 
i.e., the 68.3\% confidence intervals
for the two ``parameters of interest", are given by 
the projections onto
the $\tau$- and $A$-axes of the ``confidence region ellipse"
where the function $\chi^2(\tau, A)$ takes a value of $\chi^2_\mathrm{min} + 2.30$ (Press et al. 2010, p.815).
Figure \ref{fig:bestfit} presents an example showing such an error analysis
at the \Vlsr\ $=\,-2.6$ \kms\ channel of the spectra shown in Figure \ref{fig:spanalysis}a.
After numerically obtaining $\Delta A_\mathrm{rms}$,
we calculated $\Delta T_\mathrm{ex}^\mathrm{rms}$ through 
$\left(\Delta A_\mathrm{rms}\right)^2 = 
\left(\frac{\partial A}{\partial T_\mathrm{ex}}\right)^2 \left(\Delta T_\mathrm{ex}^\mathrm{rms}\right)^2$ where
$\frac{\partial A}{\partial T_{\rm ex}} = \left(\frac{h\nu}{kT_{\rm ex}}\right)^2
\frac{\exp(h\nu/kT_{\rm ex})\{1-\exp(-\tau)\}}{\{\exp(h\nu/kT_{\rm ex})-1\}^2}$.\par

\subsection{Error Estimates for Optical-Depth-Corrected Brightness Temperature}
\label{ass:errTmbcorr}
As described in \S\ref{s:analysis},
the optical-depth-corrected main beam brightness temperature, \Tmbcorr, 
is given by
\begin{equation}
\Tmbcorr = \frac{\tau\Tmb}{~~1-e^{-\tau}~~}.
\label{eqn:Tmbcorr}
\end{equation}
The uncertainty, \Tmbcorrms, at each velocity channel is calculated through
\begin{equation}
\Tmbcorrms = 
\left( \frac{\partial\Tmbcorr}{\partial T_{13}}\right ) \Delta T_{13}^\mathrm{rms} +
\left( \frac{\partial\Tmbcorr}{\partial\tau}\right ) \Delta\tau_\mathrm{rms}
\label{eqn:err_Tmbcorr}
\end{equation}
where
$\frac{~\partial\Tmbcorr~}{\partial \tau} = \frac{1}{1-e^{-\tau}}+\frac{\tau e^{-\tau}}{(1-e^{-\tau})^2}$ and
$\frac{~\partial\Tmbcorr~}{\partial T_{13}} = \frac{\tau}{1-e^{-\tau}}$.
Notice that the optical depth error in Eq.(\ref{eqn:err_Tmbcorr}), 
$\Delta\tau_\mathrm{rms}$, is given by Eq.(\ref{eqn:variance_tau}) 
for Cases (i) and (ii),
and is given through the error analysis described in 
Appendix \ref{ass:caseiii_err} for Case (iii).
One can see that
the mean \Tmbcorrms\ values, which are obtained by averaging \Tmbcorrms ($v$) values
along the velocity axis at each map position,
are enhanced towards the \gf\ dense cloud core.
This is most likely because $\Delta\tau_\mathrm{rms}$ values
showed relatively large ones towards the core (see Figure \ref{fig:tauerrmap}).
The mean $\langle\Tmbcorrms\rangle$ all over the observed area
was 1.5 K (median $=$ 1.4) with standard deviation of 0.66 K.

\section{Error Estimates for Spectral Momenta}
\label{as:errspmom}
In the following subsections, we present our error estimate for the spectral moment calculations
described in \S\ref{ss:mommaps}. The resultant error maps are shown in Figure \ref{fig:momerr}.

\subsection{Error Estimates for the Zeroth Moment}
\label{ass:zerothmom_err}

The zeroth spectral moment along the velocity axis,
corresponding to the integrated intensity in unit of K$\cdot$\kms,
is defined by $I = \int T(v)dv$, and is calculated by 
\begin{equation}
I = \sum_{i=1}^N \Tmbcorri\Delta v
\label{eqn:MomZero}
\end{equation}
over $N$ velocity channels where the emission exceeds 
the detection threshold.
Here \Tmbcorr\ is given by Eq.(\ref{eqn:Tmbcorr}), and $\Delta v$ denotes the width of each velocity channel.
Given the definition, the error of the zeroth moment is written by
\begin{equation}
\Delta I = \sqrt{\sum_{i=1}^N\left(\frac{\partial I}{\partial \Tmbcorri}\right)^2
\left(\Delta \Tmbcorri\right)^2},
\label{eqn:ErrMomZeroGen}
\end{equation}
and is calculated as,
\begin{equation}
\Delta I = \sqrt{\left(\Delta v\right)^2 \sum_{i=1}^N\left(\Delta \Tmbcorri\right)^2}.
\label{eqn:ErrMomZero}
\end{equation}

\begin{figure}
\begin{center}
\includegraphics[angle=0,scale=.22]{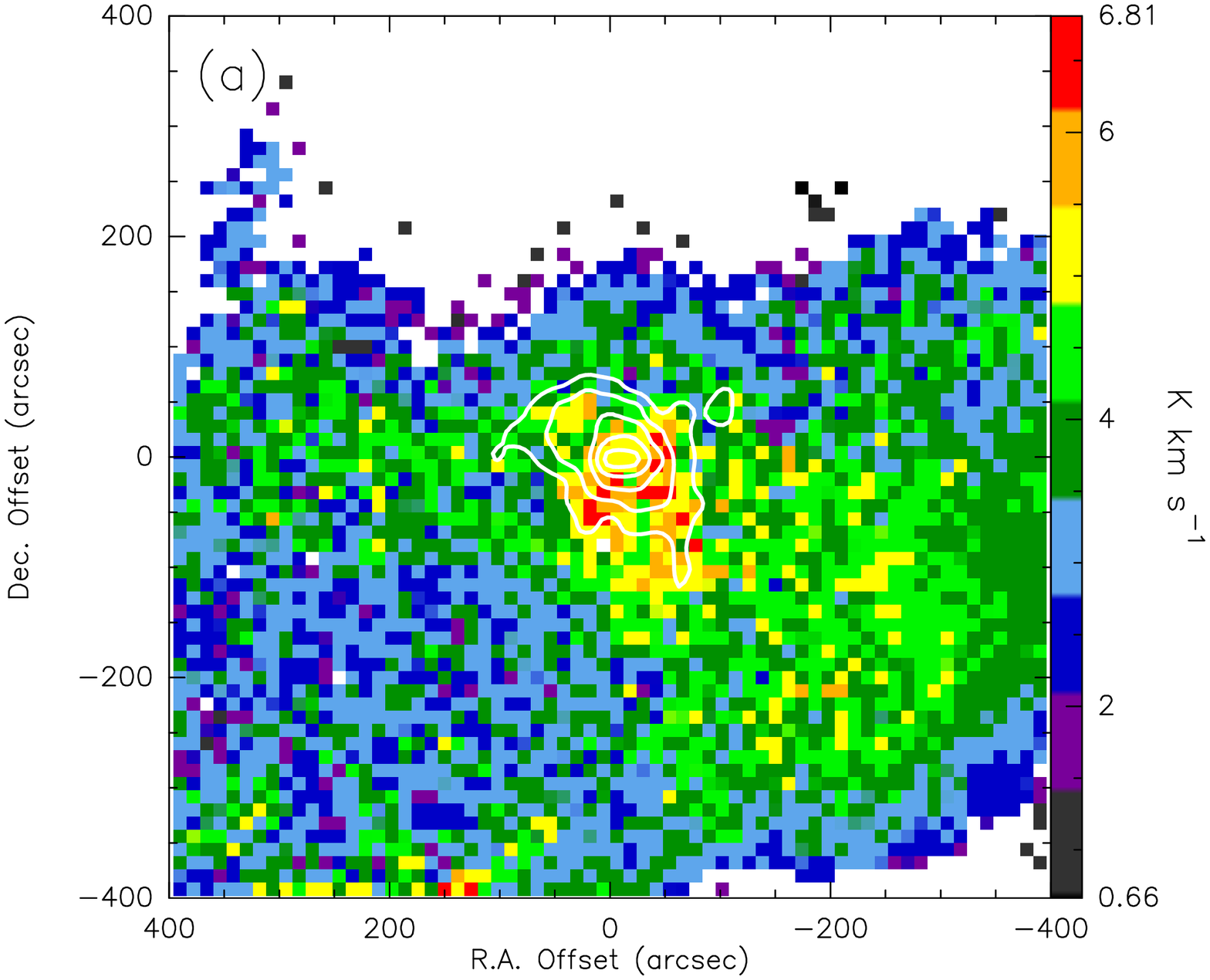}
\includegraphics[angle=0,scale=.22]{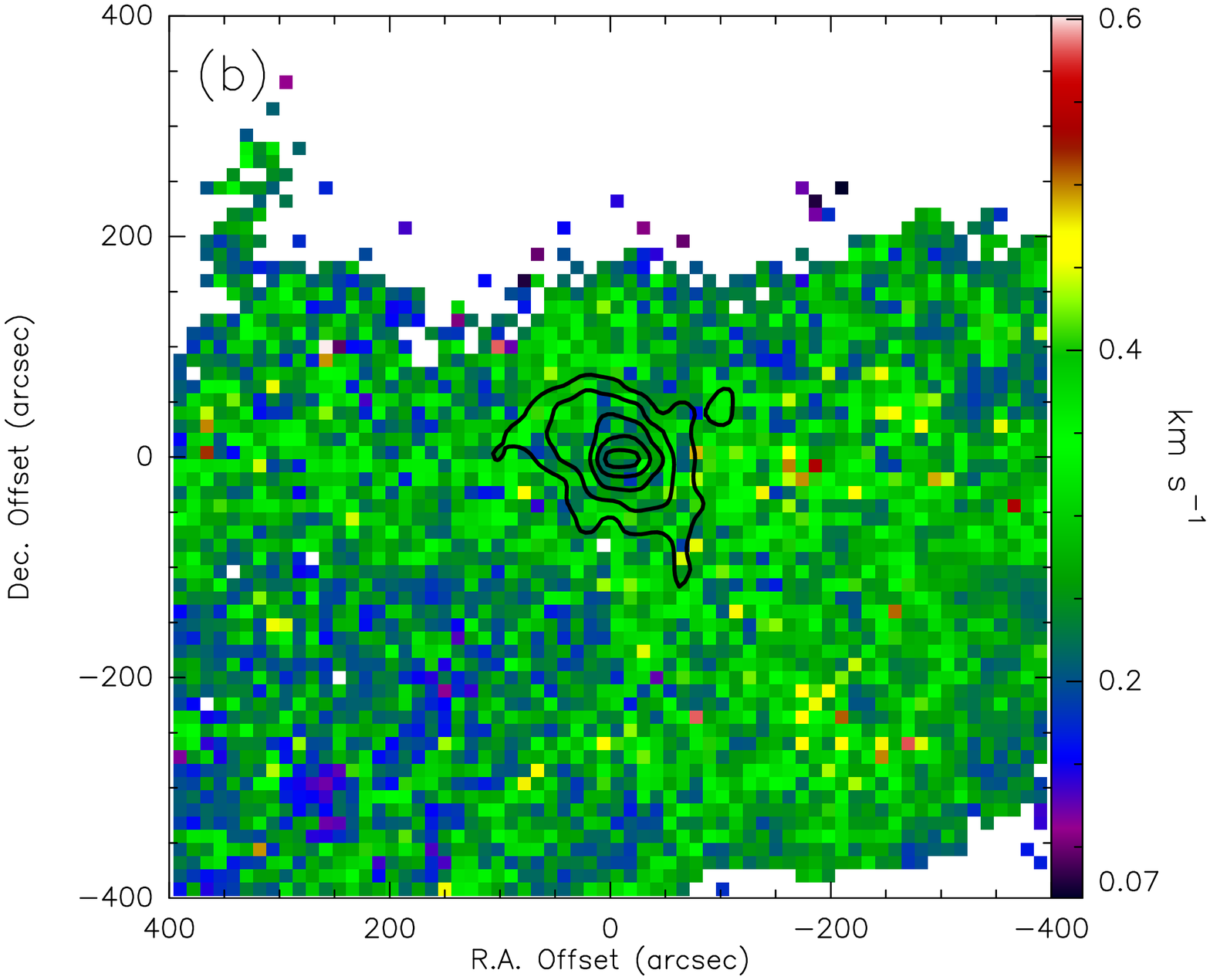}
\includegraphics[angle=0,scale=.22]{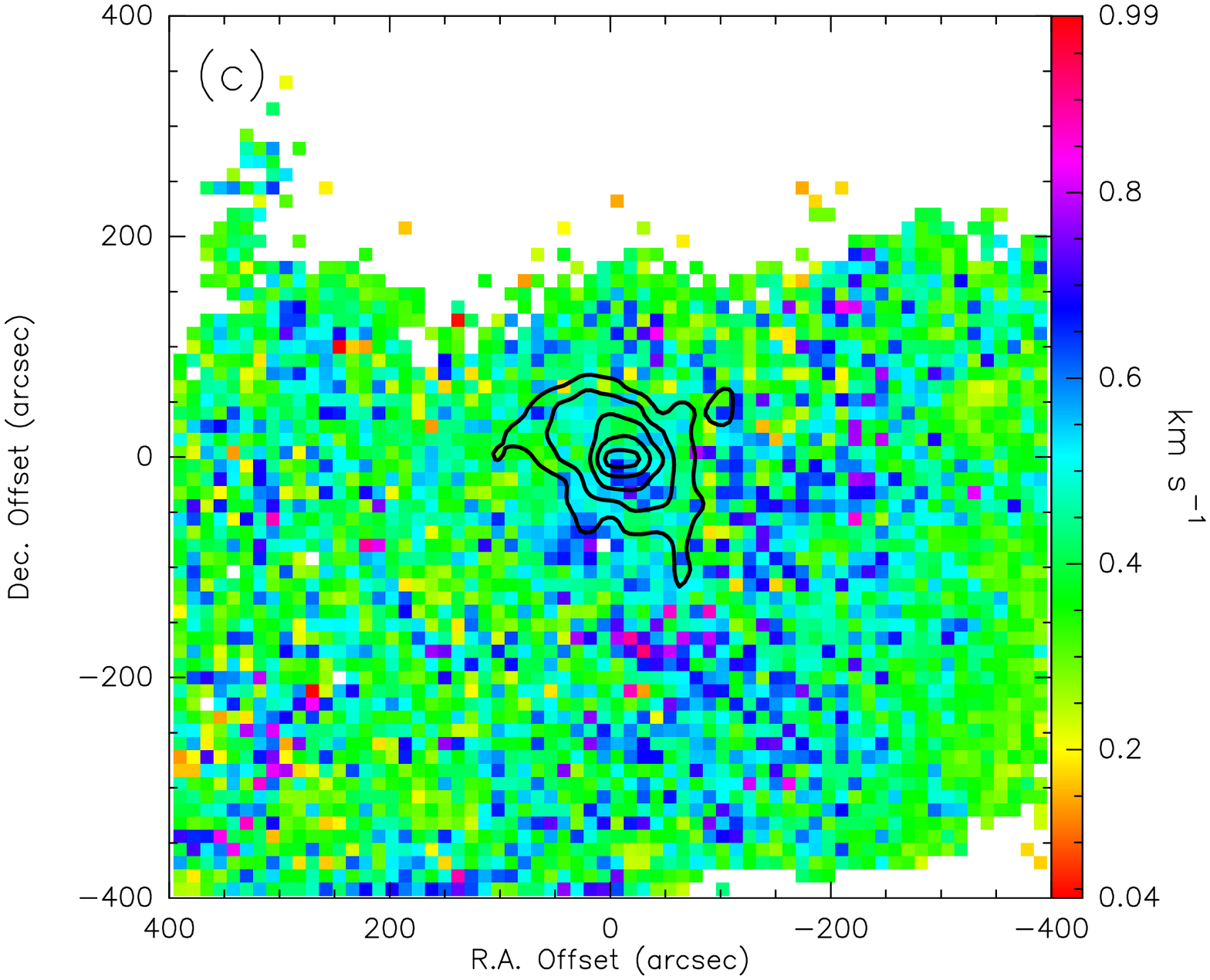}
\caption{{\bf corresponding to Appendix Fig.5. ~}
Uncertainty maps of 
(a) the total intensity in K \kms, 
(b) centroid velocity in \kms, and 
(c) velocity dispersion in \kms\ for the Component 1.
These maps are produced in the spectral moment analysis for 
the opacity-corrected \tCO\ (1--0) spectra, as described in Appendix \ref{as:errspmom}, 
and are shown in the same manner as in Figure \ref{fig:mommaps}.
\label{fig:momerr}}
\end{center}
\end{figure}

Figure \ref{fig:momerr}a presents the error map of the zeroth moment,
which is similar to the total integrated intensity map (Figure \ref{fig:totmaps}).
The mean $\Delta I$ calculated over whole the region
is 3.5 K$\cdot$\kms\ (median $=$ 3.5\, K$\cdot$\kms)
with standard deviation of 0.88\, K$\cdot$\kms.
Comparing Figure \ref{fig:mommaps}a with Figure \ref{fig:momerr}a,
the uncertainties are about 60\% with respect to the total intensities.\par

\subsection{Error Estimates for the First Moment}
\label{ass:firstmom_err}

The first moment, which gives an intensity-weighted mean velocity,
i.e., centroid velocity ($v_{\rm cent}$), is defined by 
$v_{\rm cent} = \frac{\int T(v)vdv}{\int T(v)dv}$.
This should be calculated by,
\begin{equation}
v_{\rm cent} = \frac{\sum_{i=1}^N v_i\Tmbcorri}{\sum_{i=1}^N\Tmbcorri}.
\label{eqn:MomFirst}
\end{equation}
Therefore the uncertainty of $v_{\rm cent}$ is given by,
\begin{equation}
\Delta v_{\rm cent} = \sqrt{\sum_{i=1}^N\left(\frac{\partial v_{\rm cent}}{\partial v_i}\right)^2\left(\Delta v_i\right)^2
+ \sum_{i=1}^N\left(\frac{\partial v_{\rm cent}}{\partial \Tmbcorri}\right)^2\left(\Delta \Tmbcorri\right)^2}.
\label{eqn:ErrMomFirstGen}
\end{equation}
Here $\Delta v_i$ is the uncertainty in the LSR-velocity at the $i$-th channel, 
which can be replaced by the velocity resolution of the spectrometer, $\Delta v_{\rm res}$.
Since $\Delta v_{\rm res}$ is generally represented by FWHM of the window function
of the spectrometer ($\Delta v_{\rm FWHM}$), one has to divide it by $\sqrt{8\ln 2}$ 
to obtain the uncertainty in velocity, i.e., standard deviation.
We thus obtained the equation below for computing 
error associated with the centroid velocity as,
\begin{equation}
\Delta v_{\rm cent} = \sqrt{\sum_{i=1}^N\left(\frac{\Tmbcorri}{I}\right)^2 \left(\frac{\Delta v_{\rm FWHM}}{\sqrt{8\ln 2}}\right)^2 + \sum_{i=1}^N\left(\frac{v_i-v_{\rm cent}}{I}\right)^2\left(\Delta \Tmbcorri \right)^2}.
\label{eqn:ErrMomFirst}
\end{equation}
Figure \ref{fig:momerr}b shows the error map of $\Delta v_{\rm cent}$; 
the mean value $\langle\Delta v_{\rm cent}\rangle$ over the observed region
is 0.27 \kms\ (median $=$ 0.26 \kms ) with standard deviation of 0.07 \kms.
The $\Delta v_{\rm cent}$ map seems fairly ``flat" compared with that of the $\Delta I$.
This is probably because estimating velocity width is principally sensitive to 
the dual terminal LSR-velocities at each spectrum, whereas
$\Delta I$ is generally sensitive to the peak value of the spectrum.

\subsection{Error Estimates for the Second Moment}
\label{ass:secondmom_err}
The second moment, which corresponds to the intensity-weighted velocity dispersion, 
is defined by
$\sigma = \sqrt{\frac{\int T(v)(v-v_{\rm cent})^2 dv}{\int T(v)dv}}$. 
In practice, we adopt a definition of
\begin{equation}
\sigma = \sqrt{\frac{\sum_{i=1}^N \left(v_i - v_{\rm cent}\right)^2 \Tmbcorri}{\sum_{i=1}^N\Tmbcorri}}.
\label{eqn:MomSecond}
\end{equation}
This leads to the uncertainty of the velocity dispersion as
\begin{equation}
\Delta\sigma = 
\sqrt{\sum_{i=1}^N \left(\frac{\partial\sigma}{\partial v_i} \right)^2 \left(\frac{\Delta v_{\rm FWHM}}{\sqrt{8\ln 2}}\right)^2 
+ \sum_{i=1}^N\left(\frac{\partial\sigma}{\partial \Tmbcorri}\right)^2\left(\Delta \Tmbcorri\right)^2},
\label{eqn:ErrMomSecond}
\end{equation}
where
$\frac{~\partial\sigma~}{\partial v_i} = \frac{1}{\sigma I^2}\left\{\left(v_i-v_{\rm cent}\right)\Tmbcorri I - \sum_{i=j}^N \left(v_j-v_{\rm cent}\right )\left(\Tmbcorri\right)^2\right\}$ and\\
$\frac{\partial\sigma}{~\partial \Tmbcorri ~} = \frac{1}{2\sigma I} \left\{\left(v_i-v_{\rm cent}\right)^2-3\sigma^2\right\}$.
Figure \ref{fig:momerr}c shows the $\Delta\sigma$ map whose mean over the
observed area, $\langle\Delta\sigma\rangle$, is 0.44 \kms\ 
(median $=$ 0.42 \kms ) with standard
deviation of 0.13 \kms. 
The $\Delta\sigma$ map appears to be insensitive to the uncertainty in 
determination of the terminal LSR-velocities.
This is probably because velocity dispersions are calculated over the velocity ranges having the 50\% level intensity
with respect to the peak.

\section{Calculation of the \tCO\ Column Density and Error Estimate}
\label{as:Ntot}
The column density of \tCO\ molecules ($N_\mathrm{^{13}CO}$) 
can be calculated from the total integrated intensity, i.e.,
the zeroth spectral moment,
of the transition from the $J+1$ to $J$ state, 
where $J$ is the rotational quantum number, by assuming LTE:
\begin{equation}
N_\mathrm{^{13}CO} = \frac{3k}{8\pi^3\mu^2B}~
\frac{(\meanTex + hB/3k)}{J+1}~
\frac{{\rm exp}[hBJ(J+1)/k\meanTex]}{1-{\rm exp}(-h\nu/k\meanTex)}~
\frac{\int T_\mathrm{mb}^\mathrm{corr}(v)dv}{J_\nu(\meanTex)-J_\nu(\Tbg)},
\label{eqn:Ntot}
\end{equation}
where 
$\mu$ is the dipole moment of \tCO\ (0.1101 debye),
$B$ the rotational constant (55101.0138 MHz)
(See e.g., Appendix B of paper I and references therein),
and \meanTex\ is the mean excitation temperature averaged along the LSR-velocity axis
at each pixel position (Figure \ref{fig:Texmap}).\par

Since Eq.(\ref{eqn:Ntot}) 
is a function of \meanTex\ and $I \left (= \int T_\mathrm{mb}^\mathrm{corr}(v)dv\right)$, 
its uncertainty is given by,
\begin{equation}
\Delta N_\mathrm{^{13}CO} =
\sqrt{
\left(\frac{\partial N_\mathrm{^{13}CO}}{\partial\meanTex }\right)^2 \left( \Delta\meanTex \right)^2 +
\left(\frac{\partial N_\mathrm{^{13}CO}}{\partial I}\right)^2 \left( \Delta I \right)^2
}.
\label{eqn:Ntot_err}
\end{equation}
In Appendix \ref{ass:tauTex} and \ref{ass:zerothmom_err}, 
we showed that \Tex\ has typical uncertainty of 15\%, 
while $I$ has much larger uncertainty of 60\%.
Therefore, ignoring the first term, Eq.(\ref{eqn:Ntot_err}) can be approximated by
\begin{equation}
\Delta N_\mathrm{^{13}CO} \simeq \frac{~\partial N_\mathrm{^{13}CO}~}{\partial I} \Delta I.
\label{eqn:Ntot_err1}
\end{equation}
Another method of estimating $\Delta N_\mathrm{^{13}CO}$ is as follows.
Since Eq.(\ref{eqn:Ntot_err}) becomes
a rather complicated formula given by a function of 
\meanTex\ and $I$,
we numerically estimated $\Delta N_\mathrm{^{13}CO}$
by changing \meanTex\ values by $\pm\Delta$\meanTex\ and $I$ values by $\pm\Delta I$.
Namely, we assumed 
that the real $\Delta N_\mathrm{tot}$ value would be 
smaller than an upper limit of $N_\mathrm{tot}^\mathrm{upper}$
which may be obtained when 
\meanTex\ $=$ \meanTex\ $+\Delta$\meanTex\ and $I = I+\Delta I$.
The $N_\mathrm{^{13}CO}^\mathrm{lower}$ is calculated with the similar fashion.
We thus adopted 
\begin{equation}
\Delta N_\mathrm{^{13}CO} \simeq \frac{1}{2}
(N_\mathrm{^{13}CO}^\mathrm{upper} - N_\mathrm{^{13}CO}^\mathrm{lower}).
\label{eqn:Ntot_err2}
\end{equation}
The resultant $\Delta N_\mathrm{^{13}CO}$ maps produced from the two methods
expressed by Eqs.(\ref{eqn:Ntot_err1}) and (\ref{eqn:Ntot_err2}) agree with each other within uncertainty of $\sim$\,2\%.
We therefore arbitrary adopted the map calculated by Eq.(\ref{eqn:Ntot_err2}) (Figure \ref{fig:Nmolerr}),
which is further used in the analysis discussed in \S\ref{ss:cf_scenario}.
Figure \ref{fig:Nmolerr} presents a summary of our uncertainty estimate.
The resultant $\Delta N_\mathrm{H_2}/N_\mathrm{H_2}$ map
has a mean uncertainty of
68\% with a standard deviation of 12\%.

\section{Data Selection Criteria for Producing the Maps of
Mean Excitation Temperature, 
Spectral Momenta, and Column Density}
\label{as:filters}
As summarized in the end of Appendix \ref{ass:tauTex}, 
we obtained the solutions at a total of 36622 positions in the 3D space of ($\Delta\alpha$, $\Delta\delta$, \Vlsr).
Using the results from error calculations in 
Appendices \ref{ass:tauTexerr} and \ref{ass:errTmbcorr}, 
we checked whether or not the 36622 solutions satisfy the following three conditions:
\begin{equation}
\begin{array}{cc}
& \tau > \Delta\tau,\\[6pt]
& T_\mathrm{ex} > \Delta T_\mathrm{ex},\\[6pt]
\mathrm{and} &\\[8pt]
& T_\mathrm{mb}^\mathrm{corr} > \Delta T_\mathrm{mb}^\mathrm{corr}. \\[2pt]
\end{array}
\label{eqn:filters4tauTexmaps}
\end{equation}
A total of 35916 points in the 3D space (98.1\%) passed the above test;
34664 points from Case (i),
783 points from Case (ii), and
449 points from Case (iii), 
corresponding to 2923, 304, and 328 positions in the ($\Delta\alpha$, $\Delta\delta$) space.
Furthermore, we discarded 21 points which do not satisfy a set of empirical conditions of
\begin{equation}
\begin{array}{cc}
& 0.1 < \tau < 10.0,\\[6pt]
& \Delta\tau < 0.01,\\[6pt]
\mathrm{and} &\\[8pt]
& 14.5 > T_\mathrm{ex}/\mathrm{K} > \Tbg - \langle\Delta T_\mathrm{ex}^\mathrm{rms}\rangle.\\[6pt]
\end{array}
\label{eqn:filters4tauTexmapsEmpirical}
\end{equation}
The limit values of 0.1 and 0.01 in the first and second conditions, respectively, 
were given by the accuracy in our calculations.
The numerical upper limits in the first and third conditions are 
fiducial values obtained from ``trials-and-errors".
Eventually, we obtained reliable solutions at 
a total of 35895 positions in the ($\Delta\alpha$, $\Delta\delta$, \Vlsr) space
(see Figures \ref{fig:taumap}, \ref{fig:Texmap}, \ref{fig:tauerrmap}, and \ref{fig:Texerrmap}),
yielding 3555 spatial positions in the ($\Delta\alpha$, $\Delta\delta$) space.
Namely, we eventually obtained a total of 3555 optical-depth-corrected \tCO\ spectra 
in the $72\times 72\,=$\,5184 observed positions.
Here we accepted the 20 points obtained from Case (iv) which are required
to pass only the first and third conditions in Eq.(\ref{eqn:filters4tauTexmapsEmpirical}).
This is because we were unable to estimate the uncertainties of 
$\Delta\tau$, $\Delta$\Tex, and $\Delta$\Tmbcorr\ in Case (iv) 
(see Appendix \ref{ass:tauTex}). 
Towards the central regions of the \Vlsr$=\,-2.1$ and $-1.9$ \kms\ panels in
Figures \ref{fig:taumap} and \ref{fig:Texmap}, 
we could not obtain reliable $\tau$ and \Tex\ solutions owing to the three conditions 
in Eq. (\ref{eqn:filters4tauTexmapsEmpirical}).\par

Subsequently we performed momentum analysis to the 3555 spectra,
as described in Appendix \ref{as:errspmom}.
For this purpose, one needs to define
an LSR-velocity range over which the spectrum moments are calculated.
We defined the range by two terminal velocities of 
$v_\mathrm{t, blue}$ and $v_\mathrm{t, red}$
where the \Tmbcorr $(v)$ spectra
first drop below the 3$\sigma$ levels 
in searching from the peak LSR-velocity toward the blue- and redward directions, respectively.
After several iterations, we found that the conditions of,
\begin{equation}
\begin{array}{cc}
& v_\mathrm{t, blue} <  v_\mathrm{cent} <  v_\mathrm{t, red},\\[6pt]
& |v_\mathrm{t, blue} - v_\mathrm{t, red}| > \sigma > \Delta\sigma > \frac{\Delta v_\mathrm{res}}{\sqrt{8\ln 2}},\\[6pt]
& I > \Delta I, \\[6pt]
& \langle\Tex\rangle > \Tbg,\\[6pt]
\mathrm{and} &\\[6pt]
& N_\mathrm{^{13}CO} > \Delta N_\mathrm{^{13}CO}\\[6pt]
\end{array}
\label{eqn:filters4spmaps}
\end{equation}
give the most reasonable results for 
Figures \ref{fig:meanTex}, \ref{fig:mommaps}, \ref{fig:Ncol}, \ref{fig:momerr} and \ref{fig:Nmolerr}.
In these figures, a total of 3093 spatial positions out of 3555 are presented.
Here $\frac{\Delta v_\mathrm{res}}{\sqrt{8\ln 2}}$ is calculated to be 0.0425 \kms\ (see \S\ref{s:obs}).

\section{Error Estimate of the Filament Scale Height}
\label{as:H_errors}

\begin{figure}
\begin{center}
\includegraphics[angle=0,scale=.33]{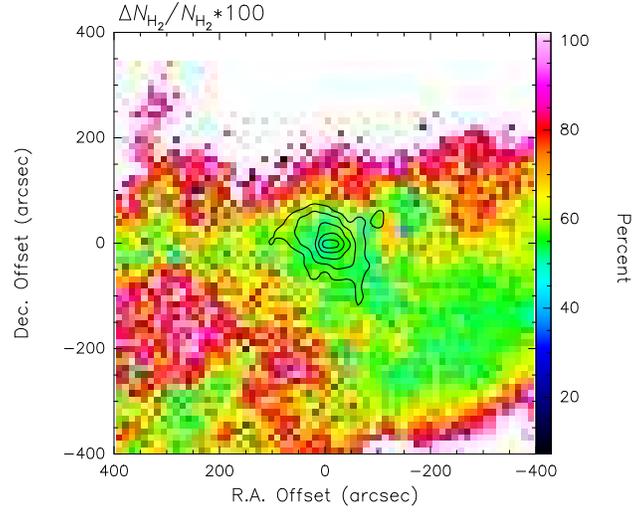}
\caption{{\bf corresponding to Appendix Fig.6. ~}Map of the uncertainty of  molecular hydrogen column density
defined by $\Delta N_\mathrm{H_2}/N_\mathrm{H_2}$.
See Appendix \ref{as:Ntot} for details.
\label{fig:Nmolerr}}
\end{center}
\end{figure}

\begin{figure}
\begin{center}
\includegraphics[angle=0,scale=.39]{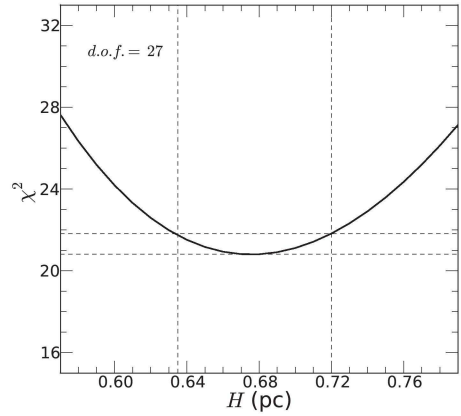}
\caption{{\bf corresponding to Appendix Fig.7. ~}
Plot of the scale height $H$ vs. $\chi^2$-values 
in the radial column density profile analysis shown
in Figure \ref{fig:NmolRadP}
(see also \S\ref{sss:Hestimate}); the degree of freedom ($d.o.f.$) is 27.
The lower and upper horizontal dashed lines present
the minimum $\chi^2$-value of $\chi^2_\mathrm{min}$
and the $\chi^2_\mathrm{min}+1.0$ value, respectively.
The interval between the two vertical dashed lines gives 
the 1$\sigma$ uncertainty in $H$ of $\Delta H\,=\,0.04$\,pc.
See Appendix \ref{as:H_errors}.
\label{fig:HvsChSq}}
\end{center}
\end{figure}

In our model analysis based on the Stodolkiewicz-Ostriker cylinder (\S\ref{sss:Hestimate}), 
the averaged radial column density profile shown in Figure \ref{fig:NmolRadP} was analyzed
considering the uncertainty of each column density,
$\Delta N_\mathrm{H_2}$, obtained from Figure \ref{fig:Nmolerr} (see also Appendix \ref{as:Ntot}). 
Then the uncertainty in the best-fit scale height $H$ of $\Delta H\, =\, 0.02$
was determined from the interval between the minimum $\chi^2$ value $\chi_\mathrm{min}^2$ (20.8) 
and $\chi_\mathrm{min}^2+1$, as shown in Figure \ref{fig:HvsChSq}.
Since the degree of freedom is 27,  
the reduced $\chi_\mathrm{min}^2$ value becomes 0.77. \par

In addition, we can analytically derive the uncertainty of $H$ from the following approximate equation of,
\begin{equation}
\Delta H\,\sim\,\frac{~\Delta{\chi}^2~}{\frac{~\partial\chi^2~}{\partial H}} \sim \frac{1}{~\frac{~N~}{H}~} \,=\, \frac{~H~}{N}~~\mathrm{for~}\,r<H,
\label{eqn:DeltaH}
\end{equation}
yielding $\Delta H\, =\, 0.02$ for the number of the data points of $N\, =\, 28$.

\begin{thebibliography}{}
\bibitem[Allen \& Shu(2000)]{Allen00} Allen, A., \& Shu, F.~H.\ 2000, \apj, 536, 368 
\bibitem[Andr{\'e} et al.(2009)]{Andre09} Andr{\'e}, P., Basu, S., \& Inutsuka, S.\ 2009, Structure Formation in Astrophysics, 254 
\bibitem[Andr{\'e} et al.(2010)]{Andre10} Andr{\'e}, P., Men'shchikov, A., Bontemps, S., et al.\ 2010, \aap, 518, L102 
\bibitem[Arzoumanian et al.(2011)]{Arzoumanian11} Arzoumanian, D., Andr{\'e}, P., Didelon, P., et al.\ 2011, \aap, 529, L6 
\bibitem[Barranco \& Goodman(1998)]{Barranco98} Barranco, J.~A., \& Goodman, A.~A.\ 1998, \apj, 504, 207
\bibitem[Bodenheimer(2011)]{Bodenheimer11} Bodenheimer, P.~H.\ 2011, Principles of Star Formation: , Astronomy and Astrophysics Library.~ISBN 978-3-642-15062-3.~Springer-Verlag Berlin Heidelberg, 2011,   
\bibitem[Bontemps et al.(1996)]{Bontemps96} Bontemps, S., Andre, P., Terebey, S., \& Cabrit, S.\ 1996, \aap, 311, 858 
\bibitem[Caselli et al.(2002)]{Caselli02} Caselli, P., Benson, P.~J., Myers, P.~C., \& Tafalla, M.\ 2002, \apj, 572, 238 
\bibitem[Ciardi et al.(1998)]{Ciardi98} Ciardi, D.~R., Woodward, C.~E., Clemens, D.~P., Harker, D.~E., \& Rudy, R.~J.\ 1998, \aj, 116, 349 
\bibitem[Ciardi et al.(2000)]{Ciardi00} Ciardi, D.~R., Woodward, C.~E., Clemens, D.~P., Harker, D.~E., \& Rudy, R.~J.\ 2000, \aj, 120, 393 
\bibitem[Chandrasekhar \& Fermi(1953a)]{CFmethod53} Chandrasekhar, S., \& Fermi, E.\ 1953a, \apj, 118, 113
\bibitem[Chandrasekhar \& Fermi(1953b)]{Chandrasekhar53} Chandrasekhar, S., \& Fermi, E.\ 1953b, \apj, 118, 116 
\bibitem[Chapman et al.(2011)]{Chapman11} Chapman, N.~L., Goldsmith, P.~F., Pineda, J.~L., et al.\ 2011, \apj, 741, 21 
\bibitem[Crutcher(2004)]{Crutcher04} Crutcher, R.~M.\ 2004, \apss, 292, 225 
\bibitem[Dickman(1978)]{Dickman78} Dickman, R.~L.\ 1978, \apjs, 37, 407 
\bibitem[Elmegreen(2007)]{Elmegreen07} Elmegreen, B.~G.\ 2007, \apj, 668, 1064 
\bibitem[Furuya et al.(2006)]{rsf06} Furuya, R.~S., Kitamura, Y., \& Shinnaga, H.\ 2006, \apj, 653, 1369 (paper I)
\bibitem[Furuya et al.(2008)]{rsf08} Furuya, R.~S., Kitamura, Y., \& Shinnaga, H.\ 2008, \pasj, 60, 421(paper II)
\bibitem[Furuya et al.(2009)]{rsf09} Furuya, R.~S., Kitamura, Y., \& Shinnaga, H.\ 2009, \apj, 692, 96 (paper III)
\bibitem[Furuya et al.(2014)]{rsf14} Furuya, R.~S., Kitamura, Y., \& Shinnaga, H.\ 2014, D. Stamatellos et al. (eds.), {\it The Labyrinth of Star Formation}, Astrophysics and Space and Science Proceedings 36, Springer International Publishing Switzerland, in press
\bibitem[Frerking et al.(1982)]{Frerking82} Frerking, M.~A., Langer, W.~D., \& Wilson, R.~W.\ 1982, \apj, 262, 590 
\bibitem[Garden et al.(1991)]{Garden91} Garden, R.~P., Hayashi, M., Hasegawa, T., Gatley, I., \& Kaifu, N.\ 1991, \apj, 374, 540 
\bibitem[Goldsmith et al.(2008)]{Goldsmith08} Goldsmith, P.~F., Heyer, M., Narayanan, G., et al.\ 2008, \apj, 680, 428
\bibitem[Goodman et al.(1998)]{Goodman98} Goodman, A.~A., Barranco, J.~A., Wilner, D.~J., \& Heyer, M.~H.\ 1998, \apj, 504, 223
\bibitem[Hartmann et al.(2001)]{Hartmann01} Hartmann, L., Ballesteros-Paredes, J., \& Bergin, E.~A.\ 2001, \apj, 562, 852   
\bibitem[Hacar \& Tafalla(2011)]{Hacar11} Hacar, A., \& Tafalla, M.\ 2011, \aap, 533, A34 
\bibitem[Hacar et al.(2013)]{Hacar13} Hacar, A., Tafalla, M., Kauffmann, J., \& Kov{\'a}cs, A.\ 2013, \aap, 554, A55 
\bibitem[Hanawa et al.(1993)]{Hanawa93} Hanawa, T., Nakamura, F., Matsumoto, T., et al.\ 1993, \apjl, 404, L83 
\bibitem[Heitsch et al.(2009)]{Heitsch09} Heitsch, F., Ballesteros-Paredes, J., \& Hartmann, L.\ 2009, \apj, 704, 1735 
\bibitem[Heyer et al.(2008)]{Heyer08} Heyer, M., Gong, H., Ostriker, E., \& Brunt, C.\ 2008, \apj, 680, 420 
\bibitem[Hunter(1977)]{Hunter77} Hunter, C.\ 1977, \apj, 218, 834 
\bibitem[Inutsuka \& Miyama(1992)]{Inutsuka92} Inutsuka, S.-I., \& Miyama, S.~M.\ 1992, \apj, 388, 392 
\bibitem[Inutsuka \& Miyama(1997)]{Inutsuka97} Inutsuka, S.-i., \& Miyama, S.~M.\ 1997, \apj, 480, 681 
\bibitem[Kainulainen et al.(2009)]{Kainulainen09} Kainulainen, J., Beuther, H., Henning, T., \& Plume, R.\ 2009, \aap, 508, L35 
\bibitem[Klessen et al.(2000)]{Klessen00} Klessen, R.~S.,  Heitsch, F., \& Mac Low, M.-M.\ 2000, \apj, 535, 887 
\bibitem[Klessen et al.(2005)]{Klessen05} Klessen, R.~S., Ballesteros-Paredes, J., V{\'a}zquez-Semadeni, E., \& Dur{\'a}n-Rojas, C.\ 2005, \apj, 620, 786 
\bibitem[Krumholz \& McKee(2005)]{Krumholz05} Krumholz, M.~R., \& McKee, C.~F.\ 2005, \apj, 630, 250 
\bibitem[Krumholz \& Tan(2007)]{Krumholz07} Krumholz, M.~R., \& Tan, J.~C.\ 2007, \apj, 654, 304 
\bibitem[Lang(1980)]{Lang80} Lang, K. R. 1980, Astrophysical Formulae. A Compendium for the Physicist and Astrophysicist., ed. K. R. Lang (Berlin: Springer), 563 
\bibitem[Larson(1969)]{Larson69} Larson, R.~B.\ 1969, \mnras, 145, 271 
\bibitem[Larson(1985)]{Larson85} Larson, R.~B.\ 1985, \mnras, 214, 379 
\bibitem[Le{\~a}o et al.(2013)]{Leao13} Le{\~a}o, M.~R.~M., de Gouveia Dal Pino, E.~M., Santos-Lima, R., \& Lazarian, A.\ 2013, \apj, 777, 46 
\bibitem[Lequeux(2005)]{Lequeux05} Lequeux, J. 2005, The Interstellar Medium (Berlin: Springer)
\bibitem[Mac Low(1999)]{MacLow99} Mac Low, M.-M.\ 1999, \apj, 524, 169
\bibitem[Mac Low \& Klessen(2004)]{MacLow04} Mac Low, M.-M., \& Klessen, R.~S.\ 2004, Reviews of Modern Physics, 76, 125 
\bibitem[Men'shchikov et al. (2010)]{Men10} Men'shchikov, A., Andr{\'e}, P., Didelon, P., et al.\ 2010, \aap, 518, L103 
\bibitem[Miyama et al.(1984)]{Miyama84} Miyama, S.~M., Hayashi, C., \& Narita, S.\ 1984, \apj, 279, 621 
\bibitem[Nagasawa(1987)]{Nagasawa87} Nagasawa, M.\ 1987, Progress of Theoretical Physics, 77, 635 
\bibitem[Nakamura \& Li(2005)]{Nakamura05} Nakamura, F., \& Li, Z.-Y.\ 2005, \apj, 631, 411
\bibitem[Nakamura \& Li(2007)]{Nakamura07} Nakamura, F., \& Li, Z.-Y.\ 2007, \apj, 662, 395 
\bibitem[Nakamura \& Li(2008)]{Nakamura08} Nakamura, F., \& Li, Z.-Y.\ 2008, \apj, 687, 354 
\bibitem[Nutter et al.(2008)]{Nutter08} Nutter, D., Kirk, J.~M., Stamatellos, D., \& Ward-Thompson, D.\ 2008, \mnras, 384, 755 
\bibitem[Ossenkopf \& Mac Low(2002)]{Ossenkopf02} Ossenkopf, V., \& Mac Low, M.-M.\ 2002, \aap, 390, 307 
\bibitem[Ostriker(1964)]{Ostriker64} Ostriker, J.\ 1964, \apj, 140, 1056 
\bibitem[Ostriker et al.(2001)]{Ostriker01} Ostriker, E.~C., Stone, J.~M., \& Gammie, C.~F.\ 2001, \apj, 546, 980 
\bibitem[Padoan \& Nordlund(1999)]{Padoan99} Padoan, P., \& Nordlund, {\AA}.\ 1999, \apj, 526, 279 
\bibitem[Padoan et al.(2001)]{Padoan01} Padoan, P., Juvela, M., Goodman, A.~A., \& Nordlund, {\AA}.\ 2001, \apj, 553, 227 
\bibitem[Padoan \& Nordlund(2002)]{Padoan02} Padoan, P., \& Nordlund, {\AA}.\ 2002, \apj, 576, 870 
\bibitem[Palmeirim et al.(2013)]{Palmeirim13} Palmeirim, P., Andr{\'e}, P., Kirk, J., et al.\ 2013, \aap, 550, A38 
\bibitem[Palla \& Stahler(2002)]{Palla02} Palla, F., \& Stahler, S.~W.\ 2002, \apj, 581, 1194 
\bibitem[Penston(1969)]{Penston69} Penston, M.~V.\ 1969, \mnras, 144, 425 
\bibitem[Phillips et al.(1979)]{Phillips79} Phillips, T.~G., Huggins, P.~J., Wannier, P.~G., \& Scoville, N.~Z.\ 1979, \apj, 231, 720 
\bibitem[Pineda et al.(2010)]{Pineda10} Pineda, J.~E., Goodman, A.~A., Arce, H.~G., et al.\ 2010, \apjl, 712, L116 
\bibitem[Poidevin \& Bastien(2006)]{PB06} Poidevin, F., \& Bastien, P.\ 2006, \apj, 650, 945 
\bibitem[Plummer(1911)]{Plummer1911} Plummer, H.~C.\ 1911, \mnras, 71, 460 
\bibitem[Sawada et al.(2008)]{Sawada08} Sawada, T., et al.\ 2008, \pasj, 60, 445 
\bibitem[Schneider \& Elmegreen (1979)]{Schneider79} Schneider, S., \& Elmegreen, B.~G.\ 1979, \apjs, 41, 87 
\bibitem[Shu(1977)]{Shu77} Shu, F.~H.\ 1977, \apj, 214, 488 
\bibitem[Shu(1992)]{Shu92} Shu, F.~H.\ 1992, The physics of astrophysics.~Volume II: Gas dynamics., by Shu, F.~H..~ University Science Books, Mill Valley, CA (USA), 1992, 
\bibitem[Spitzer(1978)]{Spitzer78} Spitzer, L., Jr. 1978, Physical Processes in the Interstellar Medium (New York: Academic) 
\bibitem[Stahler \& Palla(2005)]{SP05} Stahler, S.~W., \& Palla, F.\ 2005, The Formation of Stars, by Steven W.~Stahler, Francesco Palla, pp.~865.~ISBN 3-527-40559-3.~Wiley-VCH , January 2005  
\bibitem[Stod{\'o}lkiewicz(1963)]{Stodolkiewicz63} Stod{\'o}lkiewicz, J.~S.\ 1963, Acta Astron., 13, 30 
\bibitem[V{\'a}zquez-Semadeni et al.(1996)]{Enrique96} V{\'a}zquez-Semadeni, E., Passot, T., \& Pouquet, A.\ 1996, \apj, 473, 881 
\bibitem[V{\'a}zquez-Semadeni et al.(2005)]{Enrique05} V{\'a}zquez-Semadeni, E., Kim, J., Shadmehri, M., \& Ballesteros-Paredes, J.\ 2005, \apj, 618, 344 
\bibitem[White(1977)]{White77} White, R.~E.\ 1977, \apj, 211, 744 
\end{thebibliography}
\end{document}